\documentclass[aps,prd,superscriptaddress,tightenlines,nofootinbib,floatfix,eqsecnum]{revtex4-1}

\usepackage{graphicx}
\usepackage{dcolumn} 
\usepackage{bm}
\usepackage{amssymb}  
\usepackage{standalone}
\usepackage{color}
\usepackage{slashed}
\usepackage{booktabs}
\usepackage{multirow}
\usepackage{amsmath}
\usepackage{stackrel}
\usepackage{rotating}
\usepackage{rotfloat}
\usepackage{slashbox}
\usepackage[geometry]{ifsym}
\usepackage{natbib}
\usepackage{hyperref}
\hypersetup{
    colorlinks=true,       % false: boxed links; true: colored links
    linkcolor=blue,          % color of internal links
    citecolor=blue,        % color of links to bibliography
    filecolor=blue,      % color of file links
    urlcolor=blue           % color of external links
}
\usepackage{amsmath}
\usepackage[utf8]{inputenc}

\definecolor{red}{rgb}{0.8,0.0,0.0}
\definecolor{green}{rgb}{0.0,0.6,0.0}
\definecolor{darkblue}{rgb}{0.0,0.1,0.7}
\definecolor{brown}{rgb}{0.6,0.1,0.0}
\definecolor{grey}{rgb}{0.6,0.6,0.6}
\definecolor{darkgreen}{rgb}{0.0, 0.545098, 0.0}
\definecolor{applegreen}{rgb}{0.55, 0.71, 0.0}
\definecolor{purple}{rgb}{0.5,0.0,0.5}
\definecolor{babypink} {rgb}{0.64, 0.44, 0.44}
\definecolor{orange}{rgb}{1.0,0.5,0.0}

\definecolor{DARKBLUE}{rgb}{0.0,0.1,0.7}

\allowdisplaybreaks

\newcommand{\ie}{\emph{i.e.}}

\newcommand{\bi}{\begin{itemize}}
\newcommand{\ei}{\end{itemize}}

\newcommand{\ben}{\begin{enumerate}}
\newcommand{\een}{\end{enumerate}} 

\newcommand{\bt}[1]{\begin{table}[tb]\begin{tabular}{#1} \hline\hline  \\[-1.0em]}
\newcommand{\et}[2]{\hline\hline \end{tabular} \caption{#1} \label{#2} \end{table}}

\newcommand{\be}{\begin{equation}}
\newcommand{\ee}{\end{equation}}

\newcommand{\bea}{\begin{eqnarray}}
\newcommand{\eea}{\end{eqnarray}}

\newcommand{\order}{\ensuremath{\text{O}}} %use consistent notation for "order", whatever we decide
\newcommand{\op}{\ensuremath{\mathcal{O}}}
\newcommand{\latop}{\ensuremath{O}}

\renewcommand{\Im}{\ensuremath{\mathop{\mathrm{Im}}}}
\renewcommand{\Re}{\ensuremath{\mathop{\mathrm{Re}}}}

\newcommand{\MSbar}{\ensuremath{\overline{\rm MS}}}
\newcommand{\MSbarNDR}{\ensuremath{\overline{\rm MS}\text{-NDR}}}

\renewcommand{\case}[2]{\ensuremath{{\textstyle\frac{#1}{#2}}}}

\newcommand{\ihalf}{\ensuremath{\case{i}{2}}}
\newcommand{\quarter}{\ensuremath{\case{1}{4}}}

\newcommand{\LamQCD}{\ensuremath{\Lambda_\text{QCD}}}  % QCD scale
\newcommand{\Lambdaqcd}{\ensuremath{\Lambda_{\text{QCD}}}}
\newcommand{\LambdaNP}{\ensuremath{\Lambda_\text{NP}}}  % new-physics scale

\newcommand{\ZVcc}{\ensuremath{Z_{V^4_{cc}}}}

\newcommand{\alV}{\alpha_V}        % strong coupling, V scheme

\begin{document}

\title{Short-distance matrix elements for $D^0$-meson mixing from $N_f=2+1$ lattice QCD}

\author{A.~Bazavov}
\affiliation{Department of Computational Mathematics, Science and Engineering,\\ Department of Physics and Astronomy,\\ Michigan State University, \\ East Lansing, Michigan 48824, USA}

\author{C.~Bernard} 
\affiliation{Department of Physics, Washington University, St.~Louis, Missouri, 63130, USA}

\author{C.~M.~Bouchard}
\affiliation{School of Physics and Astronomy, University of Glasgow, Glasgow, G12 8QQ, UK}

\author{C.~C.~Chang}\email{chiachang@lbl.gov}
\altaffiliation{Present address: Lawrence Berkeley National Laboratory, Berkeley, California, 94720, USA}
\affiliation{Department of Physics, University of Illinois, Urbana, Illinois, 61801, USA}
\affiliation{Fermi National Accelerator Laboratory, Batavia, Illinois, 60510, USA}

\author{C.~DeTar} 
\affiliation{Department of Physics and Astronomy, University of Utah, \\ Salt Lake City, Utah, 84112, USA}

\author{D.~Du}
\affiliation{Department of Physics, Syracuse University, Syracuse, New York, 13244, USA}

\author{A.~X.~El-Khadra}\email{axk@illinois.edu}
\affiliation{Department of Physics, University of Illinois, Urbana, Illinois, 61801, USA}
\affiliation{Fermi National Accelerator Laboratory, Batavia, Illinois, 60510, USA}

\author{E.~D.~Freeland}
\affiliation{Liberal Arts Department, School of the Art Institute of Chicago, \\ Chicago, Illinois, 60603, USA}

\author{E.~G\'amiz}
\affiliation{CAFPE and Departamento de F\'{\i}sica Te\'orica y del Cosmos, Universidad de Granada,
18071, Granada, Spain}

\author{Steven~Gottlieb}
\affiliation{Department of Physics, Indiana University, Bloomington, Indiana, 47405, USA}

\author{U.~M.~Heller}
\affiliation{American Physical Society, Ridge, New York, 11961, USA}

\author{A.~S.~Kronfeld}
\affiliation{Fermi National Accelerator Laboratory, Batavia, Illinois, 60510, USA}
\affiliation{Institute for Advanced Study, Technische Universit\"at M\"unchen, 85748 Garching, Germany}

\author{J.~Laiho}
\affiliation{Department of Physics, Syracuse University, Syracuse, New York, 13244, USA}

\author{P.~B.~Mackenzie}
\affiliation{Fermi National Accelerator Laboratory, Batavia, Illinois, 60510, USA}

\author{E.~T.~Neil}
\affiliation{Department of Physics, University of Colorado, Boulder, Colorado 80309, USA}
\affiliation{RIKEN-BNL Research Center, Brookhaven National Laboratory, \\ Upton, New York 11973, USA}

\author{J.~N.~Simone}
\affiliation{Fermi National Accelerator Laboratory, Batavia, Illinois, 60510, USA}

\author{R.~Sugar}
\affiliation{Department of Physics, University of California, Santa Barbara, California, 93016, USA}

\author{D.~Toussaint}
\affiliation{Department of Physics, University of Arizona, Tucson, Arizona, 85721, USA}

\author{R.~S.~\surname{Van de Water}}
\affiliation{Fermi National Accelerator Laboratory, Batavia, Illinois, 60510, USA}

\author{R.~Zhou}
\affiliation{Fermi National Accelerator Laboratory, Batavia, Illinois, 60510, USA}

\collaboration{Fermilab Lattice and MILC Collaborations}
\noaffiliation

\date{\today}

\begin{abstract}
We calculate in three-flavor lattice QCD the short-distance hadronic matrix elements of all five $\Delta C=2$ four-fermion operators that contribute to neutral $D$-meson mixing both in and beyond the Standard Model. We use the MILC Collaboration's $N_f = 2+1$ lattice gauge-field configurations generated with asqtad-improved staggered sea quarks. We also employ the asqtad action for the valence light quarks and use the clover action with the Fermilab interpretation for the charm quark. We analyze a large set of ensembles with pions as light as $M_\pi \approx 180$~MeV and lattice spacings as fine as $a\approx0.045$~fm, thereby enabling good control over the extrapolation to the physical pion mass and continuum limit. We obtain for the matrix elements in the $\MSbarNDR$ scheme using the choice of evanescent operators proposed by Beneke \emph{et al.}, evaluated at 3~GeV, $\langle D^0|\op_i|\bar{D}^0\rangle = \{0.0805(55)(16), -0.1561(70)(31), 0.0464(31)(9), 0.2747(129)(55), 0.1035(71)(21)\}~\text{GeV}^4$ ($i=1$--5). The errors shown are from statistics and lattice systematics, and the omission of charmed sea quarks, respectively. To illustrate the utility of our matrix-element results, we place bounds on the scale of CP-violating new physics in $D^0$~mixing, finding lower limits of about 10--50$\times 10^3$~TeV for couplings of \order(1). To enable our results to be employed in more sophisticated or model-specific phenomenological studies, we provide the correlations among our matrix-element results. For convenience, we also present numerical results in the other commonly used scheme of Buras, Misiak, and Urban.
\end{abstract}

\maketitle

\section{Introduction}

The mixing between neutral $K$, $D$, $B$, and $B_s$ mesons and their antiparticles is loop suppressed in the Standard Model and, therefore, provides a window into new physics. Both indirect $CP$ violation in neutral kaon system ($\epsilon_K$) and the $B^0_d$ and $B^0_s$-meson oscillation frequencies ($\Delta M_d$ and $\Delta M_s$) have been measured to the subpercent level~\cite{Olive:2016xmw,Amhis:2016xyh}. Although still not as precise as experiment, the Standard Model theory for kaon and $B_{(s)}$-meson mixing is also under good control, owing to recent lattice-QCD calculations of the relevant hadronic matrix elements for kaons~\cite{Durr:2011ap,Blum:2014tka,Carrasco:2015pra} and for neutral $B_{(s)}$ mesons~\cite{Bazavov:2016nty}. Neutral $D^0$-meson mixing remains the least understood of the four mixing processes, both theoretically and experimentally, but progress is being made on both sides.

\begin{figure}[b]
	\centering
    \includegraphics[width=0.7\textwidth]{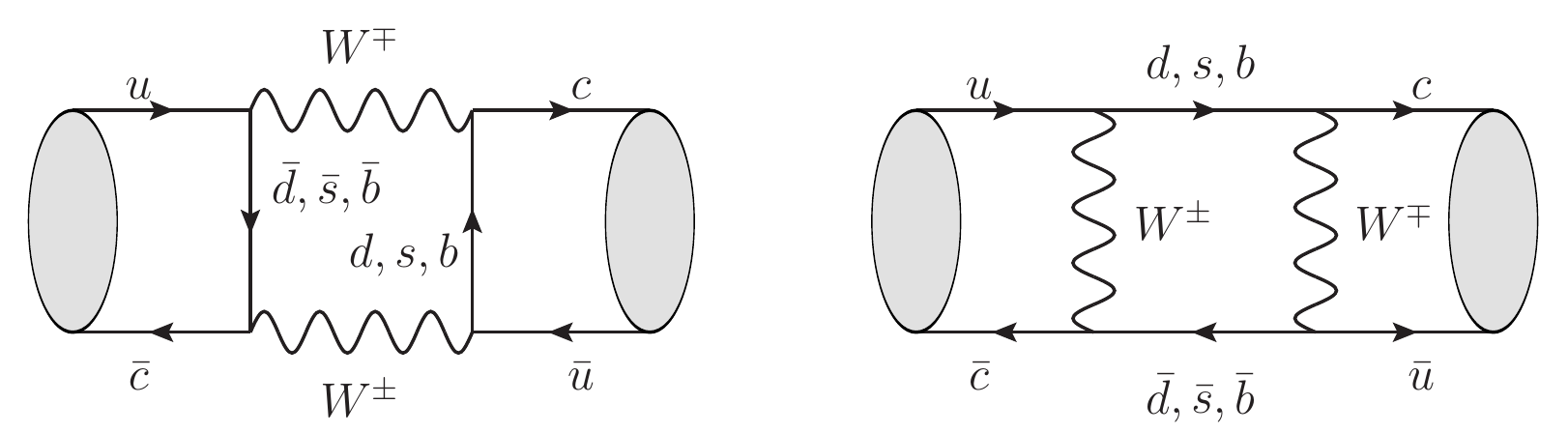}
	\caption{Feynman diagrams for the leading electroweak contributions to neutral $D$-meson mixing.}
	\label{fig:DmixBox}
\end{figure}

In the Standard Model, neutral $D$-meson mixing is mediated at leading order in the electroweak interactions by intermediate down-type quarks, as illustrated in Fig.~\ref{fig:DmixBox}.
Hence, it provides unique information on new-physics contributions to the down-quark sector that is complementary to that provided by kaons and $B_{(s)}$ mesons, in which mixing is mediated by up-type quarks. In particular, $D$-meson mixing does not receive any top-quark enhancements at leading order. Further, mixing via the bottom quark is Cabibbo suppressed by $|V_{ub}V^*_{cb}|^2/|V_{u(d,s)}V^*_{c(d,s)}|^2 \approx 0.2^8 \sim\text{few}~10^{-6}$ relative to mixing via down and strange quarks. $D$-meson oscillations are thus, to a good approximation, facilitated by only two generations of quarks, and
any observation of CP violation in $D$-meson mixing would be evidence of physics beyond the Standard Model (BSM).

At energies below the bottom quark mass, the electroweak box diagrams in Fig.~\ref{fig:DmixBox} give rise to short-distance contributions from $\Delta C=2$ interactions and long-distance contributions from two $\Delta C=1$ interactions. The hadronic matrix elements of the former can be calculated within lattice QCD using standard methods and are the focus of this work. QCD calculations of hadronic matrix elements of the latter must wait for the development of better tools; we comment on the prospects for such calculations in Sec.~\ref{sec:conclusions}. Even though these long-distance effects are a dominant contribution to neutral $D$-meson mixing in the Standard Model, knowledge of the matrix elements of all short-distance $\Delta C=2$ operators that arise in the Standard Model and beyond can provide useful BSM discrimination~\cite{Golowich:2009ii}, as described in more detail in Sec.~\ref{theory_background}.

In this paper, we provide a new calculation of the $\Delta C=2$ $D$-mixing matrix elements on the MILC Collaboration's $N_f=2+1$ gauge-field ensembles, which employ the $a^2$ tadpole-improved (asqtad) staggered action for the light quarks.
We analyze the same set of ensembles as in our previous calculation of the $B$-mixing matrix elements~\cite{Bazavov:2016nty}, and also follow an almost identical analysis procedure. Our results agree with previous $N_f=2$ and $N_f=2+1+1$ lattice-QCD calculations from the European Twisted Mass (ETM) Collaboration using twisted-mass fermions~\cite{Carrasco:2014uya,Carrasco:2015pra}, and have uncertainties commensurate with the projected experimental error on the mass difference $\Delta M$ [defined in Eq.~(\ref{eq:DeltaM})] from the LHCb and Belle~II experiments~\cite{Schwartz:2017gni,Morello:2016lnf,Bona:2016bno}.

This paper is organized as follows. First, in Sec.~\ref{theory_background}, we define the $\Delta C=2$ $D$-mixing matrix elements, and provide other important theoretical and phenomenological background. Next we present the set up of our numerical lattice-QCD calculation in Sec.~\ref{Numerical_simulation}, including the lattice actions, simulation parameters, and two- and three-point correlation functions. We describe our two- and three-point correlator fits used to obtain the bare lattice mixing matrix elements in Sec.~\ref{sec:correlator_analysis}, followed by how we match these results to a continuum renormalization scheme in Sec.~\ref{renormalization}. We adjust our matrix-element results to account for the slight difference between the simulation and physical charm-quark mass in Sec.~\ref{kappa_tuning_section} before extrapolating our results to the physical light-quark mass and continuum limit in Sec.~\ref{chiral_extrapolation}. In Sec.~\ref{error_analysis}, we discuss all sources of uncertainty in our calculation and provide a complete systematic error budget. Finally, we present our final $D$-mixing matrix-element results and discuss their phenomenological implications in Sec.~\ref{results}, and conclude with an outlook for the future in Sec.~\ref{sec:conclusions}. Several appendixes provide additional details. Appendix~\ref{app:r1acorrelations} lists the correlations between the ratio of scales $r_1/a$ used in this work. The priors used in the two-point and three-point correlator fits are given in Appendix~\ref{app:2pt3ptpriors}, and Appendix~\ref{app:MEcorrelations} lists the correlations between our matrix-element results. Final results are provided in double-precision as Supplementary Material~\cite{DmixSupplement}.

\section{Theoretical and Phenomenological Background}
\label{theory_background}

The time evolution of a neutral-meson system, such as $D^0$ and $\bar{D}^0$, can be described by a Schr\"odinger equation
\begin{align}
    i\frac{\partial}{\partial t}\begin{pmatrix} D^0 \\ \bar{D}^0 \end{pmatrix} =&
        \left(M-\frac{i}{2}\Gamma\right)\begin{pmatrix} D^0 \\ \bar{D}^0 \end{pmatrix} \nonumber \\
        =& \left(\begin{pmatrix} M_{11} & M_{12} \\ M^*_{12} & M_{11} \end{pmatrix}
        - \frac{i}{2}\begin{pmatrix} \Gamma_{11} & \Gamma_{12} \\ \Gamma^*_{12} & \Gamma_{11} \end{pmatrix}
        \right)
        \begin{pmatrix} D^0 \\ \bar{D}^0 \end{pmatrix},
    \label{eq:theory:Sch}
\end{align}
where the mass matrix $M$ and decay matrix $\Gamma$ are Hermitian. The off-diagonal term $M_{12}-\ihalf\Gamma_{12}$ mixes the flavor eigenstates into two mass (and width) eigenstates $D^0_1$ and $D^0_2$. Experiments have shown that CP violation in $D$-meson mixing is small, so the mass eigenstates are close to being CP eigenstates; usually \footnote{HFAG's charm web page~\cite{HFAG-charm-web} interchanges the labels $1\leftrightarrow2$ on the eigenstates but ends up with the same physical convention for $\Delta M$ and $\Delta\Gamma$ as in Refs.~\cite{Olive:2016xmw,Amhis:2016xyh}.} $|D^0_1\rangle$ is identified as the one with $\langle D^0_1|\text{CP}|D^0_1\rangle\approx+1$. Then, by convention, the mass and width differences between the two eigenstates are defined as~\cite{Olive:2016xmw,Amhis:2016xyh}
\begin{align}
    \Delta M     & \equiv      M_1 -      M_2, \label{eq:DeltaM} \\
    \Delta\Gamma & \equiv \Gamma_1 - \Gamma_2. \label{eq:DeltaG}
\end{align}
The signs of $\Delta M$ and $\Delta\Gamma$ are determined from experiment. The eigenvalue problem leads to the relation
\begin{equation}
    \Delta M -\frac{i}{2} \Delta \Gamma = \pm 2 Q
\end{equation}
for the mass and width differences, where the sign is the (near) CP of the heavier state, and
\begin{align}
    Q^2 &= |M_{12}|^2 - \quarter |\Gamma_{12}|^2 - i |M_{12}||\Gamma_{12}|\cos\phi_{12}
\end{align}
with 
\begin{equation}
    \phi_{12} \equiv \arg \frac{M_{12}}{\Gamma_{12}} .
    \label{eq:theory:phi}
\end{equation}
Given measurements of $\Delta M$, $\Delta\Gamma$, and CP asymmetries sensitive to $\phi_{12}$, these formulas determine $M_{12}$ and $\Gamma_{12}$ apart from a mutual, unphysical phase~\cite{Raz:2002ms,Ciuchini:2007cw,Ball:2007yz}.

The current measurements can be summarized as~\cite{Amhis:2016xyh}
\begin{align}
    y &\equiv \frac{\Delta \Gamma}{2\Gamma} = 0.61 \pm 0.07 \%,        \label{eq:y:Alan} \\
    x &\equiv \frac{\Delta M}{\Gamma}       = 0.41^{+0.14}_{-0.15} \%, \label{eq:x:Alan} 
\end{align}
and asymmetries consistent with zero. A fit then yields~\cite{Amhis:2016xyh}
\begin{align}
    \phi_{12} = -0.17^\circ \pm 1.8^\circ ,
\end{align}
and values for $y_{12}\equiv|\Gamma_{12}|/\Gamma$ and $x_{12}\equiv2|M_{12}|/\Gamma$ the same as those for
$y$ and~$x$. It is expected that future measurements by LHCb and Belle II will reduce the uncertainties on $\Delta M$
to $10\%$ or less~\cite{Schwartz:2017gni,Morello:2016lnf,Bona:2016bno}.

The interpretation of these results within the Standard Model starts with the leading electroweak contributions, shown in Fig.~\ref{fig:DmixBox}. The $W$-boson mass, $b$-quark mass, and the typical scale of QCD, $\LamQCD$, satisfy $m_W\gg m_b\gg\LamQCD$, so the mixing matrix in Eq.~(\ref{eq:theory:Sch}) can be expressed as (see, e.g., Ref.~\cite{Artuso:2008vf})
\begin{equation}
    2M_D \left( M_{12} - \frac{i}{2}\Gamma_{12} \right) =
        \langle D^0|\mathcal{H}^{\Delta C=2}|\bar{D}^0\rangle +
        \sum_n \frac{\langle D^0|\mathcal{H}^{\Delta C=1}|n\rangle
            \langle n|\mathcal{H}^{\Delta C=1}|\bar{D}^0\rangle}{M_D-E_n+i0^+},
    \label{eq:M12G12}
\end{equation}
where $\mathcal{H}^{\Delta C=1}$ ($\mathcal{H}^{\Delta C=2}$) is an effective Hamiltonian changing charm by 1 (2) unit(s), obtained by integrating out the $W$~boson and $b$~quark (and, in general, any other massive particles). The absorptive part of the second contribution is the origin of $\Gamma_{12}$; the first term and the dispersive part of the second both contribute to~$M_{12}$. The first contribution is local, stemming from processes in which all particles in Fig.~\ref{fig:DmixBox}
(and any in diagrams from new physics) propagate distances of $m_b^{-1}$ or less. In the second contribution, however, intermediate states, for example $K^+\pi^-$, can propagate a distance of order $\LamQCD^{-1}$. This second contribution is difficult to compute because several multi-hadron intermediate states enter the sum, but not so many that an appeal to quark-hadron duality is likely to be successful.

It is easy to see via the conventional parametrization of the CKM matrix that the Standard Model predicts the angle $\phi_{12}$ to be very small: the imaginary parts of $M_{12}$ and $\Gamma_{12}$, and hence their phases, come from parts of the box diagrams carrying one or two factors of $\sin\theta_{23}\sin\theta_{13}\sin\delta=1.4\times10^{-4}$, while the corresponding CKM factor in the real parts is $\sin\theta_{12}=0.225$. Because, with these conventions, both phases are small, so is the convention-independent difference~$\phi_{12}$. For the same reason, the Standard Model real part of $M_{12}$ stems mostly from the long-distance contribution, the sum in Eq.~(\ref{eq:M12G12}). An estimate from a dispersion relation based on heavy quark effective theory yields~\cite{Falk:2004wg}
\begin{align}
    y &\sim 10^{-2}, \label{eq:y:disp} \\
    x &\sim 10^{-3}~\text{to}~10^{-2}, \label{eq:x:disp}
\end{align}
which are compatible with the measurements, Eqs.~(\ref{eq:y:Alan}) and~(\ref{eq:x:Alan}).

Physics beyond the Standard Model could change $M_{12}$, $\Gamma_{12}$, or both. Many~\cite{Golowich:2009ii} extensions of the Standard Model alter only $\mathcal{H}^{\Delta C=2}$ and, hence, the magnitude and phase of~$M_{12}$, but not~$\Gamma_{12}$. In general, the $\Delta C=2$ effective Hamiltonian can be written
\begin{equation}
    \mathcal{H}^{\Delta C=2} =
        \sum_{i=1}^5 C_i \op_i + \sum_{i=1}^3 \tilde{C}_i \tilde{\op}_i,
    \label{H_OPE}
\end{equation}
where $C_i$ are the Wilson coefficients and the $\op_i$ are four-quark operators, given below. The determinations of $x_{12}$ and $\phi_{12}$ therefore can constrain the Wilson coefficients once the hadronic matrix elements $\langle D^0|\op_i|\bar{D}^0\rangle$ have been computed in (nonperturbative) QCD. Unfortunately, in view of the large range in Eq.~(\ref{eq:x:disp}), the constraint from $x$ will remain loose until new techniques for the long-distance term have been developed.

The operators in the $\Delta C=2$ effective Hamiltonian are
\begin{align}
    \op_1=&\bar{c}^\alpha \gamma_\mu L u^\alpha \bar{c}^\beta \gamma_\mu L u^\beta,  \label{eq:O1def} \\
    \op_2=&\bar{c}^\alpha L u^\alpha \bar{c}^\beta L u^\beta,  \\
    \op_3=&\bar{c}^\alpha L u^\beta  \bar{c}^\beta Lu^\alpha,  \\
    \op_4=&\bar{c}^\alpha L u^\alpha \bar{c}^\beta R u^\beta,  \\
    \op_5=&\bar{c}^\alpha L u^\beta  \bar{c}^\beta R u^\alpha, \label{intro_bsm} \\
    \tilde{\op}_1 =& \bar{c}^\alpha \gamma_\mu R u^\alpha \bar{c}^\beta \gamma_\mu R u^\beta, \\
    \tilde{\op}_2 =& \bar{c}^\alpha R u^\alpha \bar{c}^\beta R u^\beta,  \\ 
    \tilde{\op}_3 =& \bar{c}^\alpha R u^\beta  \bar{c}^\beta R u^\alpha,
\end{align}
where $\bar{c}$ and $u$ are the anticharm- and up-quark fields, with left and right Dirac projection matrices $L=\frac{1}{2}\left(1-\gamma_5\right)$ and $R=\frac{1}{2}\left(1+\gamma_5\right)$. The color indices are denoted $\alpha$ and $\beta$, and the spin indices are implied. All other Lorentz invariant four-quark operators can be reduced to this set by Fierz
rearrangement~\cite{Bouchard:2011yia}. Further, parity conservation in QCD implies $\langle D^0|\op_i|\bar{D}^0\rangle=\langle D^0|\tilde{\op}_i|\bar{D}^0\rangle$, $i=1,2,3$.

In summary, the five matrix elements $\langle D^0|\op_i|\bar{D}^0\rangle$ suffice to obtain
\begin{equation}
   2M_D M_{12}^\text{NP} = \sum_{i=1}^5 C_i^\text{NP}(\mu) \langle D^0|\op_i|\bar{D}^0\rangle(\mu),
    \label{eq:M12NP}
\end{equation}
where the $C_i^\text{NP}(\mu)$ are the Wilson coefficients of the new-physics model, subsuming $\tilde{C}_i^\text{NP}$, renormalized in the same scheme as the matrix elements. They can be calculated in lattice QCD with the same methods as for $B^0_{(s)}$-$\bar{B}^0_{(s)}$ mixing~\cite{Bazavov:2016nty}.

\begin{figure}
    \includegraphics[width=0.7\textwidth]{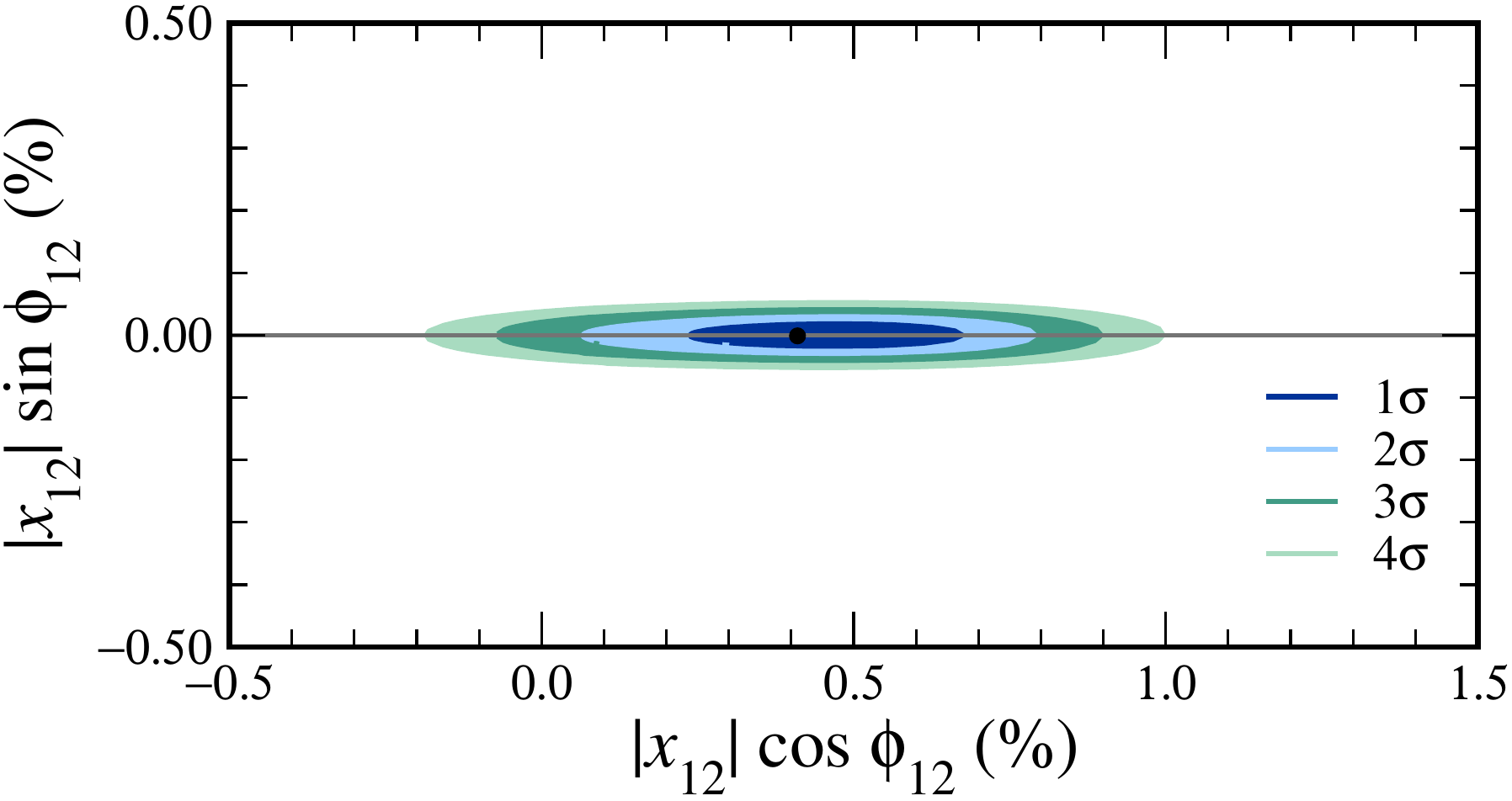} \hfill
    \caption{$|x_{12}|e^{i\phi_{12}}$ plotted as a complex number.
        The new-physics contribution should be added to the SM estimate (gray bar).
        If the total never falls inside the contours, then the new-physics model is ruled out.
        Otherwise, it remains viable.}
    \label{fig:x12-1x1}
\end{figure}

Given these matrix elements, Eq.~(\ref{eq:M12NP}) can be used to test models in which new physics does not change the phase of~$\Gamma_{12}$. \footnote{If the phase of $\Gamma_{12}$ were to change significantly, it would no longer be acceptable to treat the relative phase $\phi_{12}$ as the phase of $M_{12}$.} A~convenient way to do so is illustrated in Fig.~\ref{fig:x12-1x1}, which plots $|x_{12}|e^{i\phi_{12}}$ as a complex number.

The colored contours show the fit results to the experimental data, while the gray horizontal bar shows the range given in Eq.~(\ref{eq:x:disp}). The gray bar is close to $\sin\phi_{12}=0$ and extends well beyond the range displayed here. With the proviso that new physics does not change the phase of $\Gamma_{12}$, the new-physics calculation from Eq.~(\ref{eq:M12NP}) yields a complex number $x_{12}^\text{NP}\approx|x_{12}^\text{NP}|e^{i\phi_{12}^\text{NP}}$, which should be added to the gray bar. If the total $|x_{12}|e^{i\phi_{12}}$ lands entirely outside the contours, the new-physics model is ruled out.

\section{Lattice simulation}
\label{Numerical_simulation}

In this section, we provide details of the numerical lattice calculations of the $D$-mixing matrix elements. We begin in Sec.~\ref{subsec:EnsValParams} with an overview of the gauge-field configurations and valence light- and charm-quark actions, and then define the two- and three-point correlation functions calculated in Sec.~\ref{numericalsimulation_correlators}. We conclude in Sec.~\ref{subsec:Autocorr} with a discussion of statistical errors and autocorrelations.

\subsection{Gauge-field configurations, light-, and heavy-quark actions}
\label{subsec:EnsValParams}

\begin{figure}
	\centering
		\includegraphics[width=1.00\textwidth]{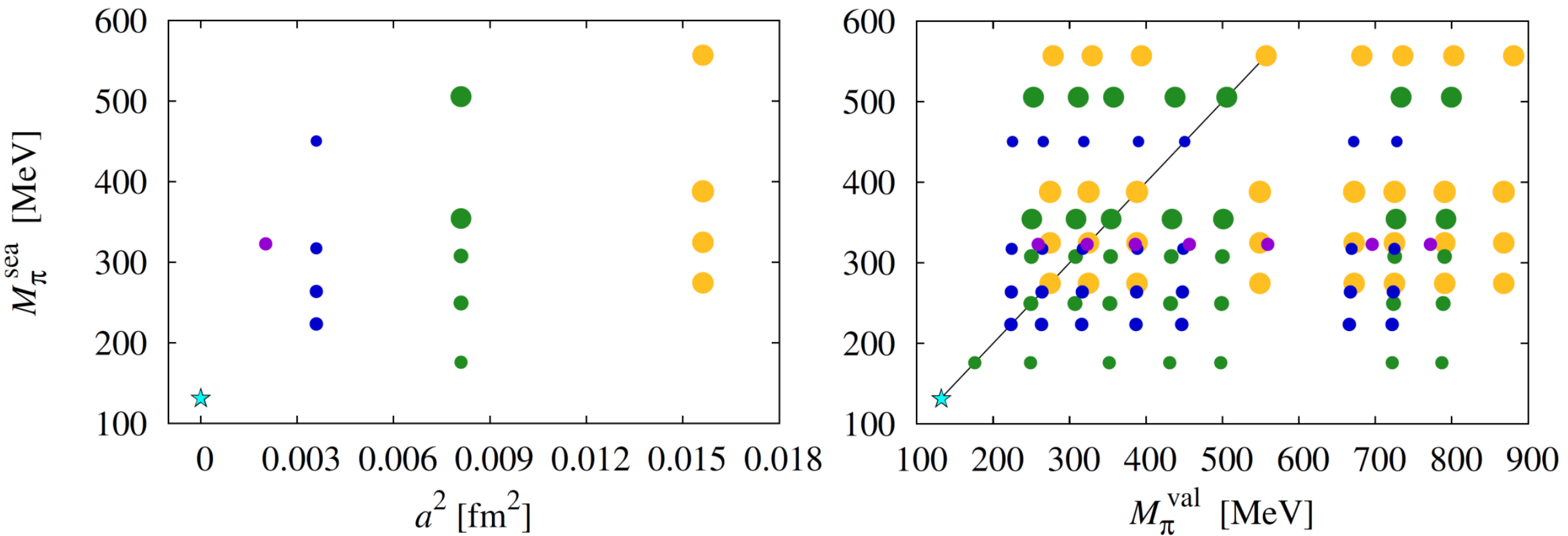}
	\caption{Left: MILC asqtad ensembles used in our analysis. The colors label the approximate lattice spacings $a \approx$ 0.12 fm (yellow), 0.09 fm (green), 0.06 fm (blue), and 0.045 fm (purple), while the symbol size is proportional to the number of configurations. The cyan star shows the physical point (continuum limit and physical pion mass.) Right: Valence pion masses used in our analysis. The diagonal line denotes full-QCD points with $M_\pi^{\rm val} = M_\pi^{\rm sea}$.} 
	\label{fig:data_visualization}
\end{figure}

We use isospin-symmetric gauge-field configurations generated by the MILC Collaboration~\cite{Bernard:2001av,Aubin:2004wf,Bazavov:2009bb} with $N_f=2+1$ dynamical quarks; the degenerate up and down sea-quark masses span a range of values from $(0.4-0.05)\times m_s$, permitting a controlled chiral extrapolation to the physical value, while the strange sea-quark mass is close to the physical value. These ensembles employ the asqtad-improved staggered quark action, and have light-quark discretization errors of $\order(\alpha_s a^2, a^4)$~\cite{Blum:1996uf,Orginos:1998ue,Lagae:1998pe,Lepage:1998vj,Orginos:1999cr,Bernard:1999xx}. The MILC asqtad ensembles were generated using the fourth-root procedure to yield a theory with one taste; both theoretical and numerical evidence indicate that the continuum limit of the rooted, staggered theory is indeed QCD~\cite{Durr:2005ax,Sharpe:2006re,Kronfeld:2007ek,Golterman:2008gt}. The gluons are simulated with the tadpole-improved L\"uscher-Weisz action and have discretization errors starting at $\order(\alpha_s a^2, a^4)$~\cite{Weisz:1982zw,Weisz:1983bn,Luscher:1984xn,Luscher:1985zq}. The lightest simulated pion mass $M_\pi = 177$~MeV is very close to the physical value; heavier pion masses up to $M_\pi = 555$~MeV are also included in the analysis to help guide the chiral extrapolation. Four lattice spacings ranging from $a \approx (0.12-0.045)$~fm provide good control over the extrapolation to the continuum. Figure~\ref{fig:data_visualization} visually summarizes the range of pion masses and lattice spacings analyzed in this work. All ensembles have sufficiently large spatial lattice volumes ($M_\pi L \gtrsim 3.8$) that finite-volume effects are expected to be at the subpercent level~\cite{Durr:2008zz}. All ensembles have at least 500 configurations, and many have over 2000 configurations, yielding small statistical uncertainties. The set of ensembles used and simulation parameters are shown in Table~\ref{tab:MILC_ensemble}.

The lattice spacing can be converted to $r_1$ units by multiplying with appropriate powers of the mass-independent ratio of scales $r_1/a$~\cite{Bazavov:2009bb} listed in Table~\ref{tab:MILC_ensemble}. Note that the ratios depend only on $\beta$, which through dimensional transmutation is linked to the lattice spacing. The scale $r_1$ is defined via the force $F(r)$ between two static quark by the condition $r_1^2 F(r_1) = 1.0$~\cite{Sommer:1993ce,Bernard:2000gd}. The relative scale is determined on each ensemble from the heavy-quark potential and then fit to a smoothing function in order to reduce sensitivity to lattice spacing~\cite{Bazavov:2009bb}, and can be obtained with tiny statistical errors. We convert our final matrix-element results to physical units using the absolute scale~\cite{Bazavov:2011aa},
\begin{equation}
	r_1=0.3117(22)~\text{fm},
\end{equation}
based on calculations of light pseudoscalar-meson decay constants from MILC~\cite{Bazavov:2009fk} and HPQCD~\cite{Davies:2009tsa}.  A detailed discussion of the smoothing procedure may be found in Sec.~IV~B of Ref.~\cite{Bazavov:2011aa}.

\begin{table*}
\caption{\label{tab:MILC_ensemble} 
Parameters of the gauge-field ensembles~\cite{Bazavov:2009bb}.  From left to right are shown the precise input value of $\beta$ which is related to the gauge coupling $\beta=10/g^2$, the approximate lattice spacing $a$, the mass-independent ratio of the scale and lattice spacing $r_1/a$, the simulated light-to-strange sea-quark mass ratio $am^\prime_l/{am^\prime_s}$, the lattice volume $(L/a)^3 \times (T/a)$, the taste-Goldstone pion mass $M_\pi$ and RMS sea-pion mass $M_\pi^{\text{RMS}}$, the spatial extent in units of the pion mass $M_{\pi}L$, and the number of configurations $N_{\text{conf}}$ in each ensemble. The primes on the light-quark masses distinguish the simulation values from the physical (unprimed) values $m_l = (m_u + m_d)/2$ and $m_s$.}.
\begin{tabular}{l@{\quad}c@{\quad}c@{\quad}c@{\quad}c@{\quad}c@{\quad}c@{\quad}c@{\quad}c}
\hline
\hline  
$\beta$ & $\approx a$~(fm) &  $r_1/a$ & $a{m}^\prime_{l}/am^\prime_{s}$  &  $(L/a)^3 \times (T/a)$  & $M_{\pi}$~(MeV) & $M_{\pi}^{\rm RMS}$~(MeV) & $M_{\pi} L$ & $N_{\rm conf}$ \\ \hline
6.790 & 0.12 & 2.8211(28) & 0.02/0.05          &$20^3 \times 64$   & 555 &  670  & 6.2 & 2052 \\ 
6.760 & 0.12 & 2.7386(33) & 0.01/0.05          &$20^3 \times 64$   & 389 &  538  & 4.5 & 2259 \\ 
6.760 & 0.12 & 2.7386(33) & 0.007/0.05        &$20^3 \times 64$   & 327 &  495  & 3.8 & 2110 \\ 
6.760 & 0.12 & 2.7386(33) & 0.005/0.05        &$24^3 \times 64$   & 277 &  464  & 3.8 & 2099 \\ \hline
7.110 & 0.09 & 3.8577(32) & 0.0124/0.031    &$28^3 \times 96$   & 494 &  549  & 5.8 & 1996 \\ 
7.090 &0.09 &  3.7887(34) & 0.0062/0.031    &$28^3 \times 96$   & 354 &  415  & 4.1 & 1931 \\ 
7.085 &0.09 &  3.7716(34) & 0.00465/0.031  &$32^3 \times 96$   & 306 &  375  & 4.1 & 984   \\ 
7.080 &0.09 &  3.7546(34) & 0.0031/0.031    &$40^3 \times 96$   & 250 &  330  & 4.2 & 1015 \\ 
7.075 &0.09 &  3.7376(34) & 0.00155/0.031  &$64^3 \times 96$   & 177 &  280  & 4.8 & 791   \\ \hline
7.480 & 0.06 & 5.399(17)   & 0.0072/0.018    &$48^3 \times 144$ & 450 &  467  & 6.3 & 593   \\
7.470 & 0.06 & 5.353(17)   & 0.0036/0.018    &$48^3 \times 144$ & 316 &  341  & 4.5 & 673   \\
7.465 & 0.06 & 5.330(16)   & 0.0025/0.018    &$56^3 \times 144$ & 264 &  293  & 4.4 & 801   \\
7.460 & 0.06 & 5.307(16)   & 0.0018/0.018    &$64^3 \times 144$ & 224 &  257  & 4.3 & 827   \\ \hline
7.810 & 0.045 & 7.208(54)   & 0.0028/0.014  &$64^3 \times 192$ & 324 &  332  & 4.6 & 801   \\
\hline\hline
\end{tabular}
\end{table*}

Although the neutral $D$-meson has a valence up quark, in our simulations we generate correlation functions with seven to eight different light valence-quark masses on each ensemble, using the same asqtad action as for the sea quarks. The valence-quark masses employed in our simulations are given in Table~\ref{tab:ValParams}, and correspond to pion masses from $M_\pi \approx 180$--880~MeV, enabling good control over the chiral extrapolation guided by SU(3) chiral perturbation theory. In Table~\ref{tab:ValParams} and throughout this work we denote the simulated valence light quark as $q$ with mass $am_q$, and reserve $m_u$ for the mass of the physical up quark.

For the heavy-quark propagators, we use the Sheikholeslami-Wohlert action with the Fermilab interpretation~\cite{Sheikholeslami:1985ij,ElKhadra:1996mp}. The couplings in the theory are smoothly bounded for arbitrary values of the heavy-quark mass $am_Q$. After tree-level matching to QCD through heavy-quark effective theory (HQET), discretization errors from the action are of $\order(\alpha_s a \Lambdaqcd, a^2 \Lambdaqcd^2)$. The bare charm-quark mass in the Fermilab action is parametrized by the hopping parameter $\kappa_c$; the values employed in our simulations are tabulated in Table~\ref{tab:ValParams}.

\begin{table*}
\centering
\caption{Parameters of the valence-quark propagators used in this work. Each ensemble is labeled by the approximate lattice spacing $a$ and the ratio of simulated light to strange sea-quark masses, $a{m}^\prime_{l}/am^\prime_{s}$. The same valence light-quark masses $am_q$ are used on all ensembles with the same approximate lattice spacing except on the
$a\approx 0.09$~fm, $a{m}^\prime_{l}/am^\prime_{s} = 0.00155/0.031$ ensemble. The next three columns list the parameters of the simulated charm-quark propagators: the clover coefficient $c_{\rm SW}$~\cite{Sheikholeslami:1985ij}, the hopping parameter $\kappa_c^\prime$, and the rotation coefficient $d_{1c}^\prime$. The primes on the hopping parameter and rotation coefficient distinguish the simulation values from the physical (unprimed) values. The last column shows the number of time sources per configuration $N_{\rm src}$.}
\label{tab:ValParams}.
\begin{tabular}{l@{\quad}c@{\quad}c@{\quad}c@{\quad}c@{\quad}c@{\quad}c}
\hline
\hline  
$\approx a$~(fm)   &  $a{m}^\prime_{l}/am^\prime_{s}$  &  $am_q$ & $c_{\rm SW}$ & $\kappa^\prime_c$ & $d^\prime_{1c}$  & $N_{\rm src}$ \\ \hline
0.12 & 0.02/0.05 & & 1.525 & 0.1259 & 0.07776 & 4 \\ 
0.12 & 0.01/0.05 & \{0.005, 0.007, 0.01, 0.02, & 1.531 & 0.1254 & 0.07900 & 4 \\ 
0.12 & 0.007/0.05 & 0.03, 0.0349, 0.0415, 0.05\} & 1.530 & 0.1254  & 0.07907 & 4 \\ 
0.12 & 0.005/0.05 & & 1.530  & 0.1254 &  0.07908 & 4 \\ \hline
0.09 & 0.0124/0.031 & &  1.473  & 0.1277  & 0.06312 & 4 \\ 
0.09 & 0.0062/0.031 & \{0.0031, 0.0047, 0.0062, & 1.476  & 0.1276  & 0.06411  & 4 \\ 
0.09 & 0.00465/0.031 & 0.0093, 0.0124, 0.0261, 0.031\} & 1.477  & 0.1275  & 0.06417 &  4\\ 
0.09 & 0.0031/0.031 & & 1.478  & 0.1275  & 0.06431 &  4 \\ 
\multirow{2}{*}{0.09} & \multirow{2}{*}{0.00155/0.031} & \{0.00155, 0.0031, 0.0062,  & \multirow{2}{*}{1.4784}  & \multirow{2}{*}{0.1275} &  \multirow{2}{*}{0.06473} &  \multirow{2}{*}{4} \\ 
        &                       &    0.0093, 0.0124, 0.0261, 0.031\}                  &              &            &                &    \\ \hline
0.06 & 0.0072/0.018 & & 1.4276  & 0.1296  & 0.04846  & 4 \\
0.06 & 0.0036/0.018 &  \{0.0018, 0.0025, 0.0036, &  1.4287  & 0.1296  & 0.04869 & 8 \\
0.06 & 0.0025/0.018 &  0.0054, 0.0072, 0.016, 0.0188\} & 1.4293  & 0.1296  & 0.04932 & 4  \\
0.06 & 0.0018/0.018 & & 1.4298  & 0.1296  & 0.04937  & 4 \\ \hline
\multirow{2}{*}{0.045} & \multirow{2}{*}{0.0028/0.014}  & \{0.0018, 0.0028, 0.004, & \multirow{2}{*}{1.3943}  & \multirow{2}{*}{0.1310}  & \multirow{2}{*}{0.03842} & \multirow{2}{*}{4} \\
	 &			    & 0.0056, 0.0084, 0.013, 0.16\} &     &             &               & \\
\hline\hline
\end{tabular}
\end{table*}

\subsection{Lattice correlation functions}
\label{numericalsimulation_correlators}
\begin{figure}
	\centering
		\includegraphics[width=0.50\textwidth]{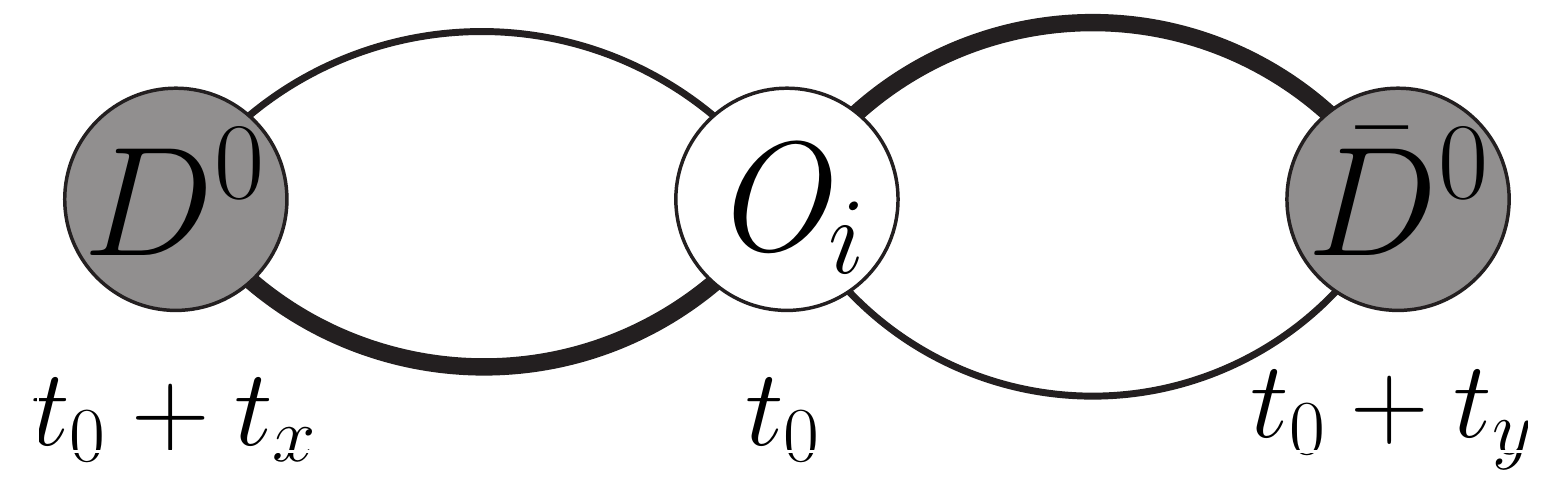}
	\caption{
	Lattice three-point correlation function $C_{\latop_i}(t_x, t_y)$. The thick and thin lines denote the charm- and light-quark propagators, respectively. The $D$-meson is created at time $t_0+t_x<t_0$, while the $\bar{D}$ meson is later annihilated at time $t_0+t_y>t_0$.	The four-quark operator $\latop_i$ is fixed at time $t_0$.}
	\label{fig:3pt_schematic}
\end{figure}

We calculate the two- and three-point correlation functions needed to obtain the matrix elements for neutral $D$-meson mixing using the same method as for $B$-meson mixing in Ref.~\cite{Bazavov:2012zs}. In particular, we first construct a specific combination of a light-quark propagator, heavy-quark propagator, and $\gamma_5$ with free spin and color indices known as an ``open-meson propagator''. We then obtain the $D$-meson two-point correlators from a single open-meson propagator contracted with $\gamma_5$, and obtain the three-point correlation functions for the five $\Delta C=2$ four-quark operators from combining two open-meson propagators contracted with the appropriate Dirac structures. Here we describe the light- and heavy-quark propagators used to construct the open-meson propagators.

The valence light-quark propagators are generated using the asqtad action. We then construct the naive field $\Upsilon(x)$ from the staggered field $\chi(x)$ following Refs.~\cite{Wingate:2002fh,Bernard:2013dfa},
\begin{equation}
    \Upsilon(x) = \Omega(x) \underline{\chi}_q(x),
    \label{eq:Upsilon}
\end{equation}
where $\Omega(x) = \gamma_1^{x_1} \gamma_2^{x_2} \gamma_3^{x_3} \gamma_4^{x_4}$ is the Kawamoto-Smit transformation and $\underline{\chi}$ denotes a vector of the four staggered copies. We remove tree-level, $\order(a)$ discretization errors from the four-fermion operator by rotating the heavy-quark field $\psi(x)$ following Ref.~\cite{Bazavov:2012zs},
\begin{equation}
\Psi(x) = \left[1+ad_{1}\boldsymbol{\gamma}\cdot \bm{D}\right]\psi(x).
\end{equation}
The simulation values of the rotation parameter $d_{1}$ are given in Table~\ref{tab:ValParams}.

We form the $D$-meson interpolating operator from the light naive field and rotated heavy field as
\begin{equation}
D^\dagger(\bm{x},t) = \sum_{\bm{x}'}\bar{\Upsilon}(\bm{x},t)S(\bm{x},\bm{x}')\gamma_5 \Psi_c(\bm{x}^\prime,t),
\label{eq:NumSim_Correlators_creation1S}
\end{equation}
where $S(\bm{x},\bm{x}')$ is a spatial smearing function. To improve the overlap with the $D$-meson ground state, we employ for the smearing function the 1S wavefunction of the Richardson potential~\cite{Richardson:1978bt,Menscher:2005kj}.

We obtain the $D$-meson mixing matrix elements from the zero-momentum three-point correlation functions,
\begin{equation}
    C_{\latop_i}(t_x, t_y) = 
    \sum_{\bm{x}, \bm{y}} \left< D^\dagger(\bm{y}, t_0+t_y)\latop_i(\bm{0}, t_0)D^\dagger(\bm{x},t_0+t_x) \right>,
    \label{eq:NumSim_Correlators_3pt}
\end{equation}
where the index $i=$1--5 labels the four-quark operator. We obtain the lattice operators, $\latop_i$, from the expressions 
(\ref{eq:O1def})--(\ref{intro_bsm}) for the continuum operators, $\op_i$, by replacing the $u$~field with $\Upsilon$ and the 
$\bar{c}$~field with $\bar{\Psi}$. As shown in Fig.~\ref{fig:3pt_schematic}, the four-quark operator location is fixed at time $t_0$, while the $D$- and $\bar{D}$-meson times $t_y$ and $t_x$ vary. The construction of the three-point correlation function, as a result of the staggered formulation, introduces mixing between wrong-spin taste-mixing terms as discussed in detail in Sec.~III~B of Ref.~\cite{Bazavov:2016nty}. This mixing is a lattice-discretization effect of $\order(a^2)$. The ``wrong-spin" contributions are included in the chiral-continuum fit function, and hence removed when we take the continuum limit.

In order to extract the hadronic matrix element from the amplitude of the three-point correlator, we normalize the three-point
correlators using the overlap function determined from the two-point correlator,
\begin{equation}
    C(t-t_0) = \sum_{\bm{x}}\left<D(\bm{x}, t)D^\dagger(\bm{0}, t_0)\right>.
    \label{eq:NumSim_Correlators_2pt}
\end{equation}

\subsection{Statistics and autocorrelations}
\label{subsec:Autocorr}

We take advantage of the large temporal extents of the MILC lattices by computing the two- and three-point correlation functions with four to eight evenly-spaced time sources per configuration. Prior to analysis, the correlators are shifted to a common $t_0=0$, then averaged. Because the correlators from different time sources are only weakly correlated, this reduces the statistical errors by approximately a factor of $\sqrt{N_{\rm src}}$.

Because the lattices were generated with periodic boundary conditions, we gain another approximate factor of two in statistics by folding the data along the temporal midpoint $T/2$ so as to include the backward propagating signal. For the two-point correlator, we identify $t$ with $T-t$ and use the range $0\leq t \leq T/2$ in our correlator fits. For the three-point correlator shown in Fig.~~\ref{fig:3pt_schematic}, we identify $-|t_{x}|$ with $-T-|t_{x}|$ and $t_{y}$ with $T-t_{y}$, and restrict our fit region to the $|t_{x}|<T/2$ and $t_{y}<T/2$ quadrant of the $|t_x|-t_y$ plane.

For the three-point correlation functions, we also exploit the parity symmetry of QCD to further increase statistics by averaging
the matrix elements of parity-equivalent operators. The operators $\mathcal{\tilde{O}}_{1,2,3}$ in Eqs.~(\ref{eq:O1def})--(\ref{intro_bsm}) differ from $\mathcal{O}_{1,2,3}$ by an interchange of $L\leftrightarrow R$, and thus transform into each other under parity inversion. Consequently, the lattice matrix elements of these operators are equal up to statistical fluctuations. We therefore generate data for both $\langle\mathcal{{O}}_{1,2,3}\rangle$ and $\langle\mathcal{\tilde{O}}_{1,2,3}\rangle$ and take their average in order to gain an approximate factor of two in statistics.

\begin{figure}
	\centering
		\includegraphics[width=1.0\textwidth]{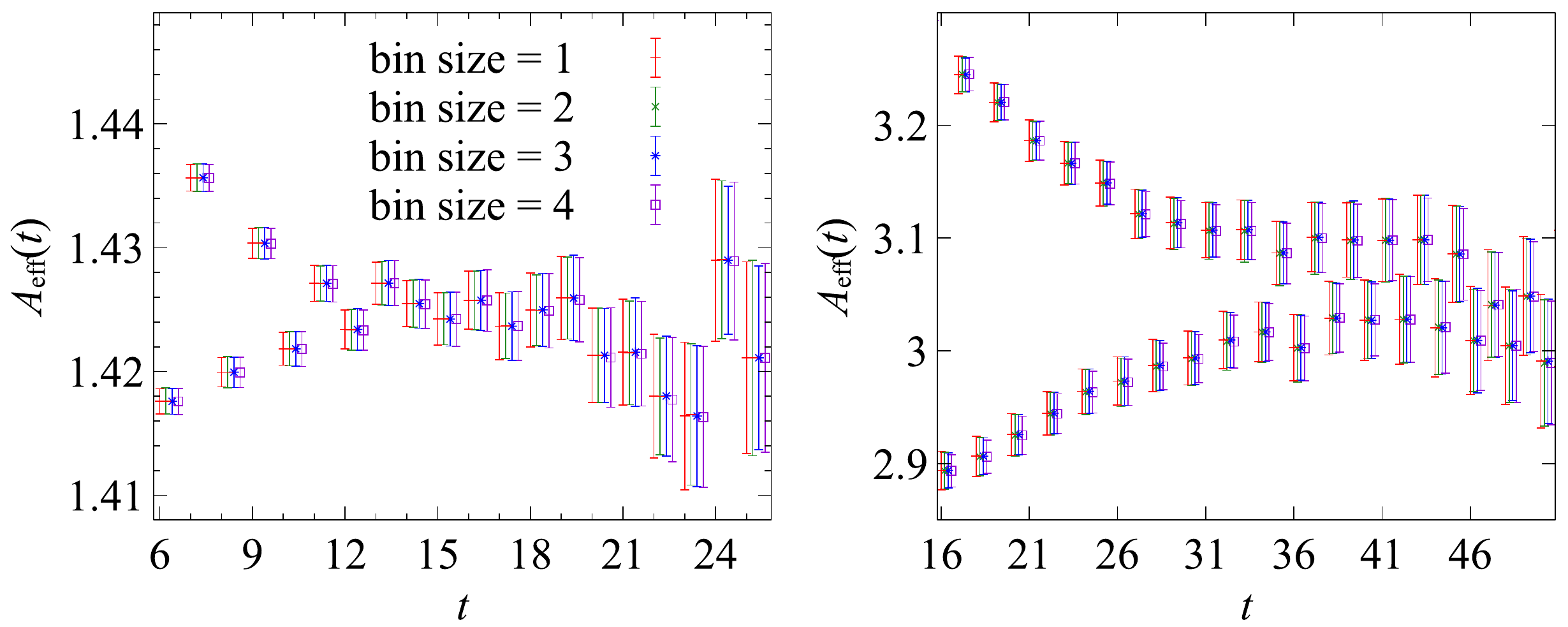}
	\caption{Scaled two-point correlator $A_{\text{eff}}$ defined in Eq.~(\ref{eq:Aeff}) as a function of bin size on the
    $a\approx0.12$~fm (left) and  $a \approx 0.045$~fm (right) ensembles with $m_l^\prime / m_s^\prime = 1/5$.}
	\label{fig:binning}
\end{figure}

We reduce autocorrelations between measurements computed on configurations close in Monte-Carlo simulation time by translating each gauge-field configuration by a random spatial shift $\vec{x}$ before calculating the valence light- and charm-quark propagators. We do not observe any remaining autocorrelations in the two- and three-point correlator data after this procedure. Figure~\ref{fig:binning} shows the scaled $D$-meson two-point correlator versus bin size on the coarsest and finest ensembles with $m_l^\prime / m_s^\prime = 1/5$. The central values and errors are stable with increasing bin size.
We also compute the integrated autocorrelation time, defined in Eq.~(4.1) of Ref.~\cite{Bazavov:2016nty}, and find it to be less than 1 on all ensembles. Based on these studies, we do not bin the data in this work.

\section{Correlator analysis}
\label{sec:correlator_analysis}

In this section, we describe how we extract the $D$-mixing matrix elements from the two- and three-point correlation functions given in Eqs.~(\ref{eq:NumSim_Correlators_3pt})~and~(\ref{eq:NumSim_Correlators_2pt}). All dimensionful quantities are given in lattice spacing units, where the factors of $a$ are suppressed for simplicity.

\subsection{Correlator fit functions}
\label{numericalsimulation_correlatorfitfunctions}

Our choice of using the naive field $\Upsilon(x)$, defined in Eq.~(\ref{eq:Upsilon}), for the valence up quark in the two- and
three-point correlation functions results in contributions from $D$-meson states with positive parity \cite{Wingate:2002fh}.
Their effects are included in the functional forms used to describe the Euclidean time-dependence of the correlation functions.
The two-point function takes the form
\begin{equation}
    C(t)=\sum_{n=0}^{N^{\text{2pt}}-1}(-1)^{n(t+1)}|Z_n|^2\left(e^{-E_nt}+e^{-E_n(T-t)}\right),
    \label{eq:twoptfitfunction}
\end{equation}
where the $Z_n$ are the overlap factors of the interpolating operators with the energy eigenstate labeled by $n$, $T$ is the
temporal extent of the lattice, and terms with even (odd) $n$ describe the effects of negative (positive) parity states.
The second term on the right-hand-side of Eq.~(\ref{eq:twoptfitfunction}) arises with periodic boundary conditions in time.

The three-point functions are described by
\begin{equation}
    C_{\latop_i}(t_x,t_y)=\sum_{m,n=0}^{N^{\text{3pt}}-1}(-1)^{n(t_x+1)}(-1)^{m(t_y+1)}
        \mathcal{Z}_{nm}^{\latop_i}\frac{Z_nZ_m^\dagger}{2\sqrt{E_nE_m}}e^{-E_n|t_x|}e^{-E_mt_y},
    \label{eq:threeptfitfunction}
\end{equation}
where the desired matrix element is related to the ground-state amplitude $\mathcal{Z}_{00}^{\latop_i}$:
\begin{equation}
    \label{eq:3ptAmpDef}
    \left<\latop_i\right>=\mathcal{Z}_{00}^{\latop_i} M_D.
\end{equation}
The factor of the $D$-meson mass $M_D$ is due to the nonrelativistic normalization of states in Eq.~(\ref{eq:threeptfitfunction}). Effects from periodic boundary conditions are negligible in our three-point data and are therefore not included in Eq.~(\ref{eq:threeptfitfunction}).

\subsection{Method}

As shown in Eq.~(\ref{eq:3ptAmpDef}), the three-point correlation function contains the desired matrix elements, however it also receives contributions from excited states. Our procedure for extracting the ground state matrix element is designed to account for the presence of excited states in the correlation functions while controlling and including the systematic errors associated with their residual contributions. We use Bayesian constraints with Gaussian priors in our fit functions in order to obtain a robust estimate of the uncertainty arising from excited state contamination. Correlations between two- and three-point functions are accounted for by performing a simultaneous fit to both data sets.

We implement the Bayesian constraints by minimizing the augmented $\chi^2$ function defined in Eq.~(B3) of Ref.~\cite{Bazavov:2016nty}, while the $Q$ parameter defined in Eq.~(B4) of Ref.~\cite{Bazavov:2016nty} is used to assess the
quality of the fits. The selection of the Bayesian priors is described in Sec/~\ref{correlator_priorselection}. In the fits to the correlation functions we use intervals $t_{\text{min}} \leq t \leq t_{\text{max}}$ that do not span the entire time range.
The time region used for the three-point functions is further restricted, as described in Sec.~\ref{correlator_fitregion}.
We limit the number of states used in Eqs.~(\ref{eq:threeptfitfunction}) and (\ref{eq:twoptfitfunction}) to $N^{\text{3pt}} =
N^{\text{2pt}} \leq 6$. These choices are designed to optimize the extraction of the desired matrix elements from the correlation function data. Section~\ref{correlator_fitstability} describes fit variations with different $t_{\text{min}}$, $t_{\text{max}}$, and $N^{\text{3pt}}$ to ensure our fit choices do not bias the fit results for the ground state parameters.

\subsection{Prior selection}
\label{correlator_priorselection}

The ground state parameters in the fit functions are well determined by the correlation function data. We therefore make sure that our selection procedure for the associated priors imposes only very loose constraints. At the same time, the data often do not provide good constraints for the excited state parameters in the fit functions. The purpose of the priors for the excited state parameters is to stabilize the fits and allow us to include the uncertainty due to residual excited state effects in the fit error. As discussed in Sec.~\ref{sec:smear}, our main analysis employs two-and three-point functions with smeared $D$-meson operators. Below we describe the selection procedure and resulting choices for the priors and widths for these correlation functions. We start with the two-point function parameters, first for the ground state, followed by the excited states, and then discuss the three-point function parameters.

\begin{figure}
	\centering
		\includegraphics[width=1.00\textwidth]{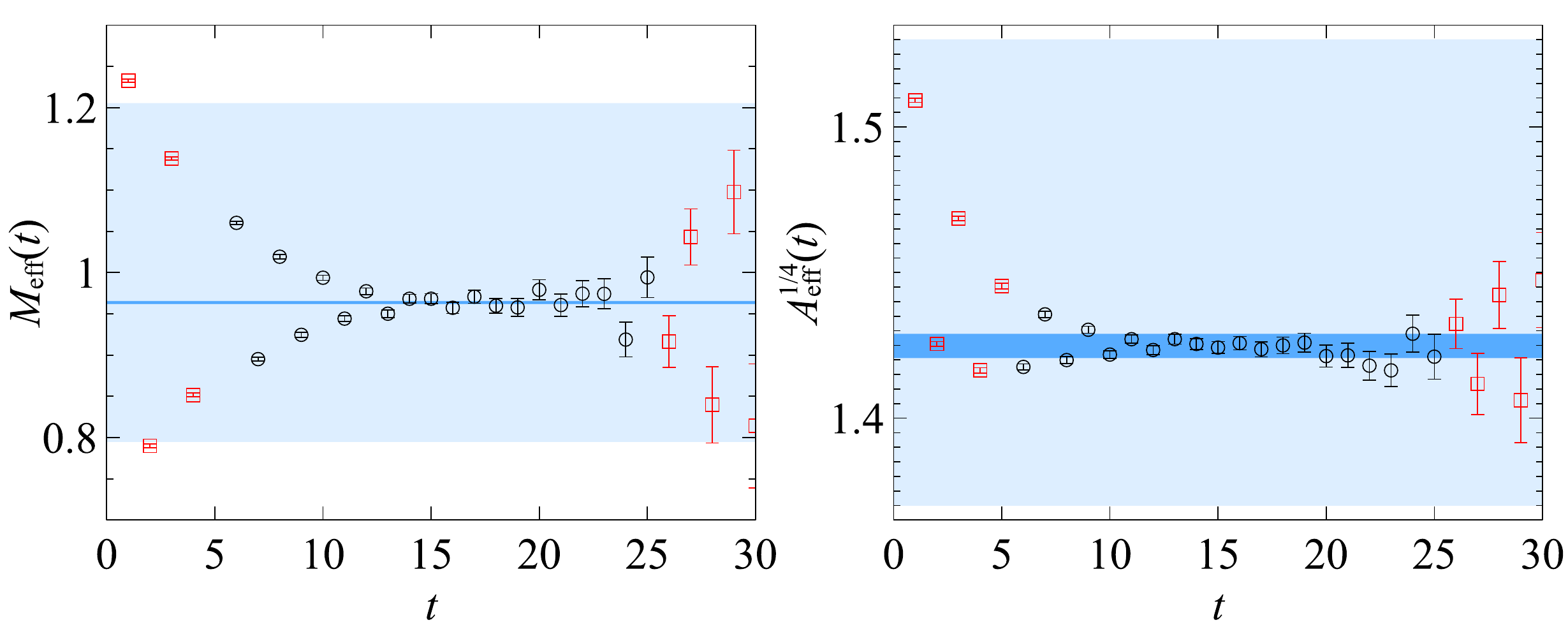}
	\caption{
	The effective mass (left) and the fourth-root of the scaled two-point correlation function (right) on the $a\approx 0.12$~fm, $m^\prime_l/m^\prime_s=0.2$ ensemble with $m_q=m^\prime_l$. At small time separations $t$, the data in both panels are affected by excited states. A plateau is reached at $t\approx 15$, while the statistical errors start to increase at $t\gtrsim 20$. The wide, light blue bands are centered at the prior central value and indicate the prior width. The narrow, dark blue bands show the central value and error of the fit posterior. The black circle data points display the data over which the fit is performed, while the red square data points are excluded in the fit.}
	\label{fig:corr_effectiveplots}
\end{figure}

The priors for the ground state energy $E_0$ are obtained by examining the effective mass at large times $t$, 
\begin{equation}
    M_{\rm eff}(t) = \text{cosh}^{-1}\left(\frac{C(t+1)+C(t-1)}{2C(t)}\right)  \xrightarrow[]{\text{large $t$}} E_0.
\end{equation}
A typical effective mass plot is shown in the left panel of Fig.~\ref{fig:corr_effectiveplots}. There is a clear plateau at large $t$ before the signal is overwhelmed by noise. We consider the different plateau values of the effective mass for the different light valence- and sea-masses, and determine a single $E_0$ prior distribution for each lattice spacing. Specifically, the central value of $E_0$ is chosen to lie in the center of the different effective mass values, and the prior width spans approximately twice the range of effective mass plateau values.
For the ground state $Z_0$ priors we examine the scaled two-point function,
\begin{equation}
    A_{\text{eff}}(t) = C(t)e^{tM_{\text{eff}}} \xrightarrow[]{\text{large $t$}} \frac{|Z_0|^2}{2E_0},
    \label{eq:Aeff}
\end{equation}
where $M_{\rm eff}$ is an estimate of $M_{\rm eff}(t)$ in the large $t$ limit. An example of the scaled two-point function is shown in the right panel of Fig.~\ref{fig:corr_effectiveplots}; the plateau region at large time separation is again used to construct the prior range. In order to constrain $Z_0$ to be positive definite, we parametrize the corresponding fit parameter as the square root of $Z_0$. As illustrated in the left (right) panel of Fig.~\ref{fig:corr_effectiveplots}, the prior widths for $E_0$ ($Z_0$) are typically more than 100 (20) times larger than the errors on the corresponding posteriors.

In our data, the first excited state corresponds to the opposite-parity (scalar) ground state. We parametrize its energy $E_1$ in terms of the splitting $\Delta_{1,0}$ from the ground state with
\begin{equation}
E_{1}\equiv E_0 +
\Delta_{1,0}
\end{equation}
and
\begin{equation}
\Delta_{k,j}\equiv \ln\left[\exp{(E_k - E_{j})}\right].
\label{eq:Esplit}
\end{equation}
This ensures that the parameter space for energy splittings is positive and enforces the ordering $E_k > E_j$. The prior central value and width for $\Delta_{1,0}$ is guided by the experimentally measured $D^*_0(2400)^0-D^0$ mass difference~\cite{Olive:2016xmw}; we use $\Delta_{1,0}\approx 450^{+400}_{-100}$~MeV. As illustrated in both panels of Fig.~\ref{fig:corr_effectiveplots}, oscillating states are clearly present in the correlation function data. We therefore expect the overlap with the opposite parity state ($Z_1$) to be nonzero. We set the prior central value for $Z_1$ to half the value of $Z_0$ because the smeared operators suppress the overlap with excited states (see Sec.~\ref{sec:smear}). The prior width of $Z_1$ is set to be about one-$\sigma$ away from zero, reflecting our expectation for nonzero overlap.

For the remaining higher-state energies, we construct two towers of prior central values (and widths) starting at the pseudoscalar and scalar ground states,
\begin{eqnarray}
	E_n\equiv E_{0} + \sum_{k=2, 4, \ldots}^{n} \Delta_{k,k-2} &&  \quad \text{for~} n = \text{even (pseudoscalar)}, \label{eq:Esplitps}\\
	E_n\equiv E_{1} + \sum_{k=3, 5, \ldots}^{n} \Delta_{k,k-2} &&  \quad \text{for~} n = \text{odd (scalar)} \label{eq:Esplits}.
\end{eqnarray}
The choice of prior for the splittings of the radial excitations is motivated from quark model estimates~\cite{Ebert:2009ua},
$\Delta_{k,k-2} \approx 650^{+2000}_{-500}$~MeV. The central values of the priors for the excited state overlap factors $Z_k$ are set to approximately half the central values of $Z_0$, again based on the expectation that smearing suppresses overlap with the excited states. The prior width is chosen to encompass $Z_k=0$, allowing for the possibility of an absence of signal in the data.

\begin{figure}
	\centering
    \includegraphics[width=1.00\textwidth]{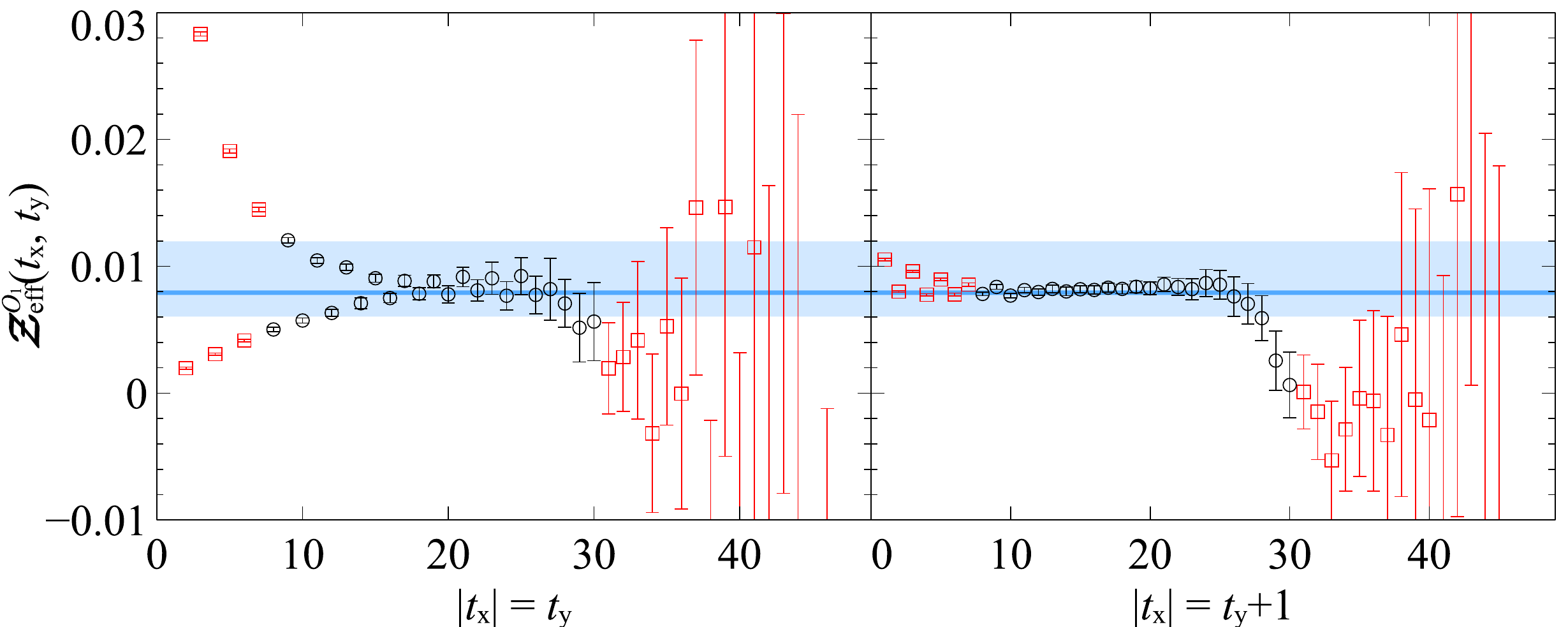}
	\caption{
	The scaled three-point correlation function along the diagonal (left) with $|t_x| = t_y$ and the off-diagonal (right) with $|t_x|=t_y+1$ on the $a\approx 0.09$~fm, $m^\prime_l/m^\prime_s=0.1$ ensemble with $m_q=0.0062$ for matrix element $\langle\latop_1\rangle$. The wide, light blue bands are centered at the prior central value and indicate the prior width. The narrow, dark blue bands show the central value and error of the fit posterior. The black circle data points display the data over which the fit is performed, while the red square data points are excluded in the fit.}
	\label{fig:corr_scaledthree}
\end{figure}

For the ground-state amplitude $\mathcal{Z}_{00}^{\latop_i}$ we examine the scaled three-point function at large times $t_x, t_y$,
\begin{equation}
    \mathcal{Z}_{\text{eff}}^{\latop_i}(t_x, t_y) = C_{\latop_i}(t_x,t_y)A_{\text{eff}}^{-1}e^{M_{\text{eff}}(|t_x|+t_y)},
\end{equation}
an example of which is shown in the two panels of Fig.~\ref{fig:corr_scaledthree}. The left panel illustrates the plateau along the diagonal where $|t_x| = t_y$, while the right panel shows the off-diagonal $|t_x|=t_y+1$. As with the two-point correlators, we obtain a prior central value and width at each lattice spacing from the variation of the scaled three-point functions with light valence- and sea-quark mass. As illustrated in both panels of Fig.~\ref{fig:corr_scaledthree}, the resulting prior widths are typically 10 times larger than the fit errors. We also increased the width of the ground state prior by up to a factor of 3 and obtained identical results for the ground state amplitudes. The exponential decay visible in both panels is the dominant signal for excited states. The priors for the excited state amplitudes $\mathcal{Z}_{mn}^{\latop_i}$ (defined in Eq.~(\ref{eq:threeptfitfunction})) are constrained to be the same order of magnitude as the ground state amplitude. The complete set of priors for the correlator analysis is listed in Table~\ref{tab:corrparams} in Appendix~\ref{app:2pt3ptpriors}.

\begin{figure}
	\centering
    \includegraphics[width=0.60\textwidth]{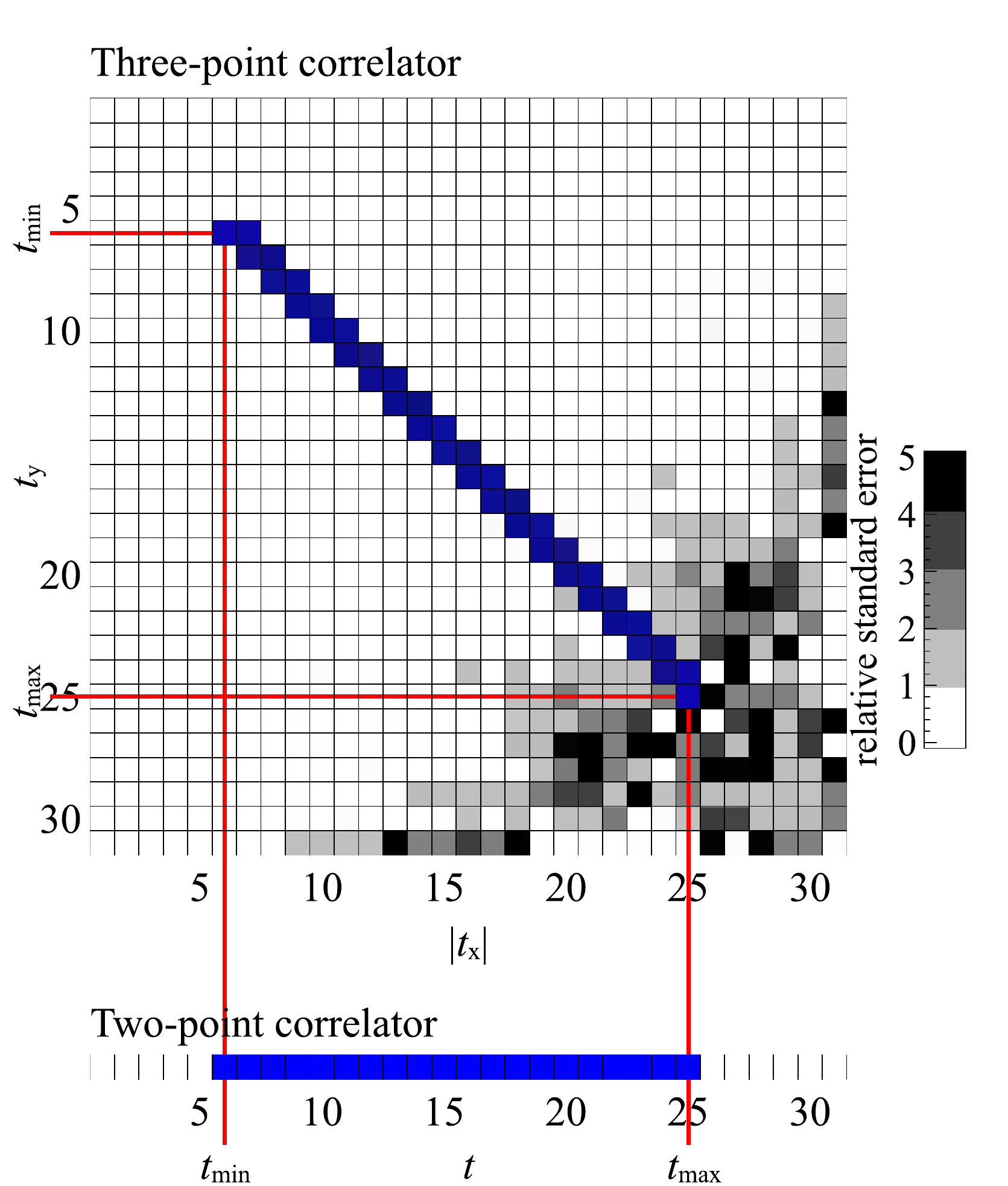}
	\caption{The preferred fit region for the $a \approx 0.12$~fm, $m^\prime_l/m^\prime_s=0.2$ ensemble, with $m_q=m_l^\prime$, overlaid on the relative errors of the three-point correlation function for $\latop_4$ in the $|t_x|-t_y$ plane. The time regions used in the preferred fits for the two- and three-point correlation functions are marked in blue. The horizontal and vertical red lines identify the extremum of the fit regions.}
	\label{fig:FitRegion}
\end{figure}

\subsection{Fit region}
\label{correlator_fitregion}

The time regions over which we fit the two- and three-point function data is illustrated by the blue-colored areas in Fig.~\ref{fig:FitRegion}, where the horizontal and vertical red lines identify the extremum of the fit regions. Because the statistical errors are small for most of the available time range, there is a large region in the $|t_x|-t_y$ plane that is useful, in principle, for the three-point function analysis. However, the number of configurations in the ensembles used in this analysis while large, is not enough to resolve the covariance matrix over the entire time region bounded by $|t_{x}|,t_{y}\in\{t_{\text{min}},\ldots,t_{\text{max}}\}$, and we therefore constrain the three-point function analysis to the data along a bi-diagonal in the $|t_x|-t_y$ plane. Figure~\ref{fig:corr_scaledthree} illustrates that information along the off-diagonal direction is needed to account for excited state contributions that would not be easily resolved if the data were limited to just the diagonal. Another choice is to reduce $t_{\text{max}}$ for the three-point function but analyze the data in the entire region bounded by $|t_{x}|,t_{y}\in\{t_{\text{min}},\ldots,t_{\text{max}}\}$, \ie\ including all the off-diagonal points in the $|t_x|-t_y$ plane. However, this would discard data at large time separations, where the ground state contribution dominates. We also consider other data reduction procedures, such as randomly drawing a certain number of data points from the $|t_{x}|,t_{y}\in\{t_{\text{min}},\ldots,t_{\text{max}}\}$ region. We find that all the choices for reducing the number of data points that we have studied yield results for the ground state parameters that are consistent with those from our main analysis. Table~\ref{tab:timeranges} lists our choices for $t_{\rm min}$ and $t_{\rm max}$ for our preferred fits. Across all four lattice spacings, we set $t_{\text{min}}=0.72$~fm while varying $t_{\text{max}}$ smoothly from 3~fm on the $a \approx 0.12$~fm ensembles, 2.7~fm on the $a \approx 0.09$~fm ensembles, 2.5~fm on the $a \approx 0.06$~fm ensembles, to 2.3~fm for the $a \approx 0.045$~fm ensemble. We choose one set of priors and time ranges for all correlation function fits at a given lattice spacing.

\begin{table*}
\caption{\label{tab:timeranges} Values of $t_{\text{min}}$ and $t_{\text{max}}$ by lattice spacing, as labeled in
Fig.~\ref{fig:FitRegion}.}.
\begin{tabular}{c@{\quad}  c@{\quad}c@{\quad} | c@{\quad}c@{\quad}}
\hline \hline
 lattice spacing	& $t_{\rm min} (a)$  & $t_{\rm min}$ (fm)  & $t_{\rm max} (a)$  & $t_{\rm max}$ (fm)   \\
\hline
$\approx 0.12$~fm 	& 6	& 0.72	& 25	& 3.0 \\
$\approx 0.09$~fm	& 8	& 0.72	& 30	& 2.7  \\
$\approx 0.06$~fm	& 12	& 0.72 	& 42	& 2.5 \\
$\approx 0.045$~fm 	& 16	& 0.72	& 50	& 2.3 \\
\hline \hline
\end{tabular}
\end{table*}

\subsection{Fit stability}
\label{correlator_fitstability}

Our preferred correlator fits are performed with the priors listed in Table~\ref{tab:corrparams}, the time ranges listed in Table~\ref{tab:timeranges}, the time region as described in Fig.~\ref{fig:FitRegion}, and with $N_{\text{state}}=2+2$ where the notation is used to denote the correlator fit ansatz with two normal parity and two opposite parity states. Here we describe fit variations that we use to test for systematic effects due to excited states, and other fit choices. We examine the dependence of the fit results for the ground state parameters under varying fit range, number of states, and operator smearing. In our simultaneous fits we vary each of the parameters ($t_{\text{min}}$, $t_{\text{max}}$, and $N^{\text{2pt}}=N^{\text{3pt}}$) for the two- and three-point functions.

\subsubsection{Fit range and number of states}

\begin{figure}[th]
	\centering
		\includegraphics[width=1.0\textwidth]{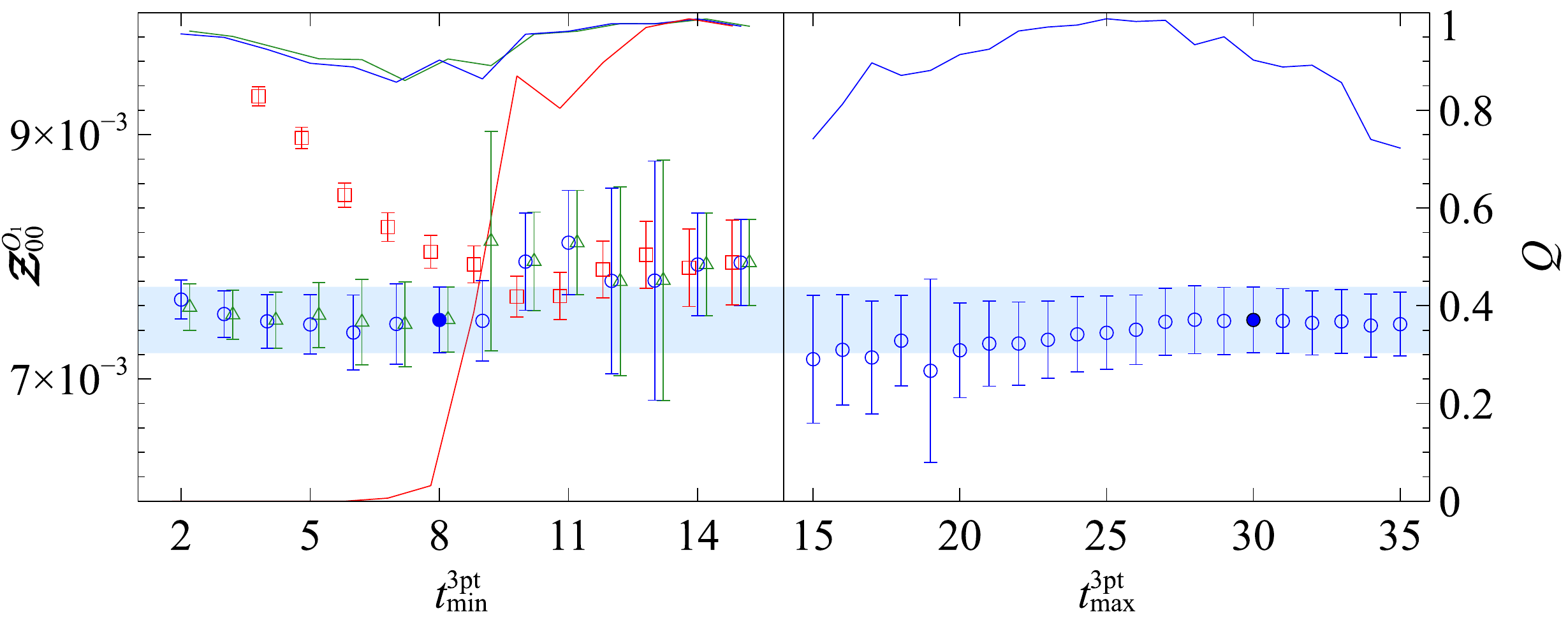}
	\caption{
	Fit results for $\mathcal{Z}_{00}^{\latop_1}$ on the $a\approx 0.09$~fm, $m^\prime_l/m^\prime_s=0.1$ ensemble, with 
    $m_q=m^\prime_l$, and with 1S operator smearing. The preferred fit is marked by the solid blue circle. In both panels the data points (symbols with error bars) indicate fit results for $\mathcal{Z}_{00}^{\latop_1}$ while the solid lines show the corresponding $Q$-values. The blue bands indicate the preferred fit and the red squares, blue circles, and green triangles represent fit results with 1+1, 2+2, and 3+3 states, respectively. The corresponding $Q$-values are indicated by the solid lines with matching colors. (Left) $\mathcal{Z}_{00}^{\latop_1}$ vs $t_{\text{min}}$ at fixed $t_{\text{max}}=30$ for different numbers of states. (Right) $\mathcal{Z}_{00}^{\latop_1}$ vs $t_{\text{max}}$ at fixed $t_{\text{min}}=8$ with 2+2 states. } \label{fig:corrfit_tstabplt}
\end{figure}

Figure~\ref{fig:corrfit_tstabplt}~(left) shows a typical example of a $t_{\text{min}}$ stability plot for $\mathcal{Z}_{00}^{\latop_1}$ with the preferred fit displayed with a solid blue point. The fits that include only 1+1 states show a strong $t_{\text{min}}$ dependence before reaching a plateau at $t_{\text{min}} \ge 9$, while fits with 2+2 or 3+3 states reach a plateau for $t_{\text{min}} \ge 3$. In addition, the 1+1 fit results have significantly smaller error bars than the 2+2 and 3+3 fits, while the errors going from the 2+2 to the 3+3 fits are essentially unchanged. We conclude that 2+2 fits with $5 \le t_{\text{min}} \le 9$ are necessary and sufficient to account for excited state effects.

A typical example of a $t_{\text{max}}$ stability plot is shown Fig.~\ref{fig:corrfit_tstabplt}~(Right) for $\mathcal{Z}_{00}^{\latop_i}$ where the preferred fit is marked with a solid blue point. The fit results (central values and error bars) do not change as $t_{\text{max}}$ is increased, indicating that contributions to the three-point function from periodic boundary conditions are negligibly small. However, the drift in the central values and increase in the error bars as $t_{\text{max}}$ is decreased to $t_{\text{max}}\lesssim26$ indicate that the correlation function data at large time separations still contribute to the ground state signal and help stabilize the ground state posteriors. This informs our choices for $t_{\text{max}}$ in Table~\ref{tab:timeranges}.

\subsubsection{Operator smearing}
\label{sec:smear}

\begin{figure}[th]
	\centering
		\includegraphics[width=1.0\textwidth]{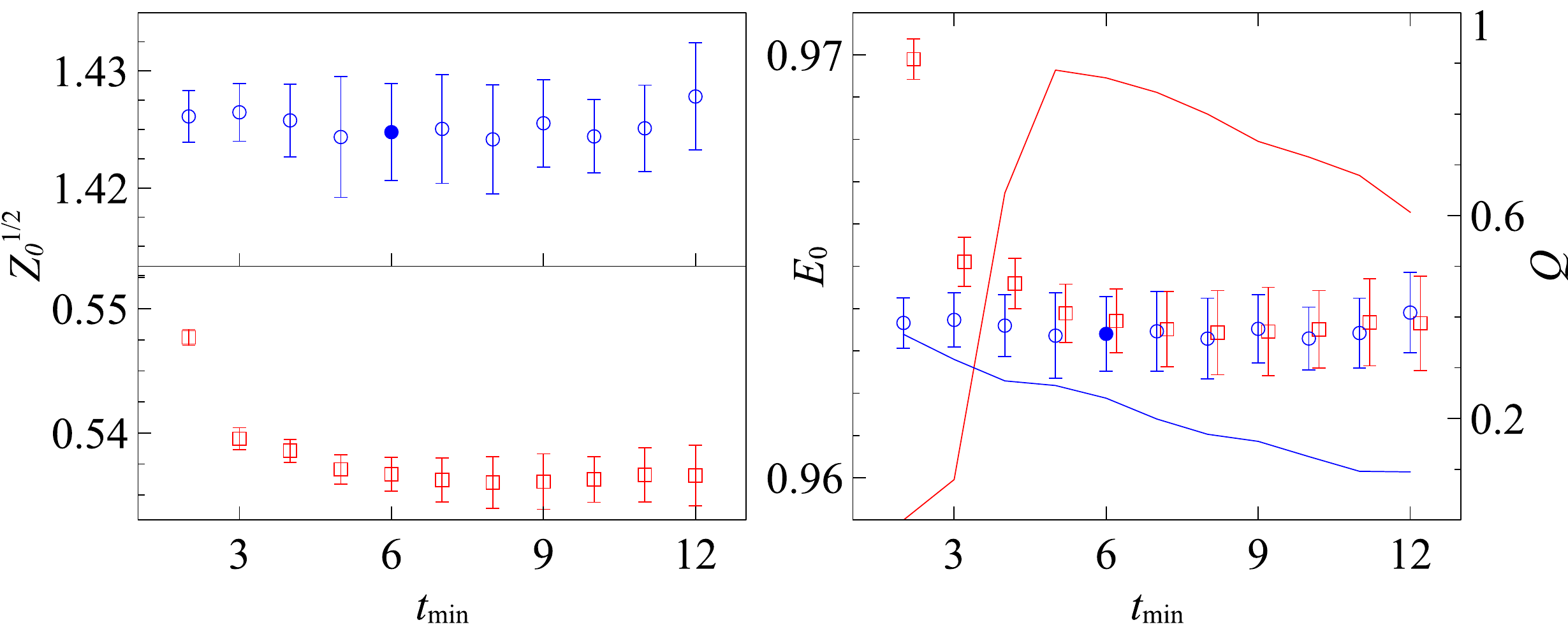}
	\caption{
	Two-point function fit results on the $a\approx 0.12$~fm, $m^\prime_l/m^\prime_s=0.2$ ensemble with $m_q=m^\prime_l$ and $t_{\text{max}}=25$ for 1S smeared (blue circles) and local (red squares) source and sink operators. The preferred fit is indicated by the solid blue circle. (Left)~Ground state overlap factor $Z_0^{1/2}$ vs $t_{\text{min}}$, (Right)~Ground state energy $E_0$ vs $t_{\text{min}}$. The solid lines indicate the corresponding $Q$-values.} \label{fig:corrfit_smearing}
\end{figure}

Figure~\ref{fig:corrfit_smearing}~(left) shows an example comparing fit results for the two-point function ground state overlap
factor $Z_0^{1/2}$ with 1S-smeared (blue circles) and local (red squares) source and sink operators. The 1S-smeared operator has better overlap with the ground state as evidenced by the larger central values for the corresponding overlap factor. In conjunction, the 1S-smeared operator has smaller overlap with excited states than the local operator, since the latter yields fit
results that exhibit significant dependence on $t_{\text{min}}$ not seen with the former. A sample comparison of the ground state energies $E_0$ from fits to the same two-point functions is shown in Fig.~\ref{fig:corrfit_smearing}~(right). The results are similar. The ground state energies obtained from the 1S-smeared two-point function do not vary with $t_{\text{min}}$, and have errors that stabilize for $t_{\text{min}}\gtrsim 5$. By comparison, the ground state energies from the local two-point function show strong $t_{\text{min}}$ dependence and differ at small $t_{\text{min}}$ from the ground state energies with 1S-smearing before they become consistent with them. The larger range of stability in $t_{\text{min}}$ observed for fit results from correlation functions with 1S-smeared source and sink operators makes it easier to obtain consistent fit regions across all ensembles and valence masses. We therefore use the correlation functions with 1S smeared sources and sinks in our main analysis.

\subsection{Error propagation}
\label{correlator_errorpropagation}

\begin{figure}
	\centering
		\includegraphics[width=0.5\textwidth]{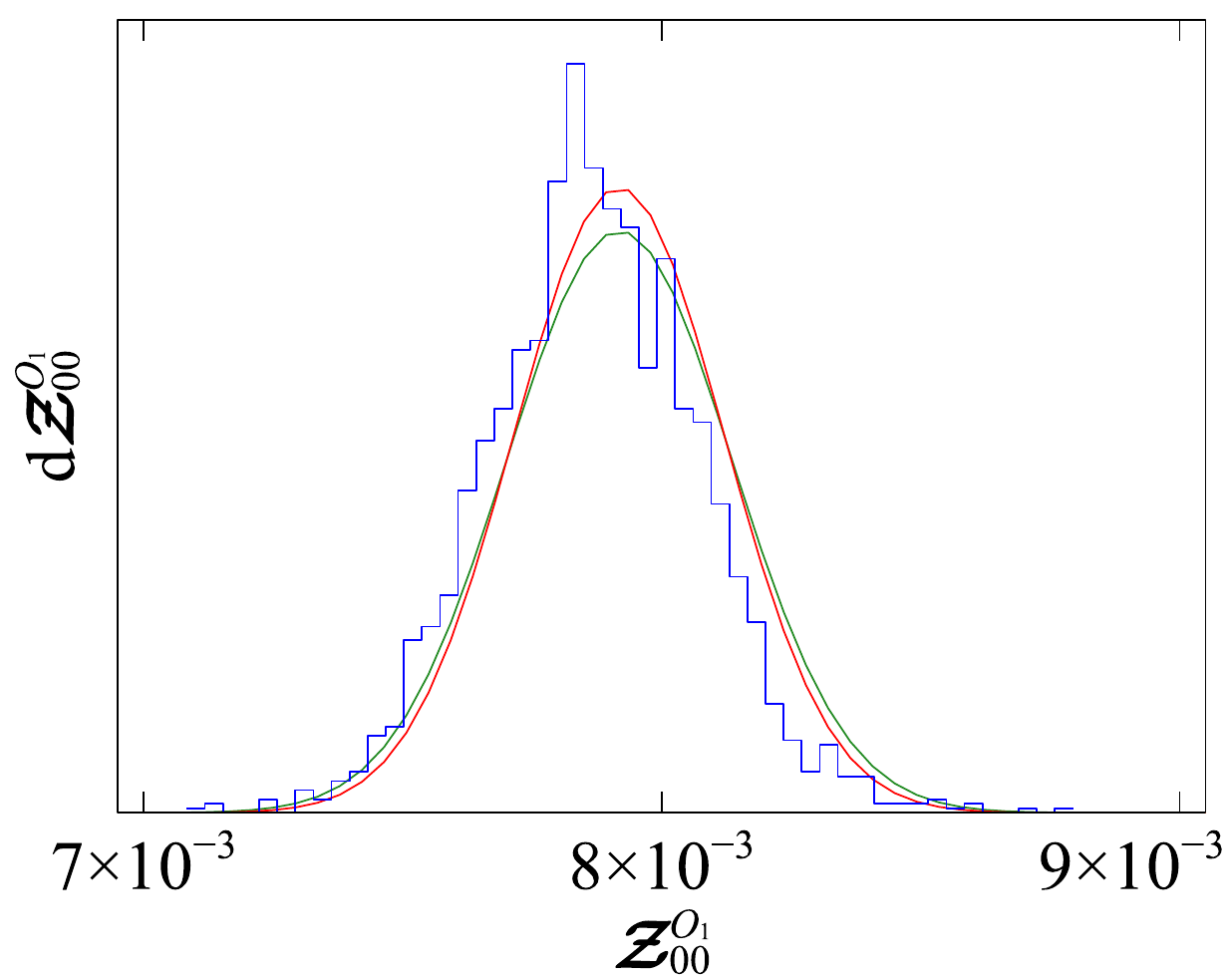}
	\caption{Posterior distributions are presented for the $\latop_1$ matrix element calculated on the 0.12~fm lattice at 
    $m^\prime_l/m^\prime_s=0.2$ and at the partially quenched point $m_q=0.05$ with 1S-smeared source and sink operators. Bootstrap (blue), jackknife (green) and Hessian (red) distributions of the preferred fit demonstrate nearly Gaussian posterior distributions. The normalized bootstrap histogram is generated by 2000 resamples. The mean of the single-elimination jackknife distribution is consistent with the Hessian central value up to round-off errors.}
	\label{fig:correlatordistribution}
\end{figure}

We propagate the distribution of the matrix element $\mathcal{Z}_{00}^{\latop_i}$ by bootstrap resampling. The distributions of the priors are included by randomizing the prior central value over the prior width under the bootstrap sampling~\cite{Lepage:2001ym}. We generate 2000 bootstrap samples for each ensemble included in our analysis. An example comparing the bootstrap, single elimination jackknife, and Hessian posterior distributions of the ground state amplitude
$\mathcal{Z}_{00}^{\latop_1}$ is shown in Fig.~\ref{fig:correlatordistribution}. This example is representative of the consistency seen amongst the bootstrap, jackknife, and Hessian distributions of our fit results and demonstrates that the posterior distributions are approximately Gaussian.  

\section{Renormalization}
\label{renormalization}
\label{sec_renorm}

In lattice-QCD calculations, the nonzero lattice spacing provides an ultraviolet cutoff. As a result, the matrix elements computed at different lattice spacings are regulated at different energy scales. In order to take the continuum limit, the matrix elements must be run to the same energy scale, and in order to combine lattice matrix-element results with continuum Wilson coefficients, they must be matched to a continuum renormalization scheme. In this paper we convert the bare lattice operators $\latop_i(a)$ evaluated at scale $a$, to the renormalized operators $\bar{\latop}_i(\mu)$ in the continuum $\overline{\text{MS}}$-scheme evaluated at the scale $\mu=3~\text{GeV}$. In noninteger dimensions, the Dirac algebra is infinite dimensional and is fully defined only after choosing a basis of evanescent operators~\cite{Dugan:1990df}. This choice, which here only affects the renormalization of operators $\op_2$ and $\op_3$, is not unique; here we consider the schemes of Beneke \emph{et al.} (BBGLN)~\cite{Beneke:1998sy, Beneke:2002rj} and of Buras, Misiak, and Urban (BMU)~\cite{Buras:2000if}, reporting results for both.

The $\Delta C = 2$ four-fermion operators mix under renormalization. At $\order{(\alpha_s)}$, the renormalized operators are given by~\cite{Evans:2009du,GamizPT}
\begin{equation}
    \bar{\latop}_i(\mu) = Z_{ij}(a\mu)\latop_j(a) + ab_{ip}(a\mu)P_p(a) \doteq \bar{\op}_i + \order(a^2),
    \label{eq:matchingmNPR}
\end{equation}
where the $P_p$ are dimension-seven operators that are not needed in this paper, because with the $\latop_i$ described after
Eqs.(\ref{eq:NumSim_Correlators_3pt}), the coefficient matrix $b_{ip}$ starts at order~$\alpha_s$. Neglecting this contribution leads to a discretization error of order $\alpha_sa\Lambda_\text{QCD}$, the same as that from our choice of $c_\text{SW}$, which is the coefficient of the clover term in the Sheikholeslami-Wohlert action~\cite{Sheikholeslami:1985ij}.

We calculate the renormalization coefficients $Z_{ij}(a\mu)$ using the ``mostly nonperturbative matching'' (mNPR) method introduced in Refs.~\cite{ElKhadra:2001rv,Harada:2001fi}. We factor out the flavor-conserving renormalization coefficients $Z^4_{V_{qq}}$ using
\begin{align}
    Z_{ij} &= Z_{V^4_{cc}}Z_{V^4_{ll}}\rho_{ij},
    \label{eq:hhllZ} \\
    \rho_{ij} &= \delta_{ij} + \sum_{l=1}\alpha_s^l \rho^{[l]}_{ij}(a\mu).
    \label{eq:mNPR_rho}
\end{align}
The $Z_{V^4_{qq}}$ are then calculated nonperturbatively from equal-mass vector current correlation functions, as discussed in Ref.~\cite{Bazavov:2011aa}. The flavor-changing coefficients $\rho_{ij}$ are calculated to one-loop in lattice perturbation theory via
\begin{equation}
    \rho^{[1]}_{ij}= Z^{[1]}_{ij} - \delta_{ij} \left(Z^{[1]}_{V^4_{cc}} + Z^{[1]}_{V^4_{ll}} \right),
    \label{eq:mNPR_rho_PT}
\end{equation}
where the coefficients $Z^{[1]}_{V^4_{ll(cc)}}$ are defined through
\begin{equation}
    Z_{V^4_{ll(cc)}} = \mathcal{C}_{ll(cc)}\left[1+\alpha_sZ_{V^4_{ll(cc)}}^{[1]}+ \order(\alpha_s^2)\right].
\end{equation}
and the tree-level matching factors are $\mathcal{C}_{ll} = 2 u_0$ and $\mathcal{C}_{cc} = 2\kappa'_cu_0(1+m_0/u_0)$ for our conventions for the staggered and clover fermion fields, respectively. The one-loop heavy-light four-fermion-operator-renormalization correction factor $\rho_{ij}$ is close to unity for $i=j$ because wavefunction renormalization diagrams cancel in the ratio $Z_{ii}/(Z_{V^4_{cc}}Z_{V^4_{ll}})$ at this order.
 
Table~\ref{tab:ZV} lists the renormalization coefficients for flavor-conserving vector currents. The same value for $Z_{V_{ll}^4}$ is used for all valence-quark masses on a given ensemble because the observed quark-mass dependence is smaller than the statistical errors.

We calculate the $\rho$ factors in tadpole-improved lattice perturbation theory taking $\alpha_s=\alpha_V(q^*)$ in the ``$V$~scheme''~\cite{Lepage:1992xa} obtained from the static-quark potential~\cite{PhysRevLett.95.052002}. We fix the scale to be $q^*=2/a$, which is the typical scale of the gluon loop momenta. Table~\ref{tab:rhos} lists the one-loop coefficients $\rho_{ij}^{[1]}$ for the BBGLN choice of evanescent operators. The matching coefficients for the BMU evanescent operator prescription can be obtained from the BBGLN coefficients via~\cite{Monahan:2014xra,Becirevic:2001xt}
\begin{subequations}\label{eq:BBGLN2BMU}
\begin{align}
    \rho_{22}^{[1]} \rightarrow & \rho_{22}^{[1]} - \frac{1}{\pi}, \\
    \rho_{21}^{[1]} \rightarrow & \rho_{21}^{[1]} - \frac{1}{24\pi}, \\
    \rho_{33}^{[1]} \rightarrow & \rho_{33}^{[1]} + \frac{1}{3\pi}, \\
    \rho_{31}^{[1]} \rightarrow & \rho_{31}^{[1]} - \frac{1}{244\pi}.
\end{align}
\end{subequations}

We also calculate the heavy-light matching coefficients $Z_{ij}$ at one loop in tadpole-improved perturbation theory. The tadpole-improved coefficients $\zeta^{[1]}_{ij}$ are related to the coefficients $Z^{[1]}_{ij}$ in Eq.~(\ref{eq:mNPR_rho_PT}) via
\begin{equation}
    \zeta^{[1]}_{ij}= Z^{[1]}_{ij} - \delta_{ij} u_0^{[1]} \left( \frac{9}{4} + \frac{1}{1+ m_0/u_0} \right).
    \label{eq:TadPT_zeta}
\end{equation}
We compute the renormalization coefficients taking the tadpole-improvement factor $u_{0P}^{[1]} = -0.76708(2)$ from the fourth root of the plaquette or $u_{0L}^{[1]} = -0.750224(3)$ from the link in Landau gauge. The numerical differences between the diagonal coefficients $\zeta^{[1]}_{ii} - \rho^{[1]}_{ii}$ obtained with one-loop perturbation theory using $u_{0P}$ and $u_{0L}$ and mNPR are given in the last two columns in Table~\ref{tab:rhos}, respectively. This difference is the same for all operators $i = 1$--5.

\begin{table}[tp]
    \centering
    \caption{Strong coupling in the $V$~scheme at the scale $\mu=2/a$, and heavy-heavy and light-light vector-current renormalization factors $Z_{V_{cc}^4}$ and $Z_{V_{ll}^4}$ with statistical errors for the listed simulation $\kappa_c'$ 
values. Also shown are the nonperturbatively determined critical hopping parameter $\kappa_{\text{crit}}$ and the tadpole-improvement factors $u_{0P}$ and $u_{0L}$ obtained from the fourth root of the plaquette and the link in Landau gauge, respectively.}   
    \label{tab:ZV}
\setlength{\tabcolsep}{5pt}
\begin{tabular}{cclllcccc}
\hline\hline
$\approx a$~(fm) & $am'_l/am'_s$&~~$\kappa'_c$&~~$\kappa_{\text{crit}}$~~&~~$u_{0P}$&$u_{0L}$&$\alV(2/a)$& \ZVcc    & $Z_{V^4_{ll}}$    \\
\hline
0.12          &  0.02/0.05  & 0.1259 & 0.14073  & 0.8688     & 0.837 & 0.3047 & 0.2912(1) & 1.734(3) \\
                 &  0.01/0.05  & 0.1254 & 0.14091  & 0.8677     & 0.835 & 0.3108 & 0.2947(1) & 1.729(3) \\
                 &  0.01/0.05  & 0.1280 & 0.14091  & 0.8677     & 0.835 & 0.3108 & 0.2786(1) & 1.729(3) \\
                 &  0.007/0.05  & 0.1254 & 0.14095  & 0.8678   & 0.836 & 0.3102 & 0.2946(1) & 1.730(3) \\
                 &  0.005/0.05  & 0.1254 & 0.14096  & 0.8678   & 0.836 & 0.3102 & 0.2946(1) & 1.729(3) \\
\hline
0.09          &  0.0124/0.031 & 0.1277 & 0.139052 & 0.8788       & --- & 0.2582 & 0.2761(2) & 1.768(4) \\ 
                 &  0.0062/0.031 & 0.1276 & 0.139119 & 0.8782       & 0.854 & 0.2607 & 0.2769(2) & 1.766(4) \\ 
                 &  0.00465/0.031 & 0.1275 & 0.139134 & 0.8781     & --- & 0.2611 & 0.2776(2) & 1.766(4) \\
                 &  0.0031/0.031 & 0.1275 & 0.139173 & 0.8779       & --- & 0.2619 & 0.2777(2) & 1.765(4) \\
                 &  0.00155/0.031 & 0.1275 & 0.139190 & 0.877805 & --- & 0.2623 & 0.2777(2) & 1.765(4) \\
\hline
0.06          &  0.0072/0.018 & 0.1295 & 0.137582 & 0.8881    &--- &0.2238 & 0.2614(2) & 1.798(5) \\  
                 &  0.0036/0.018 & 0.1296 & 0.137632 & 0.88788  &--- &0.2245 & 0.2611(2) & 1.797(5) \\
                 &  0.0025/0.018 & 0.1296 & 0.137667 & 0.88776  & 0.869 &0.2249 & 0.2612(2) & 1.797(5) \\
                 &  0.0018/0.018 & 0.1296 & 0.137678 & 0.88764  & 0.869 &0.2253 & 0.2610(2) & 1.796(5) \\
\hline
0.045        &  0.0028/0.014 & 0.1310 & 0.136640 & 0.89511  & 0.8797 &0.2013 & 0.2498(2) & 1.818(8) \\
\hline\hline
\end{tabular}
\end{table}

\begin{sidewaystable}
    \centering
    \caption{One-loop renormalization coefficients defined in Eqs.~(\ref{eq:hhllZ})--(\ref{eq:mNPR_rho}) for the heavy-light 
    $\Delta C = 2$ local four-fermion operators at the renormalization scale $\mu = 3$~GeV. The renormalization coefficients are defined in Eqs.~(\ref{eq:mNPR_rho_PT}) and $(\ref{eq:TadPT_zeta})$ and calculated for the BBGLN~\cite{Beneke:1998sy,Beneke:2002rj} choice of evanescent operators; their relation to the BMU~\cite{Buras:2000if} scheme is given in Eqs.~(\ref{eq:BBGLN2BMU}). The last two columns show the differences between the one-loop coefficients obtained in perturbation theory and mNPR for two choices of the tadpole-improvement factor $u_{0P}$ and $u_{0L}$. Renormalization coefficients not listed are vanishing.} 
    \label{tab:rhos}
    \setlength{\tabcolsep}{3pt}
 \begin{tabular}{cc rrrrr rrrrr rr}
        \hline\hline
        $\approx a~\text{(fm)}$ & $\begin{matrix}am^\prime_l\\ \hline am^\prime_s \end{matrix}$ &$\rho^{[1]}_{11}$~~ & $\rho^{[1]}_{12}$~~ & $\rho^{[1]}_{22}$~~ & $\rho^{[1]}_{21}$~~ & $\rho^{[1]}_{33}$~~ & $\rho^{[1]}_{31}$~~ & $\rho^{[1]}_{44}$~~ & $\rho^{[1]}_{45}$~~ & $\rho^{[1]}_{55}$~~ & $\rho^{[1]}_{54}$~~ & $\zeta^{[1]}_{P,ii}-\rho^{[1]}_{ii}$ & $ \zeta^{[1]}_{L,ii} -\rho^{[1]}_{ii}$ \\
        \hline
        0.12  & 0.4 & $-0.2447$ & $-0.0887$ &  $0.0107$ &  $0.0403$ &  $0.4103$ & $-0.0097$ & $-0.1728$ & $-0.2820$ &  $0.0097$ & $-0.3134$ &  $0.1382$ & $0.0893$ \\
      $(\kappa^\prime_c=0.1254)$  & 0.2&$-0.2602$ & $-0.0927$ &  $0.0353$ &  $0.0423$ &  $0.3927$ & $-0.0025$ & $-0.1344$ & $-0.2837$ &  $0.0000$ & $-0.2974$ &  $0.1447$ & $0.0960$\\
      $(\kappa^\prime_c=0.1280)$ & 0.2 & $-0.2341$ & $-0.0766$ &  $0.0369$ &  $0.0406$ &  $0.4133$ & $-0.0065$ & $-0.1396$ & $-0.2767$ &  $0.0205$ & $-0.3047$ &  $0.1178$ & $0.0682$ \\
              & 0.14 & $-0.2605$ & $-0.0929$ &  $0.0353$ &  $0.0423$ &  $0.3924$ & $-0.0025$ & $-0.1343$ & $-0.2837$ & $-0.0003$ & $-0.2975$ &  $0.1450$ & $0.0963$ \\
              & 0.1 & $-0.2606$ & $-0.0930$ &  $0.0353$ &  $0.0423$ &  $0.3925$ & $-0.0025$ & $-0.1341$ & $-0.2838$ & $-0.0004$ & $-0.2973$ &  $0.1451$  & $0.0963$ \\
        \hline
        0.09  & 0.4 & $-0.1077$ & $-0.0689$ & $-0.2540$ &  $0.0212$ &  $0.5724$ & $-0.0822$ & $-0.5806$ & $-0.2723$ &  $0.0884$ & $-0.4724$ &  $0.1018$ & $0.0519$\\
              & 0.2 & $-0.1156$ & $-0.0698$ & $-0.2385$ &  $0.0223$ &  $0.5629$ & $-0.0780$ & $-0.5571$ & $-0.2728$ &  $0.0839$ & $-0.4631$ &  $0.1038$ & $0.0538$ \\
              & 0.14 & $-0.1183$ & $-0.0706$ & $-0.2350$ &  $0.0226$ &  $0.5601$ & $-0.0768$ & $-0.5510$ & $-0.2732$ &  $0.0821$ & $-0.4606$ &  $0.1051$ & $0.0552$ \\
              & 0.1 & $-0.1202$ & $-0.0708$ & $-0.2310$ &  $0.0229$ &  $0.5577$ & $-0.0758$ & $-0.5450$ & $-0.2733$ &  $0.0810$ & $-0.4584$ &  $0.1055$ & $ 0.0556$ \\
              & 0.05 & $-0.1218$ & $-0.0709$ & $-0.2274$ &  $0.0232$ &  $0.5558$ & $-0.0748$ & $-0.5392$ & $-0.2733$ &  $0.0801$ & $-0.4562$ &  $0.1057$ & $0.0558$ \\
        \hline
        0.06  & 0.4 &  $0.0445$ & $-0.0499$ & $-0.5407$ & $-0.0003$ &  $0.7495$ & $-0.1615$ & $-1.0218$ & $-0.2611$ &  $0.1752$ & $-0.6433$ &  $0.0598$ & $0.0089$ \\
              & 0.2 &  $0.0425$ & $-0.0496$ & $-0.5335$ &  $0.0001$ &  $0.7464$ & $-0.1599$ & $-1.0111$ & $-0.2609$ &  $0.1745$ & $-0.6394$ &  $0.0590$ & $0.0082$ \\
              & 0.14 &  $0.0407$ & $-0.0498$ & $-0.5297$ &  $0.0003$ &  $0.7442$ & $-0.1590$ & $-1.0055$ & $-0.2610$ &  $0.1734$ & $-0.6373$ &  $0.0595$ & $0.0086$ \\
              & 0.1 &  $0.0392$ & $-0.0498$ & $-0.5261$ &  $0.0006$ &  $0.7422$ & $-0.1579$ & $-1.0000$ & $-0.2610$ &  $0.1726$ & $-0.6352$ &  $0.0597$ & $ 0.0088$ \\
        \hline
        0.045 & 0.2 &  $0.1771$ & $-0.0355$ & $-0.7880$ & $-0.0192$ &  $0.9027$ & $-0.2306$ & $-1.4020$ & $-0.2514$ &  $0.2504$ & $-0.7902$ &  $0.0227$ & $-0.0290$ \\
        \hline\hline
    \end{tabular}
\end{sidewaystable}

\section{Charm-quark mass correction}
\label{kappa_tuning_section}
The mass of the charm quark is set by the hopping parameter $\kappa$ in the Fermilab action. We determine the appropriate $\kappa_c$ by requiring that the $D_s$-meson kinetic mass obtained in our lattice simulations agree with the PDG value as described in Ref.~\cite{Bernard:2010fr,Bailey:2014tva}. In practice, initial low-statistics runs with several values of $\kappa$ were performed and used to determine the simulation values $\kappa_c^\prime$ for high-statistics data-production runs. The tuned value of the hopping parameter corresponding to the physical-charm quark mass, $\kappa_c$, was then determined using the high-statistics data. The physical $\kappa_c$ values on each ensemble are listed in Table~\ref{tab:m2diff}.  

\begin{figure}
	\centering
		\includegraphics[width=0.50\textwidth]{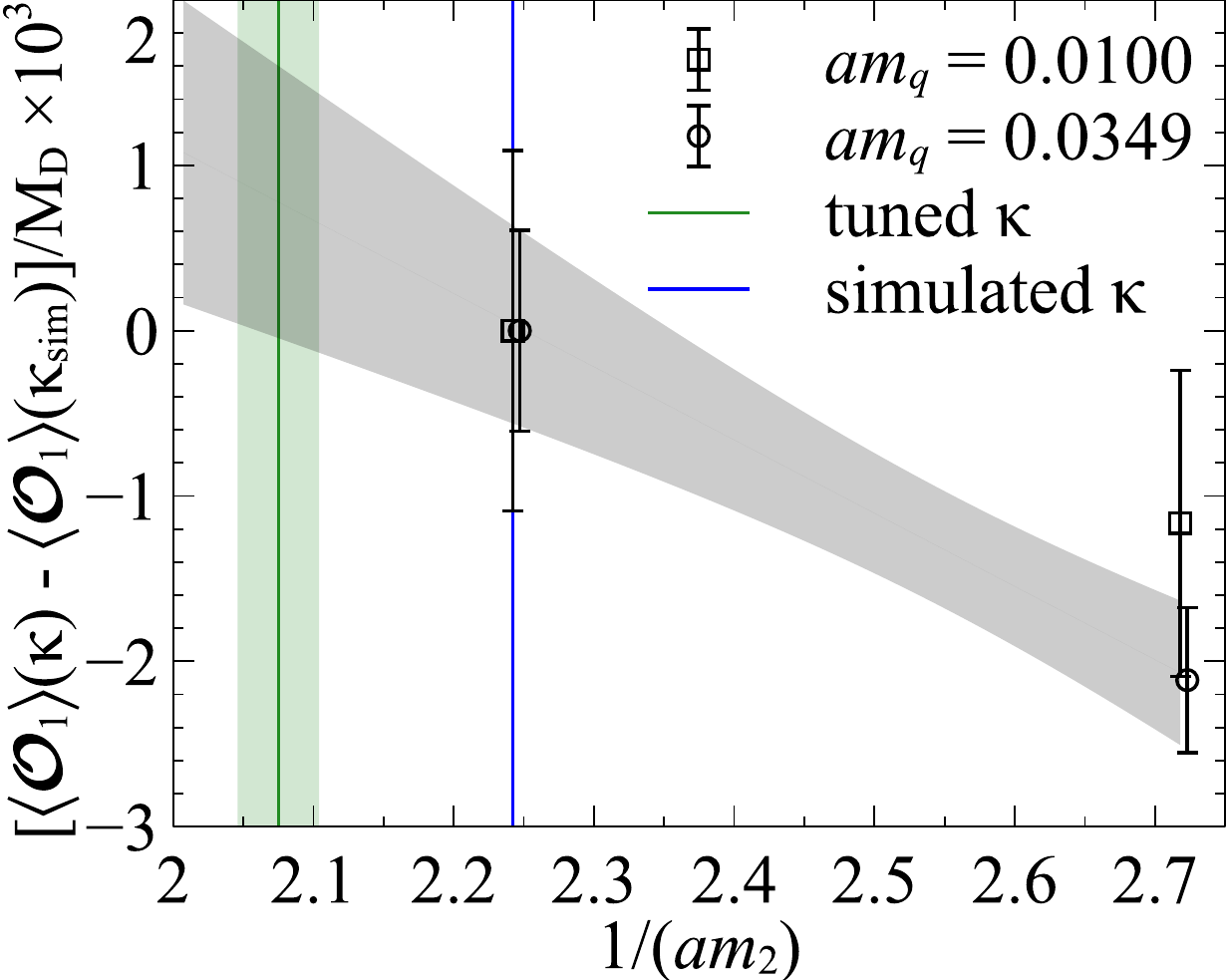}
	\caption{Determination of the slope $\mu_1$ from an unconstrained linear fit in $1/m_2$ to the renormalized matrix element $\mathcal{O}_1$ at two values of $\kappa_c^\prime$. The simulated $\kappa_c^\prime$ value employed in our matrix-element calculation is indicated by the blue vertical line, while the green vertical band shows the tuned $\kappa_c$ with statistical, fit, and $r_1$ uncertainties included.}
	\label{fig:chipt_hqslope}
\end{figure}

We account for the slight difference between our simulation $\kappa_c^\prime$ and the physical $\kappa_c$ by incorporating a charm-quark mass correction into the chiral-continuum fit.  To estimate this correction term, we start with the quark kinetic mass $am_2$, which is related to the hopping parameter as~\cite{ElKhadra:1996mp} 
\begin{equation}
\frac{1}{am_2}=\frac{2}{am_0(2+am_0)}+\frac{1}{1+am_0}
\label{chipt_ktuneM2}
\end{equation}
where $am_0$ is the tadpole-improved bare-quark mass,
\begin{align}
\label{chipt_m0}
am_0=&\frac{1}{2u_0}\left(\frac{1}{\kappa}-\frac{1}{\kappa_{\text{crit}}}\right),
\end{align}
and $\kappa_{\text{crit}}$ corresponds to the value of $\kappa$ where the mass of the pseudoscalar-meson mass vanishes. The nonperturbatively determined values of $\kappa_{\rm crit}$ are listed in Table~\ref{tab:ZV}. From heavy-quark power-counting, we expect the matrix elements to depend upon the heavy-quark mass as $1/m_Q$, which is identified with the kinetic mass in the Fermilab interpretation.  We therefore adjust the matrix elements using a function linear in the inverse kinetic quark mass.  We first compute on each ensemble the difference between the simulated and tuned inverse kinetic mass
\begin{equation}
\Delta (1/(am_2)) = \left.(1/am_2)\right|_{\kappa=\kappa_c}-\left.(1/am_2)\right|_{\kappa=\kappa^\prime_c};
\end{equation}
these values are given in Table~\ref{tab:m2diff}. We then determine the slope of the matrix elements with respect to $1/m_2$ using data with $\kappa_c' = 0.1254$ and 0.1280 on the 0.12~fm, $m_l/m_s=0.2$ ensemble with valence-quark masses $m_q  = 0.0100$ and 0.0349. Table~\ref{tab:slopes} gives the slopes
 \begin{equation} \label{eq:slope}
 \mu_i \equiv \left(\frac{r_1}{a}\right)^4\frac{\Delta(a^3\langle \mathcal{O}_i\rangle/M_D)}{\Delta(1/am_2)}.
\end{equation}
obtained for $i$=1--5 from an unconstrained linear fit of the renormalized matrix elements $\left<\mathcal{O}_i\right>/M_D$ in $1/m_2$, while here $\Delta(1/am_2)$ is the difference between the two simulated inverse kinetic masses.  Figure~\ref{fig:chipt_hqslope} shows an example fit for $\left<\mathcal{O}_1\right>$.

We add the charm-quark mass correction to the chiral-continuum fit as a constrained fit parameter.  This allows us to propagate the error stemming from the uncertainties in $\Delta(1/(am_2))$ and $\mu_i$ into the matrix-element uncertainties reported by the fit. We introduce priors for $\Delta(1/(am_2))$ with the central values and widths given in Table~\ref{tab:m2diff}, where the errors are obtained by propagating the error from $\kappa_c$ through Eq.~(\ref{chipt_ktuneM2})~and~(\ref{chipt_m0}). We take the priors for $\mu_i$ from the linear fits described above and illustrated in Fig.~\ref{fig:chipt_hqslope}; the prior values employed for $\mu_i$ are tabulated in Table~\ref{tab:slopes}.

\begin{table}
\centering
\setlength{\tabcolsep}{4pt}
\caption{Tuned $\kappa_c$ values from Ref.~\cite{Bailey:2014tva}, and differences $\Delta(1/(am_2))$ between the simulated and physical inverse charm-quark kinetic masses on each ensemble. For $\kappa_c$, the first error is from statistics and fitting, and the second is from the uncertainty in the lattice scale $r_1$. For $\Delta(1/(am_2))$ the error is from the uncertainty in the tuned $\kappa_c$.}
\label{tab:m2diff}
\begin{tabular}{l c c r}\hline\hline
 & & \vspace{-1.0em}\\
$\approx a$~(fm)   &  $a{m}'_l/am'_s$  &  ~~$\kappa_c$ & $\Delta(1/(a m_2))\,$   \\ 
 & & \vspace{-1.0em}\\ \hline
0.12 & 0.02/0.05 &  0.12452(15)(16) & $-$0.212(31) \\ 
0.12 & 0.01/0.05 &  0.12423(15)(16)  & $-$0.168(29)\\ 
0.12 & 0.007/0.05 &  0.12423(15)(16) & $-$0.167(29))\\
0.12 & 0.005/0.05 &  0.12423(15)(16) & $-$0.167(29) \\ \hline
0.09 & 0.0124/0.031 &  0.12737(9)(14) & $-$0.089(43)\\
0.09 & 0.0062/0.031 &  0.12722(9)(14) & $-$0.099(42)\\
0.09 & 0.00465/0.031 &  0.12718(9)(14) & $-$0.082(42)\\
0.09 & 0.0031/0.031 &  0.12714(9)(14) & $-$0.092(41)\\
0.09 & 0.00155/0.031 &  0.12710(9)(14) & $-$0.101(41)\\ \hline 
0.06 & 0.0072/0.018 &  0.12964(4)(11) & 0.075(64)\\
0.06 & 0.0036/0.018 & 0.12960(4)(11) & 0.000(63)\\
0.06 & 0.0025/0.018 &  0.12957(4)(11)  & $-$0.016(62)\\
0.06 & 0.0018/0.018 &  0.12955(4)(11)  & $-$0.026(61)\\ \hline 
0.045 & 0.0028/0.014  & 0.130921(16)(7) & $-$0.083(75)\\ \hline\hline
\end{tabular}
\end{table}

\begin{table}
\centering
\caption{Slopes $\mu_i$ defined in Eq.~(\ref{eq:slope}) for the $D$-mixing matrix elements renormalized in the continuum \MSbar-NDR scheme and the BBGLN~\cite{Beneke:1998sy,Beneke:2002rj} and BMU~\cite{Buras:2000if} choices of evanescent
operators in $r_1$ units. The uncertainties of $r_1/a$ are not included in this table, but instead propagated while performing the chiral-continuum extrapolation as discussed in Sec.~\ref{Sec:Xpt_Xfitparameters_constrainedfitparams}. The $r_1/a$ values are precise enough to not affect uncertainties of the slopes at the reported level of precision.}
\label{tab:slopes}
\begin{tabular}{l@{\quad}c@{\quad}cc@{\quad}cc@{\quad}c@{\quad}c}\hline\hline
 & \vspace{-1.0em}\\
$i$ & 1 & \multicolumn{2}{c}{2} & \multicolumn{2}{c}{3} & 4 & 5 \\ 
	& & BMU & BBGLN & BMU & BBGLN \\
& \vspace{-1.0em} \\ \hline
$\mu_i$ & $-$0.248(90)  	& 0.084(45) 	& 0.073(45)	& $-$0.017(22)	& $-$0.011(22)	& $-$0.12(21)	& $-$0.141(90) \\ \hline\hline  
\end{tabular}
\end{table}

\section{Chiral and continuum extrapolation}
\label{chiral_extrapolation}

\subsection{Chiral fit function}
\label{chiralfitfunction}
We extrapolate our renormalized lattice matrix-element results to the physical light-quark mass and continuum limit using  SU(3), partially-quenched, heavy-meson, rooted staggered chiral perturbation theory (HMrS$\chi$PT)~\cite{Aubin:2005aq, Bernard:2013dfa}.  HMrS$\chi$PT describes the dependence of the matrix elements on the light-quark masses and on the lattice spacing from taste-symmetry breaking in the staggered action.  To incorporate additional systematics into the chiral-continuum fit, we supplement the HMrS$\chi$PT expression with terms to parametrize discretization errors from the heavy-quark, light-quark, and gluon actions, the uncertainty in the adjustment from the simulation to physical charm-quark mass, and higher-order terms in the operator matching procedure.  Schematically, the fit function takes the form
\begin{equation}
    F_i = F_i^{\text{logs.}} + F_i^{\text{analytic}} + F_i^{\text{HQ disc.}} +
        F_i^{\alpha_s a^2\text{ gen}} + F_i^{\kappa} + F_i^{\text{\,renorm}}.
\label{eq:schematic_fitfcn}
\end{equation}
Because the lattice spacings on all ensembles differ, we bring all lattice masses and matrix elements into $r_1$ units during the physical point extrapolation. We discuss each term in turn in the following subsections.

\subsubsection{Chiral logarithms}
We work at next-to-leading order (NLO) in HMrS$\chi$PT.  The one-loop chiral logarithms describe nonanalytic dependence on the light-quark masses and lattice spacing, and are
\begin{equation}
\label{eq:schematic_chipt}
    F^{\text{logs}}_i = \beta_i\left(1+\frac{\mathcal{W}_{u\bar{c}}+\mathcal{W}_{c\bar{u}}}{2}+\mathcal{T}_u^{(i)}\right)+\beta^\prime_i\mathcal{Q}_u^{(i)}+\beta_i^{(\xi)}\tilde{\mathcal{T}}_u^{(i,\xi)}+\beta_i^{\prime(\xi)}\tilde{\mathcal{Q}}_u^{(i,\xi)},
\end{equation}
where repeated indices $\xi$ are summed. The coefficients $\beta_i$ and $\beta_i^\prime$ are the leading-order low energy constants (LECs) for the mixing matrix elements $\left<\bar{D}|\op_i|D\right>$ and $\left<\bar{D}^*|\op_i|D^*\right>$, respectively. The terms, $\mathcal{W}$, $\mathcal{T}$, and $\mathcal{Q}$ are the one-loop wavefunction renormalization, tadpole, and sunset corrections; their explicit expressions are given in Eqs.~(63), (82)--(83), and (89) of Ref.~\cite{Bernard:2013dfa}, respectively. The implementation of the staggered action employed in our simulations introduces mixing between spin and taste degrees of freedom.  To account for this, we include the wrong-spin taste-mixing terms $\tilde{\mathcal{T}}$ and $\tilde{\mathcal{Q}}$ in our chiral-continuum fit with coefficients $\beta_i^{(\xi)}$ and $\beta_i^{\prime(\xi)}$, where $\xi=\{\text{P, A, T, V, I}\}$ labels the different taste contributions. For the wrong-spin terms $\tilde{\mathcal{T}}$ and $\tilde{\mathcal{Q}}$, we follow a different notation than in Ref.~\cite{Bernard:2013dfa} in order to separate the LECs $\beta_i^{(\prime)}$ from the one-loop diagram functions.  The relationships between $\tilde{\mathcal{T}}$ and $\tilde{\mathcal{Q}}$ in Eq.~(\ref{eq:schematic_chipt}) and the chiral-logarithm functions in Ref.~\cite{Bernard:2013dfa} are given in Eqs.~(C1a)--(C2e) of Ref.~\cite{Bazavov:2016nty}. The coefficients $\beta^{(\xi)}_i$ and $\beta^{\prime(\xi)}_i$ are not all independent; for convenience, their relationships are given in Table~\ref{tab:betakappa}. Inspection of Table~\ref{tab:betakappa} shows that the matrix elements
$\{\left<\op_1\right>, \left<\op_2\right>,
\left<\op_3\right>\}$ mix, as do $\{\left<\op_4\right>,
\left<\op_5\right>\}$.

To account for the discrete momentum spectrum dictated by the finite lattice spatial size and periodic boundary conditions, we use the finite-volume expressions for the NLO chiral logarithms, which are obtained by replacing the integrals over loop momenta by discrete sums. The explicit expressions are given in Eqs.~(63) and (64) of Ref.~\cite{Bernard:2013dfa}.

We work at leading order in the heavy-meson expansion, but include the largest $1/M_D$ effects from the hyperfine splitting $\Delta^*\equiv M_{D^*}-M_D$ and the SU(3)-flavor splitting $\delta_{su}\equiv M_{D_s}-M_{D^0}$.  The parameter that characterizes the flavor splitting in HMrS$\chi$PT is $\lambda_1 \equiv \delta_{su} /\left(M^2_{\eta_s}-M^2_{\pi^0}\right)$, where $M_{\eta_s}$ is the mass of a theoretical $\bar{s}s$ bound state.

\begin{table*}
\caption{\label{tab:betakappa} Relationships between the coefficients of the wrong-spin terms in Eq.~(\ref{eq:schematic_chipt}). The relationship between the primed LECs are analogous to those between the unprimed LECs, but with all entries in the table $\beta_j\rightarrow \beta_j^\prime$ in $\beta^{\prime(\xi)}_i$.}
\begin{tabular}{c@{\quad}c@{\quad}|c@{\quad}c@{\quad}c@{\quad}c@{\quad}c@{\quad}}
\hline
\hline
& & \multicolumn{5}{c}{$\xi$} \\
 & & P & A & T & V & I \\
\hline
&$1$ & $\beta_1$ & $-8\beta_2-8\beta_3$ & $-6\beta_1$ & $8\beta_2+8\beta_3$ & $\beta_1$ \\
&$2$ & $\beta_2$ & $-\beta_1$ & $-2\beta_2+4\beta_3$ & $\beta_1$ & $\beta_2$\\
$i$ &$3$ & $\beta_3$ & $-\beta_1$ & $-4\beta_2-2\beta_3$ & $\beta_1$ & $\beta_3$\\
&$4$ & $-\beta_4$ & $-2\beta_5$ & $0$ & $-2\beta_5$ & $\beta_4$\\
&$5$ & $-\beta_5$ & $ -2\beta_4$ & $0$ & $-2\beta_4$ & $\beta_5$\\
\hline
\hline
\end{tabular}
\end{table*}

\subsubsection{Analytic terms in the chiral expansion}

The analytic terms
\begin{equation}
\label{eq:ChPTAnalTerms}
	F_i^{\rm analytic} = F_i^{\rm NLO} + F_i^{\rm NNLO} + F_i^{\rm N^3LO}
\end{equation}
are simple polynomials in the light-quark masses and lattice spacing.   At nonzero lattice spacing, taste-symmetry breaking splits the masses of pions with different taste representations $\xi = \{P, A, T, V, I\}$.  The tree-level staggered $\chi$PT relationship between the mass $M^2_{ab,\xi}$ of a pion with taste $\xi$ and the constituent staggered quark masses $m_a$ and $m_b$ is
\begin{equation}
M^2_{ab,\xi}=B_0(m_a + m_b)+a^2\Delta_\xi,
\label{squared_pion_mass}
\end{equation}
where the $a^2\Delta_\xi$ are the taste splittings, and the leading-order LEC $B_0$ is related to the chiral condensate.
Following Ref.~\cite{Bazavov:2011aa}, we construct the analytic terms in Eq.~(\ref{eq:ChPTAnalTerms}) using the dimensionless variables
\begin{equation}
	x_q \equiv \frac{M_{qq}^2}{8\pi^2 f^2_\pi} ,
\end{equation}
where $M_{qq}$ is the mass of the taste-pseudoscalar $\bar{q}q$ meson ($q = u, l, s$) and $f_\pi$ is the pion decay constant, and
\begin{equation}
	x_{\bar{\Delta}}\equiv \frac{a^2\bar{\Delta}}{8\pi^2 f_\pi^2 } ,
\end{equation}
where $a^2\bar{\Delta} = \frac{1}{16}\sum_{\xi} w_\xi a^2\Delta_\xi$ is the average taste splitting, and the weight factors for $\xi = \{P, A, T, V, I\}$ are $w_\xi = \{1, 4, 6, 4, 1\}$, respectively.  The coefficients of the analytic terms are expected to be of $\mathrm{O}(1)$ from chiral power counting when written in terms of $x_q$ and $x_{\bar{\Delta}}$.

The NLO, NNLO, and NNNLO analytic terms are obtained by forming all combinations of $x_i$ ($i=u,d,s,\bar{\Delta}$) linear, quadratic, and cubic in $x_i$, respectively. We include the full set of NLO and NNLO analytic terms in our base fit used to obtain our matrix-element central values.  They are:
\begin{align}
F_i^\text{NLO} = &\big [ c_{0,i} x_u + c_{1,i}(2x_l+x_s) + c_{2,i} x_{\bar{\Delta}} \big]  \beta_i,   \label{chipt_NLO}  \\
F_i^\text{NNLO} = & \big[ d_{0,i} x_u x_{\bar{\Delta}} + d_{1,i} (2x_l+x_s) x_{\bar{\Delta}} + d_{2,i} (2x_l+x_s) x_u  \nonumber\\
& + d_{3,i} x^2_u +d_{4,i} (2x_l+x_s)^2 + d_{5,i} x^2_{\bar{\Delta}} + d_{6,i}(2x_l^2 + x_s^2) \big] \beta_i ,\label{chipt_NNLO} 
\end{align}
where $c_i$ and $d_i$ are LECs of the theory. The inclusion of NLO terms is needed to absorb the dependence of the nonanalytic one-loop terms in Eq.~(\ref{eq:schematic_chipt}) on the scale $\Lambda_\chi$ in the chiral logarithms, while the NNLO terms capture higher-order effects that might contaminate the lower-order LECs. Although the NNLO terms are not needed to obtain an acceptable fit, they improve the $\chi^2_{\rm aug}/{\rm dof}$. We also perform fits including NNNLO analytic terms to check for fit stability and look for truncation errors.  The N$^3$LO terms are:
\begin{align}
F^{\text{N$^3$LO}}_i=&\big[  e_{0,i} x_q^2x_{\bar{\Delta}} + e_{1,i} x_q (2x_l+x_h)x_{\bar{\Delta}} + e_{2,i} x_q x_{\bar{\Delta}}^2+e_{3,i} x_q^2 (2x_l+x_h) \nonumber \\
&+ e_{4,i} x_q^3 + e_{5,i} x_q(2x_l+x_h)^2+ e_{6,i} (2x_l+x_h)^2x_{\bar{\Delta}} + e_{7,i}(2x_l+x_h)x_{\bar{\Delta}}^2\nonumber\\
&+ e_{8,i}(2x_l+x_h)^3 + e_{9,i}(2x_l+x_h)(2x_l^2+x_h^2)  + e_{10,i}x_{\bar{\Delta}}^3 \nonumber \\
&+ e_{11,i}(2x_l^2+x_h^2)x_{\bar{\Delta}}+ e_{12,i}(2x_l^3+x_h^3) + e_{13,i}x_q(2x_l^2+x_h^2)\big] \beta_i  .\label{chipt_NNNLO}
\end{align}

\subsubsection{Heavy quark discretization effects}
We parametrize the leading heavy-quark discretization errors of $\order(\alpha_s a, a^2,a^3)$ in our data by adding the terms $F_i^{\text{HQ disc.}}$ to our fit function:
\begin{equation}
\label{chiralfit_hqet}
F_i^\text{HQ disc.} = F_i^{\alpha_s a \text{ HQ}}+F_i^{a^2 \text{ HQ}}+F_i^{a^3 \text{ HQ}},
\end{equation}
where
\begin{align}
F_i^{\alpha_s a\text{\,HQ}} = & \left[z_{B}^i \left(a\Lambda_{\text{HQ}}\right)f_B(m_0a) + z^i_{3} \left(a\Lambda_{\text{HQ}}\right)f_3(m_0a)\right]\beta_i\,,\label{eq:HQ_a} \\
F_i^{a^2\text{\,HQ}} = & \left[z^i_E \left(a\Lambda_{\text{HQ}}\right)^2f_E(m_0a) + z^i_X \left(a\Lambda_{\text{HQ}}\right)^2f_X(m_0a) + z^i_Y \left(a\Lambda_{\text{HQ}}\right)^2 f_Y(m_0a)\right] \beta_i\,, \\
F_i^{a^3\text{\,HQ}} = & \left[z^i_2 \left(a\Lambda_{\text{HQ}}\right)^3 f_2(m_0a)\right]\beta_i\,, \label{eq:HQ_a3}
\end{align} 
and the $\beta_i$ are the same LECs as in Eq.~(\ref{eq:schematic_chipt}). The fit parameters $z_h^i$ combined with powers of $a\Lambda_{\text{HQ}}$ represent the HQET matrix elements. The ``mismatch functions" $f_h(m_0a)$ with $h\in\left\{B,3,E,X,Y,2\right\}$ are smoothly varying functions of the bare lattice heavy-quark mass that encapsulate the short-distance differences between the lattice and continuum-QCD action and operator descriptions.

The tree-level $\order(a^2)$ and $\order(a^3)$ mismatch functions of the action were calculated in Ref.~\cite{Oktay:2008ex} by performing a matching calculation between lattice HQET and continuum HQET. The tree-level $\order(a^2)$ mismatch functions of the spinor, and consequently, the 4-quark operators, were worked out in Ref.~\cite{ElKhadra:1996mp}. For $\order(\alpha_s a)$ errors, mismatches of the action and operator are modeled, using the Fermilab interpretation, by a smooth function that has the correct $am_0\rightarrow 0$ and $am_0\rightarrow \infty$ limits. A complete list of the functional forms of the $f_h(m_0a)$ employed here is given in Appendix A of Ref.~\cite{Bazavov:2011aa}.

We also consider generic $\order{(\alpha_s a^2)}$ discretization errors when studying the stability of our fits by including the term
\begin{equation}
	F_i^{\alpha_s a^2\,\text{gen}} = h_{0,i} \frac{\alpha_s a^2}{r_1^2} \beta_i. \label{chipt_generic}
\end{equation}
This term receives one-loop corrections from the light-quark and gluon actions, and from heavy-quark discretization errors of higher order than those in Eqs.~(\ref{eq:HQ_a})--(\ref{eq:HQ_a3}).

\subsubsection{Heavy-quark mass adjustment}
Following Sec.~\ref{kappa_tuning_section}, we adjust the matrix elements for the slight difference between the simulation and physical charm-quark masses within the chiral-continuum fit by adding the correction term
\begin{equation}
\label{eq:FiKappa} 
F_i^\kappa =\mu_i \Delta\left(\frac{1}{r_1m_2}\right) \,.
\end{equation}
We propagate the uncertainties in the slopes $\mu_i$ and differences $\Delta\left(\frac{1}{r_1m_2}\right)$ to the final fit error by including them as constrained fit parameters with Gaussian priors.

\subsubsection{Renormalization errors}

As discussed in Sec.~\ref{sec_renorm}, we calculate the renormalization coefficients at one loop; therefore, truncation errors start at $\order(\alpha_s^2, \alpha_s \Lambda_{\text{QCD}}/m_c)$. The $\order{(\alpha_s^2)}$ truncation errors are estimated in the chiral-continuum fit by adding the terms
\begin{align}
F^{\alpha_s^2\text{\,renorm}}_i = \alpha_s^2(q^*)\left(\rho^{[2]}_{ii} \beta_i + \rho^{[2]}_{ij} \beta_j \right)
\label{chipt_renorm2}
\end{align}
where the coefficients $\rho^{[2]}_{ii}$ and $\rho^{[2]}_{ij}$ are free parameters (with $i \neq j$), and the $\beta_i$ are the leading-order LECs for matrix elements $\left<\op_i\right>$ defined in Eq.~(\ref{eq:schematic_chipt}). The second term in Eq.(\ref{chipt_renorm2}) parametrizes the mixing between operators $\op_i$ under renormalization. We evaluate the renormalized coupling $\alpha_s$ at $q^* = 2/a$.
In tests of fit stability, discussed below, we consider effects of $\order{(\alpha_s^3)}$ by adding the terms
\begin{equation}
F^{\alpha_s^3\text{\,renorm}}_i = \alpha_s^3(q^*)\left(\rho^{[3]}_{ii} \beta_i + \rho^{[3]}_{ij} \beta_j \right).
\label{chipt_renorm3}
\end{equation}
The coefficients are set to $\rho^{[3]}_{ii} = \rho^{[3]}_{ij} = 0$ in our base fit.

Errors from omitted $\order{(\alpha_s \Lambda_{\text{QCD}}/m_c)}$ terms in the perturbative calculation are absorbed by the coefficients $z_{\{B,3\}}$ of the mismatch functions $f_{\{B,3\}}(m_0a)$, which also scale as $a\Lambda_{\text{QCD}}/m_c$

\subsection{Chiral-continuum fit parameters}
\label{chiralparameters}
In the following section, we discuss the priors chosen for the chiral-continuum fit parameters. Briefly, we employ loose constraints based on power counting for the coefficients associated with the chiral and perturbative expansions in $\alpha_s$, and for those of the heavy-quark discretization terms.  We incorporate the parametric uncertainties from our fit inputs into the final fit error by including them as constrained fit parameters with Gaussian prior widths corresponding to the errors on the inputs.  We fix the values of a small number of inputs for which the uncertainty contribution to the final fit is negligible.

\subsubsection{Loosely constrained fit parameters}
The LECs of HMrS$\chi$PT in Eqs.~(\ref{eq:schematic_chipt}) and~Eqs.~(\ref{chipt_NLO})--(\ref{chipt_NNNLO}) are expected to be of $\order{(1)}$.  We constrain them only loosely to allow their values to be determined by the data; the priors improve fit stability when adding higher-order terms in the chiral expansion. The priors for the coefficients of the chiral-logarithms $\beta_i$ and $\beta^\prime_i$ are set to
\begin{align}
\beta_{1,3,4,5} = 1(1)\,, \quad \beta_2 = -1(1)\,, \quad \beta_{2,3,4,5}^\prime = 0(1),
\end{align}
where the central values are rough guesses based on the correlator fits. We take very wide prior widths for the coefficients of the NLO analytic terms, which are well determined by the data:
\begin{align}
c_{n,i} & = 0(10), \quad n\in\{0,1,2\}\,,
\end{align}
and use prior widths of $\order{(1)}$ as motivated by chiral power counting for the coefficients of the NNLO (and N$^3$LO) analytic terms:
\begin{align}
d_{m,i}&=0(1), \quad m\in\{0,1,2,3,4,5,6\}, \\
e_{m,i}&=0(1), \quad m\in\{0,1,2,3,4,5,6,7,8,9,10,11,12,13\}.
\end{align}
Recall that our base fit includes terms only through NNLO.

When we include the generic discretization term $F_i^{\alpha_s a^2\,\text{gen}} $ in our fit, we constrain its coefficient to be $h_{0,i} = 0(1)$.

We choose priors for the heavy-quark discretization terms based on HQET power-counting.  We expect the individual coefficients $z_i$ to be of $\order(1)$.  In some cases, however, more than one operator shares the same mismatch function.  We therefore choose the width for each prior such that the width-squared equals the number of terms sharing the corresponding mismatch function.  The priors for the $z_i$ are given in Table~\ref{tab:hqpriors}.  The heavy-quark discretization terms also depend upon the scale $\Lambda_{\rm HQ}$, which is the cutoff of the effective theory.  We use $\Lambda_{\text{HQ}}=500~\text{MeV}$ based on studying the lattice-spacing dependence of our full-QCD matrix-element data adjusted to the same sea-quark masses via the chiral-continuum fit.  Our physical continuum-limit matrix-element results are insensitive to reasonable variations in $\Lambda_{\rm HQ}$.

\begin{table*}
\caption{\label{tab:hqpriors} Priors for heavy-quark discretization terms in Eqs.~(\ref{eq:HQ_a})--(\ref{eq:HQ_a3}). }
\begin{tabular}{c@{\quad}c@{\quad}c@{\quad}c@{\quad}c@{\quad}c@{\quad}}
\hline\hline
 $z_B$ & $z_3$ & $z_E$ & $z_X$ & $z_Y$ & $z_2$\\
\hline
$0(2)$ & $0(\sqrt{5})$ & $0(2\sqrt{2})$ & $0(2\sqrt{2})$ & $0(2)$ & 0(2)\\
\hline\hline
\end{tabular}
\end{table*}

The priors for the unknown two-loop coefficients $\rho^{[2]}_{ii}$ and $\rho^{[2]}_{ij}$ are set to
\begin{align}
\rho^{[2]}_{ii}=0(1), \quad \rho^{[2]}_{ij}=0(1).
\end{align}
When $\order{(\alpha_s^3)}$ terms are included, we use the same constraints for the higher-order coefficients $\rho^{[3]}_{ii}=0(1)$ and $\rho^{[3]}_{ij}=0(1)$.  These values are consistent with the observation that the one-loop coefficients in Table~\ref{tab:rhos} are at most of $\order{(1)}$.

\subsubsection{Constrained fit parameters}
\label{Sec:Xpt_Xfitparameters_constrainedfitparams}
As discussed in Sec.~\ref{subsec:EnsValParams}, we bring our renormalized matrix-element results on all ensembles into the same units before the chiral-continuum extrapolation using the intermediate scale $r_1/a$. We also convert all fit inputs taken from experiment into $r_1$ units using the physical scale $r_1$ in fm. We propagate the uncertainties in $r_1/a$ values to our fit parameter by including them as constrained fit parameters with prior central values and widths given by their values and errors in Table~\ref{tab:MILC_ensemble}. The values of $r_1/a$ are correlated between ensembles because they are obtained from a fit of the data on all MILC asqtad ensembles to a smooth function of the coupling $\beta$~\cite{Allton:1996kr,Bazavov:2009bb} of Appendix~\ref{app:r1acorrelations}. We include the correlations between $r_1/a$ values in our fit; the correlation matrix is given in Table~\ref{tab:r1a_correlation}, while double-precision values for $r_1/a$ and its correlations are provided as supplementary material. For the physical scale $r_1$, we use the prior $r_1 = 0.3117(22)$~fm taken from Ref.~\cite{Bazavov:2009bb}.

The coefficients of the one-loop chiral logarithms depend upon the light pseudoscalar-meson decay constant $f_\pi$ and the $D^*$-$D$-$\pi$ coupling $g_{D^*D\pi}$. We constrain $f_\pi$ to the PDG value of the pion decay constant~\cite{Rosner:2015wva}
\begin{align}
f_\pi = 130.50(1)(3)(13)~\text{MeV}\,, \quad \left(r_1f_\pi\right)^2 = 0.04249(61) \label{convertedfpi}
\end{align}
where the uncertainties on $f_\pi$ are from $\Gamma$, $|V_{ud}|$ and higher-order radiative corrections, respectively. The error on the fit input $r_1 f_\pi$ includes the uncertainties from both $f_\pi^{\rm PDG}$ and $r_1$ added in quadrature.
The coupling $g_{D^*D\pi}$ has been studied in unquenched lattice QCD with 2 flavors, yielding $g_{D^*D\pi}^{N_f = 2} = 0.53(3)(3)$~\cite{Becirevic:2012pf}, and with three flavors, yielding $g_{D^*D\pi}^{N_f = 2+1} = 0.55(6)$~\cite{Can:2012tx}. Based on these results, we constrain the coupling in our fit to be
\begin{equation}
g_{D^*D\pi}=0.53(8) \,,
\end{equation}
which covers the 1$\sigma$ ranges of both.

The one-loop HMrS$\chi$PT chiral logarithms also depend upon the parameters $a^2\delta'_{A,V}$, which multiply contributions from quark-disconnected ``hairpin" diagrams.  Because the hairpin contributions arise from taste-symmetry breaking, the coefficients $a^2\delta'_{A,V}$ scale with lattice spacing approximately as $\alpha_s^2 a^2$.  We constrain their values in our fit at $a\approx 0.12$~fm to the determinations from the MILC Collaboration's chiral-continuum fit of pion and kaon masses and decay constants in Ref.~\cite{Aubin:2004fs}:
\begin{align}
r_1^2a^2\delta'_V = 0.00(7)\,, \quad r_1^2a^2\delta'_A = -0.28(6) \,.
\end{align}
For the remaining lattice spacings, we scale these values by the weighted average of the taste splittings $\bar\Delta_a / \bar\Delta_{0.12~{\rm fm}}$.

We constrain the hyperfine and flavor splittings in our fit to the experimentally-measured values.  For the $D$-meson system, $\Delta^*\equiv M_{D^*}-M_D = 142.020(71)~\text{MeV}$~\cite{Olive:2016xmw}, which corresponds to
\begin{align}
r_1  \Delta^*=& 0.2243(16) 
\end{align}
including the uncertainty on $r_1$. The SU(3) flavor splitting $\delta_{su} = M_{D_s}-M_{D_\pm} = 98.69(13)$~MeV~\cite{Olive:2016xmw}.  Combining this with $M_{\pi^0} = 134.9766(6)$~MeV~\cite{Olive:2016xmw} and $M_{\eta_s} = 685.8(4.0)$~MeV~\cite{Davies:2009tsa},  
we obtain
\begin{align}
\lambda_1 = 0.2183(28)~\text{GeV}^{-1}\,, \rm~or~ \lambda_1/r_1 = 0.1382(20).
\end{align}

We incorporate the uncertainty from the charm-quark-mass correction into the fit error by constraining the
slopes $\mu_i$ and differences in inverse kinetic masses $\Delta (1/(r_1m_2))$ that enter the correction
term $F_i^\kappa$, Eq.~(\ref{eq:FiKappa}), with Gaussian priors.
We fix the prior central values and errors to the results obtained from our fits of the charm-quark-mass
dependence of the matrix elements in Sec.~\ref{kappa_tuning_section}.
The values are listed in Tables~\ref{tab:m2diff}--\ref{tab:slopes}, and include the errors from statistics,
fitting, and $r_1$.

We extrapolate the matrix elements to the physical light-quark masses given in Table~VIII of~\cite{Bailey:2014tva},
\begin{align}
r_1m_u = 2.284(97)\times 10^{-3}\,, \quad r_1\hat{m} = 3.61(12)\times 10^{-3}\,, \quad r_1m_s = 99.2(3.0)\times10^{-3}
\end{align}
by evaluating the fit function with the valence-light quark mass fixed to $m_u$, and the light and heavy sea-quark masses fixed to $\hat{m} \equiv (m_u + m_d)/2$ and $m_s$, respectively.  The errors on the physical light-quark masses include statistics and the dominant systematic uncertainties from the chiral-continuum extrapolation $r_1$. Finally, we convert the chiral-continuum extrapolated matrix elements to the relativistic normalization by dividing by the experimental $D^0$-meson mass~\cite{Olive:2016xmw}
\begin{equation}
    M_{D^0} = 1864.83(05)~\text{MeV}.
\end{equation}

\subsubsection{Fixed inputs}

We fix the pseudoscalar-meson taste splittings and the leading-order LEC $B_0$ in the tree-level expression for the squared pion mass, Eq.~(\ref{squared_pion_mass}), in the fit because their uncertainties are negligible compared to other contributions to the error. We use the values given in Table~\ref{tbl:tastesplittings}, which were obtained from an analysis of the staggered light pseudoscalar-meson spectrum.

\begin{table}
\caption{Taste splitting and leading-order LEC $r_1 B_0$ for the lattice spacings analyzed in this work, and in the continuum~\cite{Bailey:2014tva}. The labels $A, T, V, I$ denote the axial-vector, tensor, vector, and scalar tastes respectively.}
\label{tbl:tastesplittings}
\begin{tabular}{c  c c c c  c}
\hline\hline
    $\approx a$ (fm) & $r_1^2a^2\Delta_A$ & $r_1^2a^2\Delta_T$ & $r_1^2a^2\Delta_V$ & $r_1^2a^2\Delta_I$ &  $r_1B_0$ \\ 
    \hline 
    0.12    &  0.2270               & 0.3661                & 0.4803                & 0.6008                & 6.832 \\       
    0.09    &  0.0747               & 0.1238                & 0.1593                & 0.2207                & 6.639  \\
    0.06    &  0.0263               & 0.0430                & 0.0574                & 0.0704                & 6.487  \\
    0.045   &  0.0104               & 0.0170                & 0.0227                & 0.0278                & 6.417  \\
    continuum & 0 & 0 & 0 & 0 & 6.015\\
\hline\hline    
\end{tabular}
\end{table}

We fix the scale $\Lambda_\chi$ in the chiral logarithms to the experimental $\rho$-meson mass $M_\rho^{\rm PDG} = 775~\text{MeV}$~\cite{Olive:2016xmw}, which corresponds to a value in $r_1$ units of $\left(\Lambda_\chi/r_1\right)^2 =  1.5$. We have checked that our physical continuum-limit matrix-element results are insensitive to reasonable variations in $\Lambda_\chi$.

\subsection{Chiral-continuum fit results}

\begin{figure}
	\centering
    \includegraphics[width=0.97\textwidth]{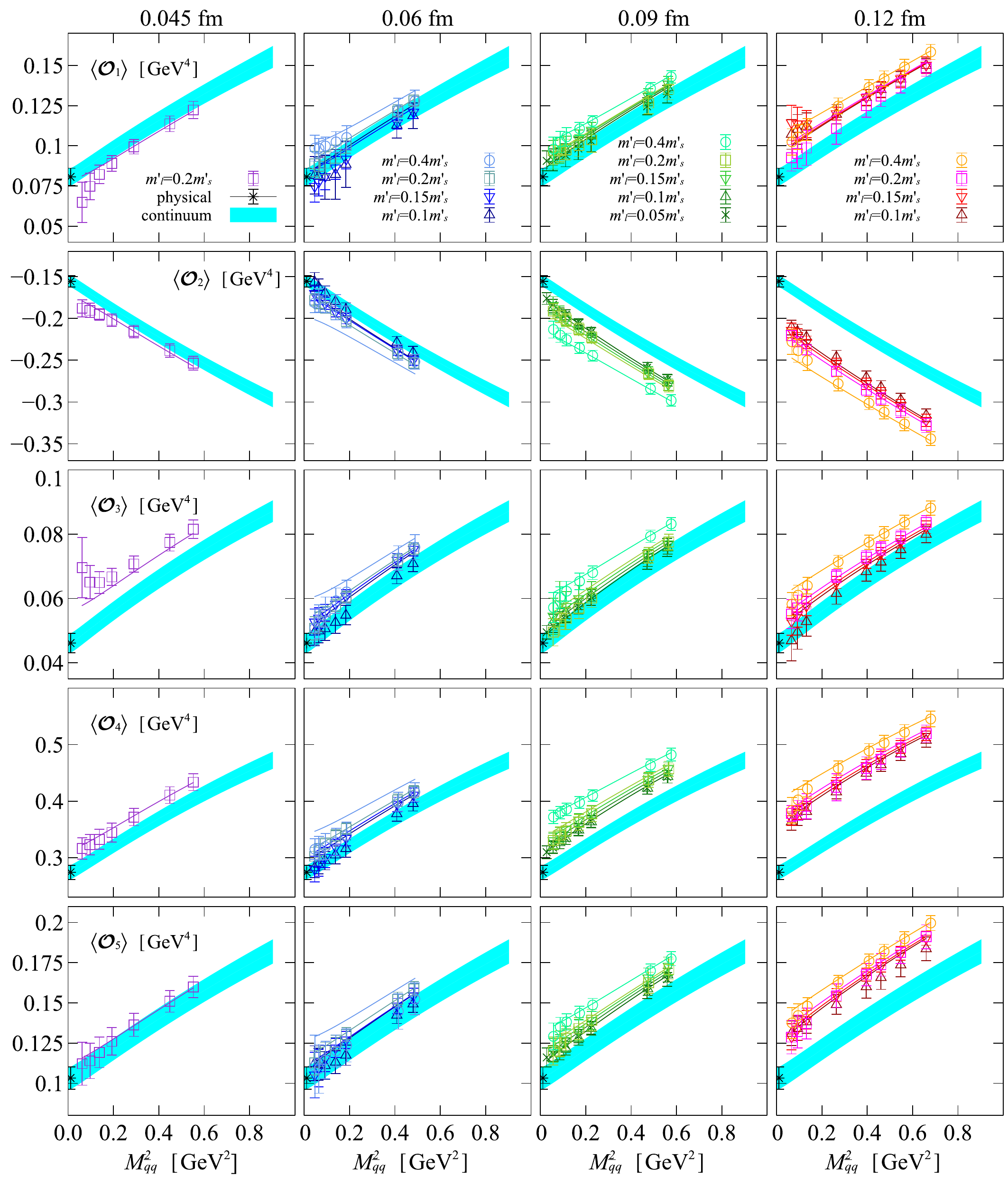}
	\caption{({\it Color online.}) Chiral-continuum extrapolation of the neutral $D$-mixing matrix elements from a combined, correlated fit to all data.  The panels show from left-to-right the data on lattice spacings $a \approx$ 0.045--0.12 fm, and from top-to-bottom the matrix elements of operators $\langle\op_1\rangle$--$\langle\op_5\rangle$.  The solid lines are the fit results evaluated at the sea-quark-mass ratio $m_l^\prime / m_s^\prime$ of the corresponding color in the legend.  The cyan band shows the continuum fit curve, while the physical-mass results, at the lower-left corner of each panel, are denoted by the black stars with errors.  The correlated $\chi^2_{\text{aug}}/\text{dof}=122.4/510$.}
	\label{fig:chipt_fitresult}
\end{figure}

Our base fit used to obtain our matrix-element central values is to the function
\begin{equation}
F_i^{\text{base}} = F_i^{\text{logs.}} + F_i^{\text{NLO}} + F_i^{\text{NNLO}} + F_i^{\text{HQ disc.}} + F_i^{\kappa} + F_i^{\alpha_s^2\text{\,renorm}} \,,
\label{eq:chipt_fitfcn}
\end{equation}
and includes terms to account for errors from truncating the chiral expansion, discretization errors from taste-symmetry breaking, heavy-quark discretization errors, errors from omitted higher order terms in the renormalization factors, and errors in the charm-quark-mass correction factors. We fit the renormalized lattice matrix elements for all five operators simultaneously including statistical correlations between data on the same ensembles. This reduces the error on the physical continuum-limit matrix elements, which share common LECs in HMrS$\chi$PT and mix under renormalization. Figure~\ref{fig:chipt_fitresult} shows our preferred chiral-continuum extrapolation as a function of the squared valence-meson mass $M_{\bar{q}q}^2$, which is approximately linear in the valence light-quark mass $m_q$. We obtain a good fit with a correlated $\chi^2_{\text{aug}}/\text{dof}=122.4/510$, where the quantity $\chi^2_{\text{aug}}/\text{dof}$, which is suitable for assessing the quality of constrained fits, is defined in Eq.~(7.27) of Ref.~\cite{Bazavov:2016nty}.

\section{Systematic error analysis}

\label{error_analysis}
We now discuss all sources of systematic error that contribute to our $D$-meson-mixing matrix-element uncertainties and provide complete error budgets. We begin, in Sec.~\ref{subsec:FitErrors}, with a discussion of errors that are included in the chiral-continuum fit. Next, in Sec.~\ref{subsec:OtherErrors}, we estimate the remaining contributions that must be added to the fit error {\it a posteriori} to obtain the total error. The one exception is the error due to the omission of charmed sea quarks, which we present separately in Sec.~\ref{subsec:CharmSea}. Throughout these sections we refer to alternate fits that we used to test error saturation. Figures in Sec.~\ref{subsec:ErrorSummary} show the results of these alternate fits and additional consistency checks of our fit results and error estimates. Complete error budgets for all five $D$-mixing matrix elements are listed in Table~\ref{tbl:ME_total_error}.

\subsection{Base chiral-continuum fit errors}
\label{subsec:FitErrors}

As described previously, we constrain the parameters in our chiral-continuum fit with Gaussian priors.  This enables us to account for the uncertainties in our input parameters, as well as to include higher-order terms in the chiral and heavy-quark expansions thereby incorporating possible truncation errors.  We use the dependence of the best-fit parameters on each piece of information, including correlations, to separate the total fit error into approximate suberrors.  The approximate breakdown of the total fit error into the suberrors for each matrix element is shown in Table~\ref{tbl:ME_fit_error}.  The first column shows the statistical error, which is obtained from the quadrature sum of the errors from all data points.  The other suberrors are discussed in the following Secs.~\ref{subsec:InputErrs}--\ref{subsec:RenormErrs}. The three dominant sources of error for all matrix elements are statistics, matching, and heavy quark discretization effects, each of which contribute a similar amount to the total errors. The errors from tuning the simulation $c$-quark masses and other inputs, from the extrapolation to the physical light-quark mass and the continuum, and from the relative lattice-spacing determination all contribute at the percent or subpercent level.

\begin{table}[tp]
\caption{Breakdown of the chiral-continuum fit error.  The labels and estimation procedure are described in the text.
Entries are in percent.
\label{tbl:ME_fit_error}}
\begin{tabular}{c@{\quad}c@{\quad}c@{\quad}c@{\quad}c@{\quad}c@{\quad}c@{\quad}c@{\quad}c@{\quad}c@{\quad}c@{\quad}c@{\quad}c@{\quad}c}\hline\hline
                          & stat. & inputs &$\kappa$ tuning & matching & chiral & LQ disc & HQ disc & $r_1/a$ & fit total \\
\hline
$\langle\op_1\rangle$ & 3.5 & 0.6    & 1.5                    & 3.8         & 1.3    & 0.6      & 3.1        &   0.4           & 6.4 \\
$\langle\op_2\rangle$ & 1.8 & 0.5    & 0.4                    & 2.2         & 0.8    & 0.4      & 2.4        &   0.5           & 4.0\\
$\langle\op_3\rangle$ & 3.1 & 0.3    & 0.6                    & 3.8         & 1.3    & 0.5      & 3.6        &   0.4           & 6.3\\ 
$\langle\op_4\rangle$ & 2.2 & 0.6    & 0.5                    & 2.0         & 0.9    & 0.3      & 2.6        &   0.5           & 4.2\\
$\langle\op_5\rangle$ & 3.0 & 0.7    & 0.5                    & 4.1         & 1.5    & 0.5      & 3.5        &   0.3           & 6.5\\ 
\hline\hline
\end{tabular}
\end{table}

\subsubsection{Parametric inputs}
\label{subsec:InputErrs}
The ``inputs" column of Table~\ref{tbl:ME_fit_error} is given by the quadrature sum of the error contributions from most of the input parameters (the error from $r_1/a$ is considered separately), which are constrained with Gaussian prior widths given by their estimated uncertainties. The largest contribution to the ``inputs" error is from the uncertainty in the coupling $g_{D^*D\pi}$.   The uncertainties from the parameters in the chiral logarithms $f_\pi, \Delta^*, \lambda_1, \delta_V^\prime$, and $\delta_A^\prime$ are subdominant.  The input errors also include the uncertainties from the physical $u$-quark and $D^0$-meson masses, which are used to fix the physical-point in the chiral extrapolation and to convert the matrix-element results to physical units. 

The parametric error from the pion decay constant is already included in the ``inputs" uncertainty.  We also check for stability of the chiral-continuum extrapolation against reasonable changes in the decay constant, which provides a measure of the error due to truncating the chiral expansion.  We replace $f_\pi$ in the coefficient of the chiral logarithms with the PDG value of $f_{K^{\pm}} = 155.6(4)$~\cite{Rosner:2015wva}, which corresponds to
\begin{align}
r_1f_{K^\pm} = & 0.2458(18).
\end{align}
This leads to only a tiny shift in the matrix elements, as shown by the fit variation labeled ``$f_K$ vs. $f_\pi$'' in Figs.~\ref{fig:chipt_fit_variationO1} and~\ref{fig:chipt_fit_variationO2O5}, indicating that the fit error indeed encompasses the chiral truncation error.

\subsubsection{Charm-quark mass uncertainty}
We adjust the simulation charm-quark masses to the physical tuned values before the chiral-continuum fit as described in Sec.~\ref{kappa_tuning_section}.  We propagate the uncertainty in the charm-quark mass correction by including the matrix-element slopes $\mu_i$ ($i$=1--5) and  the shift in the kinetic mass $\Delta(1/(r_1m_2))$ as constrained parameters with prior widths given by the uncertainties listed in Tables~\ref{tab:m2diff} and~\ref{tab:slopes}. The sum of uncertainty contributions from the fit parameters associated with the charm-quark mass adjustment are listed in the column ``$\kappa$ tuning'' in Table~\ref{tbl:ME_fit_error}.

\subsubsection{Renormalization and matching uncertainty}
We include terms of $\order{(\alpha_s^2)}$ in our base chiral-continuum fit with unknown coefficients $\rho^{[2]}_{ij}$ constrained to be of $\order{(1)}$ to incorporate the uncertainty due to omitted higher-order renormalization and matching terms.  The sum of uncertainty contributions from the fit parameters $\rho^{[2]}_{ij}$ are listed in the ``matching'' column in Table~\ref{tbl:ME_fit_error}.  The renormalization and matching uncertainties estimated from fitting our lattice simulation data range from 2.0\% to 4.1\%.  Their values are compatible with the na\'ive estimate obtained from taking $\alpha_s = 0.2$ from our finest  lattice spacing and $\rho^{[2]}_{ij} = 1$, which yields 6.5\% for all operators.

We check that the inclusion of generic $\mathrm{O}(\alpha_s^2)$ terms captures the uncertainty from truncating the perturbative expansion in $\alpha_s$ by performing two alternate fits: one with only the known renormalization and matching terms of $\order{(\alpha_s)}$, and another with terms through $\order{(\alpha_s^3)}$ and coefficients constrained to be of $\order{(1)}$. The results of these fits are labeled ``mNPR'' and ``mNPR+$\alpha_s^3$'', respectively, in Figs.~\ref{fig:chipt_fit_variationO1} and~\ref{fig:chipt_fit_variationO2O5}.  In both cases, the shift in central value is small. Without the  $\order{(\alpha_s^2)}$ terms, the errors on the matrix elements are underestimated.   The errors do not increase from those of the base fit, however, with the inclusion of  $\order{(\alpha_s^3)}$ terms.  We therefore conclude that the base fit includes the uncertainty from omitted higher-order matching and renormalization terms.

We also compare the results of the base fit, in which the matrix elements are renormalized with the mNPR approach, to those from fits in which the matrix elements are renormalized using 1-loop tadpole-improved perturbation theory, and $\alpha_s$ is obtained either from the fourth-root of the plaquette or from the gauge-fixed Landau link.  In these fits, labeled ``$\text{PT}_{\rm P}+\alpha_s^2$''  and ``$\text{PT}_{\rm L}+\alpha_s^2$'', respectively, in Figs.~\ref{fig:chipt_fit_variationO1} and~\ref{fig:chipt_fit_variationO2O5}, we include $\order{(\alpha_s^2)}$ terms constrained as in the base fit. Here we observe larger changes in the matrix-central values, which are still less than 2$\sigma_{\rm fit}$ away from the base-fit results.  The ``$\text{PT}_{\rm P}+\alpha_s^2$''  and ``$\text{PT}_{\rm L}+\alpha_s^2$'' results themselves differ by almost 1$\sigma_{\rm fit}$, indicating a systematic uncertainty in the perturbative matching associated with the choice of tadpole-improvement factor.  This supports our expectation that the mNPR matching approach is more reliable.

\subsubsection{Truncation of the chiral and heavy-meson expansion}
We estimate the uncertainty from truncating the chiral expansion by summing contributions to the matrix-element errors from the NLO LECs $\{\beta_i, \beta_i^\prime\}$ and all of the analytic LECs $\{c_n, d_n\}$ that do not depend on lattice spacing. The results are given in the column labeled ``chiral'' in Table~\ref{tbl:ME_fit_error}.

We check the robustness of the fit error and look for residual truncation effects by considering two variations of the chiral fit function with different sets of analytic terms.  In the first fit, we remove all NNLO analytic terms.  For this fit, we also restrict the matrix-element data included to only those points for which the valence-quark mass $m_q<0.65m_s$, since we expect heavier-mass data to be outside the validity of NLO $\chi$PT.  In the second fit, we add all possible analytic terms of \order{(N$^3$LO)}.  The results of these fits are labeled ``NLO ($m_q<0.65m_s$)'' and ``N$^3$LO'', respectively, in Figs.~\ref{fig:chipt_fit_variationO1} and~\ref{fig:chipt_fit_variationO2O5}.   In both cases, the central values for the matrix elements shift only slightly.  The NLO $\chi$PT fit without NNLO analytic terms underestimates the errors on the matrix elements and also has a significantly larger $\chi^2_{\rm aug}/{\rm dof}$ than the other variations shown.   The errors do not increase from those of the base fit, however, with the inclusion of N$^3$LO analytic terms.   We also consider a fit variation in which we set the hyperfine and flavor splittings  in the chiral logarithms, which are the leading corrections in the $1/M_B$ expansion, to zero. The result is labeled ``no splitting'' in Figs.~\ref{fig:chipt_fit_variationO1} and~\ref{fig:chipt_fit_variationO2O5}.  
Again, the changes in matrix-element central values and errors are small.  All of these tests demonstrate that the base fit includes the uncertainty from higher-order terms in the chiral and heavy-meson expansions.

We also study the impact of our prior constraints on the $\chi$PT LECs, which are based on expectations from chiral power counting.  We perform three fits in which we double the prior widths of (1) the LO LECs; (2) the NLO LECs; or (3)  the NNLO LECs.  The results are labeled ``LO x 2, NLO x 2, NNLO x 2'', respectively,  in Figs.~\ref{fig:chipt_fit_variationO1} and~\ref{fig:chipt_fit_variationO2O5}.  The errors on the matrix elements are stable against increasing the prior widths by a reasonable amount, indicating that the priors are sufficiently unconstraining to allow the data to determine the fit results.

\subsubsection{Light quark discretization errors}
We estimate the uncertainties from light-quark discretization via the uncertainty in the coefficients $\{c_n, d_n\}$ of all analytic terms which depend on the lattice spacing. This error is labeled ``LQ disc'' in Table~\ref{tbl:ME_fit_error}.

The base chiral-continuum fit function includes taste-symmetry breaking effects in the expressions for the chiral logarithms.  Therefore corresponding analytic terms proportional to $a^2 \bar\Delta$ are needed to maintain invariance under variation of the chiral scale.  These terms scale as $\alpha_s^2 a^2$. Generic one-loop contributions from improving the gluon and light-quark actions at tree-level, however, give rise to discretization terms of $\order{(\alpha_s a^2)}$.  We therefore perform an alternate fit including such a term to account for generic light-quark and gluon discretization effects.  The result is labeled ``generic $\mathrm{O}(\alpha_s a^2)$'' in Figs.~\ref{fig:chipt_fit_variationO1} and~\ref{fig:chipt_fit_variationO2O5}, and is indistinguishable from the base-fit result. This indicates that the terms already included in the base chiral-continuum fit function are sufficient to describe the lattice-spacing dependence of the data, and that the base-fit error properly includes light-quark discretization errors.

\subsubsection{Heavy quark discretization errors}
We estimate the uncertainties from heavy-quark discretization via the uncertainty in the coefficients $z_i$ of the $\order{(\alpha_s a, a^2, a^3)}$ terms in Eqs.~(\ref{eq:HQ_a})--(\ref{eq:HQ_a3}).  This error is labeled ``HQ disc'' in Table~\ref{tbl:ME_fit_error}.
We also perform two fits, each of which includes fewer heavy-quark terms than in the base fit.
 These are labeled ``HQ $\mathrm{O}(\alpha_s a)$ only'' and ``HQ $\mathrm{O}(\alpha_s a, a^2)$ only'' in Figs.~\ref{fig:chipt_fit_variationO1} and~\ref{fig:chipt_fit_variationO2O5}, the label referring to the type of terms included in the fit.  The tiny changes in matrix-element central values and errors confirm that the base-fit errors properly include the uncertainty from truncating the heavy-quark expansion.

As a consistency check, we can compare the heavy-quark discretization errors estimated from the data with those based on power counting. We evaluate Eqs.~(\ref{eq:HQ_a})--(\ref{eq:HQ_a3}) taking $\Lambda_{\text{HQET}}=500$~MeV for the heavy-quark scale, and coefficients $z_i$ set by the combinatoric factors $z_E=2\sqrt{2}, z_{B}=2, z_{3}=\sqrt{5}, z_X=2\sqrt{2}, z_Y=2, z_2=2$.  Assuming that all contributions enter with the same sign, this leads to a conservative 5\% estimate for all matrix elements.   This is larger than the data-driven heavy-quark-discretization-error estimates in Table~\ref{tbl:ME_fit_error}, which range from 2.4\% to 3.6\%, but the similar size suggests that errors obtained from the fit are reasonable.

\subsubsection{Relative scale ($r_1/a$) uncertainty}
\label{subsec:RenormErrs}
The relative scale $r_1/a$ is used to convert lattice data on each ensemble to $r_1$ units before the chiral-continuum fit.
We incorporate the uncertainty from $r_1/a$  through the use of prior constraints in the same manner as the parametric ``input'' errors.  The relative scale errors are given in the column labeled ``$r_1/a$" in Table~\ref{tbl:ME_fit_error}.

\subsection{Additional errors}
\label{subsec:OtherErrors}

\subsubsection{Absolute scale ($r_1$) uncertainty}
The absolute lattice scale $r_1$ in fm enters the matrix-element analysis during conversions between lattice-spacing and physical units.  The scale $r_1$ is used to convert the PDG average meson masses and pion decay constant, which are parametric inputs to the chiral-continuum fit, from GeV to $r_1$ units. We account for the error on $r_1$ during the unit conversion.  The scale $r_1$ is also needed to obtain the $D$-mixing matrix elements in GeV.  Because the absolute scale does not affect the minimization of the $\chi^2$ statistic, we add the error from $r_1$ due to this final unit conversion in quadrature to the fit error {\it a posteriori}; the results are listed in column ``$r_1$'' of Table~\ref{tbl:ME_total_error}.

\subsubsection{Finite-volume effects}
We employ the finite-volume expressions for the 1-loop chiral logarithms in the base chiral-continuum fit.  To estimate the systematic uncertainty from finite-volume effects, we perform a second fit using the infinite-volume expressions.  The results are labeled ``no FV" in Figs.~\ref{fig:chipt_fit_variationO1} and~\ref{fig:chipt_fit_variationO2O5}.  The observed shifts in the central value of the matrix elements are approximately half a percent. We take half the value of the matrix-element shifts as the error due to finite-volume effects, noting that this is conservative because we are in fact including NLO finite-volume corrections, and the omitted terms are of NNLO and hence even smaller. The finite-volume errors are listed in the ``FV" column of Table~\ref{tbl:ME_total_error}.

\subsubsection{Isospin breaking and electromagnetism}
We obtain the $D$-meson matrix elements in the chiral-continuum limit by evaluating the valence light-quark mass at the physical $m_u$ and the light sea-quark mass at the isospin average $\hat{m}=(m_u+m_d)/2$.   This accounts for the dominant isospin-breaking effects from the valence sector, but isospin-breaking effects from the sea sector must be included as a systematic uncertainty. Following the analysis of Sec. ~III.B.4 in Ref.~\cite{Bazavov:2016nty}, we estimate isospin-breaking effects to enter at $\mathrm{O}((m_d^{\text{sea}}-m_u^{\text{sea}})^2)$, leading to a negligible $\sim 0.01\%$ uncertainty. The MILC asqtad ensembles employed do not include electromagnetism.  Contributions from dynamical photons would enter at the one-loop level, and we estimate their size to be of $\order{(\alpha_{\text{EM}}/\pi)} \sim 0.2\%$, again following Sec.~VIII.B.4 of Ref.~\cite{Bazavov:2016nty}. We add this error from the omission of electromagnetism in quadrature to the fit error and list it in the ``EM" column of Table~\ref{tbl:ME_total_error}.

\subsection{Omission of the charmed sea quark}
\label{subsec:CharmSea}
The MILC asqtad ensembles do not include dynamical charm quarks in the sea.  The effects of ignoring the charmed sea quark are discussed in detail in Sec.~VIII.C of Ref.~\cite{Bazavov:2016nty}, and are estimated from power counting to be $\sim$1--2\% for reasonable choices of $\alpha_s$ and $\Lambda_{\rm QCD}$.   To be conservative, we take the upper end of the range, or 2\%, for the ``Charm sea" error in Table~\ref{tbl:ME_total_error}. We note, however, that for the decay constants $f_\pi$, $f_K$, and $f_{D_{(s)}}$ where both 3- and 4-flavor simulation results are available, the observed differences are consistent with zero within errors.

\subsection{Other consistency checks and error summary}
\label{subsec:ErrorSummary}

\begin{figure}[t]
	\centering
		\includegraphics[width=0.90\textwidth]{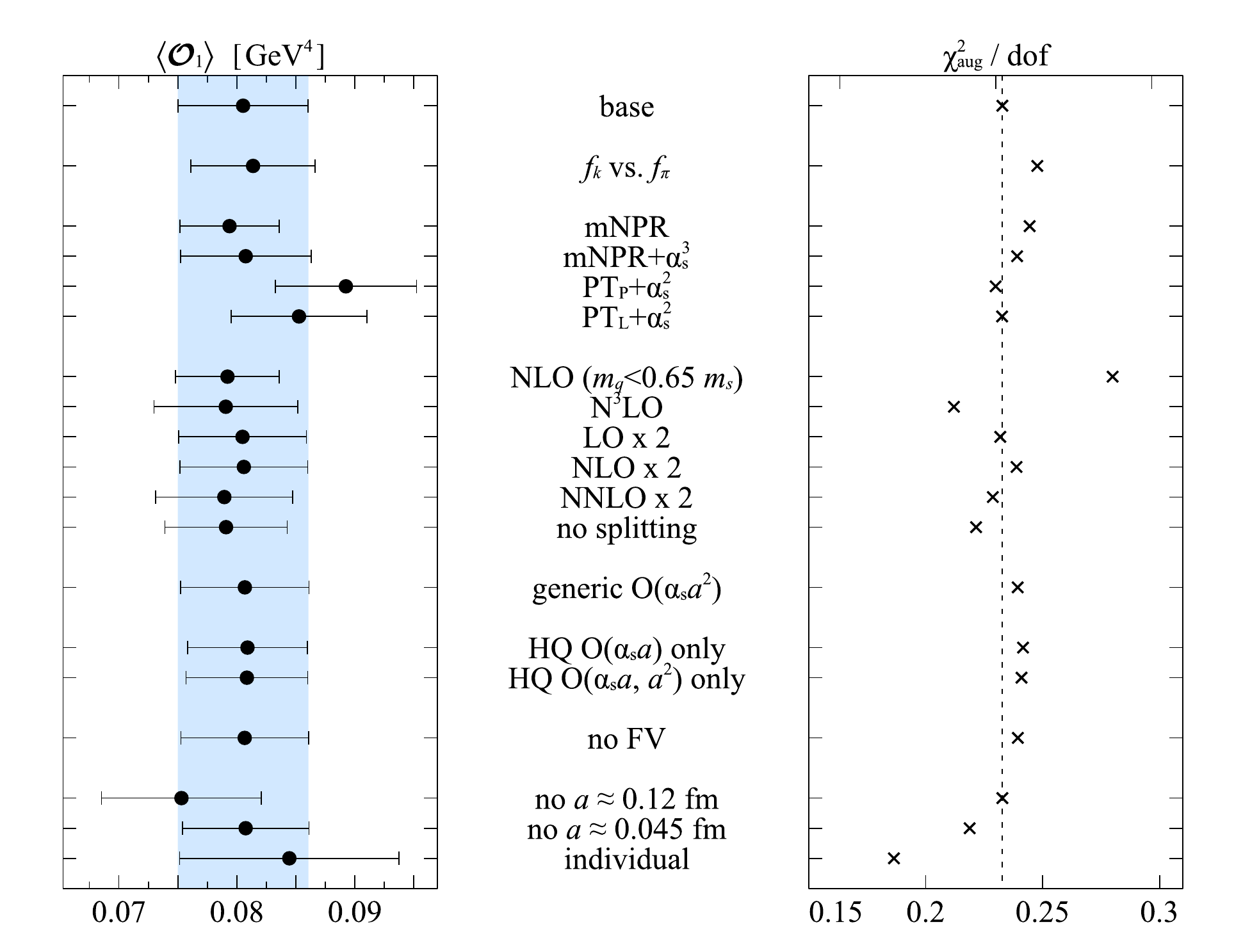}
	\caption{({\it left}) Results for the matrix element $\langle {\mathcal O}_1 \rangle$ from the chiral-continuum fit variations discussed in Secs.~\ref{subsec:FitErrors}--\ref{subsec:ErrorSummary}, and ({\it right}) the corresponding goodness-of-fit. On the left, the base-fit result and error are denoted by the vertical blue band, while on the right, the base-fit $\chi^2_{\rm aug}/{\rm dof}$ is shown by the vertical dashed line. For the ``individual" fit, the  $\chi^2_{\rm aug}/{\rm dof}$ shown is for $\langle {\mathcal O}_1 \rangle$; those for operators $\langle {\mathcal O}_2 \rangle$--$\langle {\mathcal O}_5 \rangle$ are similar.}
	\label{fig:chipt_fit_variationO1}
\end{figure}

\begin{figure}[t]
	\centering
		\includegraphics[width=0.90\textwidth]{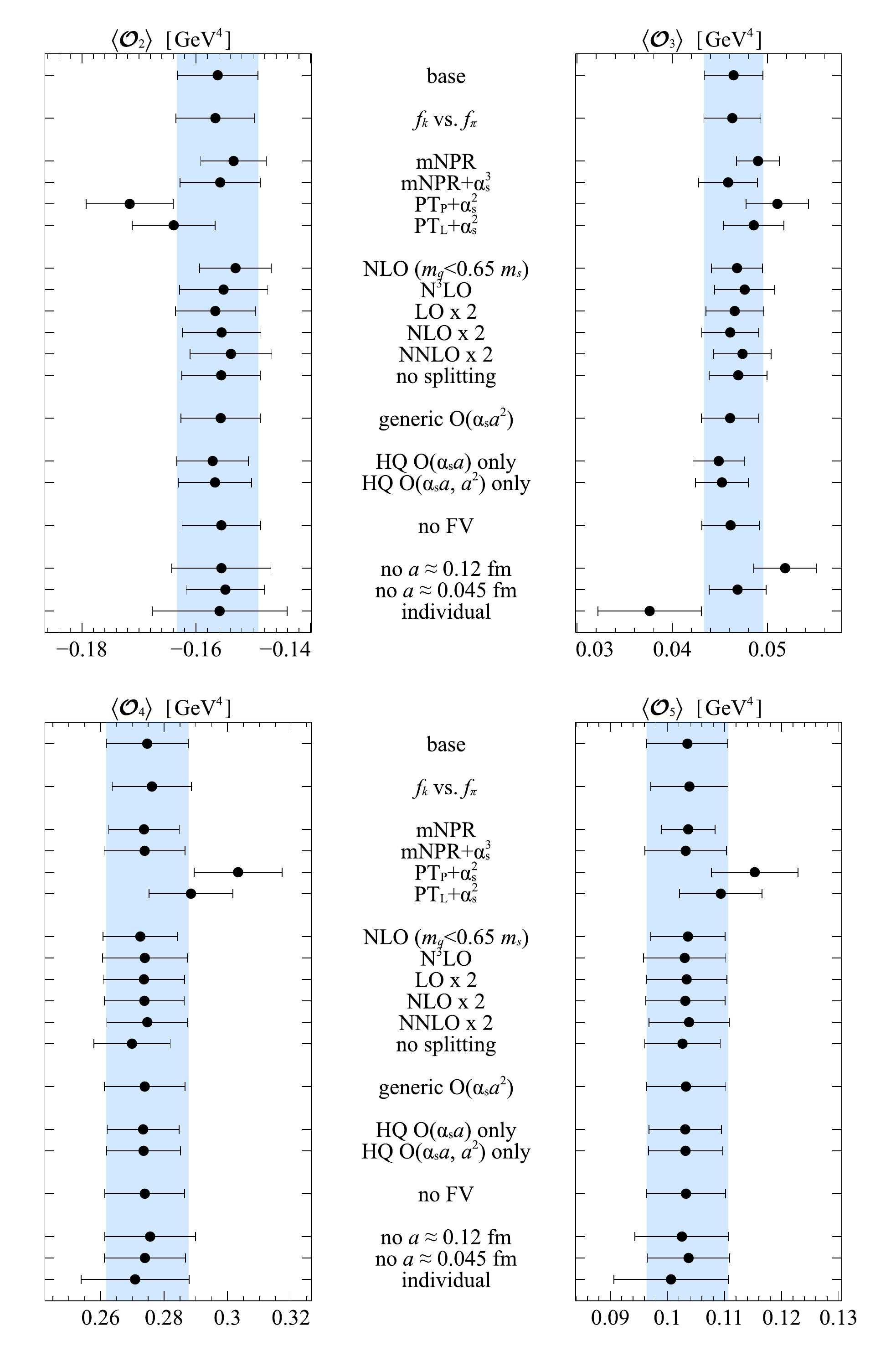}
	\caption{Same as Fig.~\ref{fig:chipt_fit_variationO1} for matrix elements $\langle {\mathcal O}_2 \rangle$--$\langle {\mathcal O}_5 \rangle$.}
	\label{fig:chipt_fit_variationO2O5}
\end{figure}

Finally, we perform fits over various subsets of our data to check for overall consistency and further verify that our base-fit results are reasonable. These are included in Figs.~\ref{fig:chipt_fit_variationO1} and~\ref{fig:chipt_fit_variationO2O5}.
First we perform fits omitting data from the largest or smallest lattice spacing, which are labeled ``$\text{no }a\approx 0.12\text{~fm}$'' and ``$\text{no~}a\approx 0.045~\text{fm}$'', respectively. The resulting matrix-element central values agree with the base fit within 1$\sigma_{\rm fit}$, providing further evidence that our wide range of lattice spacings is sufficient to control discretization errors.  The resulting matrix-element uncertainties are larger, however, which is to be expected because a smaller data set is employed.

Our base-fit results are obtained from a single chiral-continuum fit to the matrix elements of all five operators, including correlations, to optimally constrain the shared LECs and parametric inputs.  To test the impact of the correlations we perform five separate fits of the individual matrix elements; the results are labeled as ``individual".  We observe large shifts in the matrix-element central values, as much as 2$\sigma_{\rm fit}$ in some cases, which are covered by equally substantial increases in the uncertainties for all matrix elements except $\langle {\mathcal O}_3 \rangle$. In this case, the individual-fit error does not quite overlap that of the combined fit. The large errors obtained from the individual fits are associated with the uncertainties on the NLO LECs $\beta^{(\prime)}_j$. The majority LO, NLO and even NNLO LECs are well constrained by the data, while the parametric inputs are tightly constrained by priors.  The LECs $\beta^{(\prime)}_j$, however, cannot be well determined using data for only a single matrix element, and the errors obtained in the individual fits are governed by the loose prior widths.  The base fit resolves the $\beta^{(\prime)}_j$ because these coefficients multiply terms that mix under operator renormalization.

After considering all possible significant sources of uncertainty in the previous sections, Table~\ref{tbl:ME_total_error} presents complete systematic error budgets for all five $D$-meson matrix elements.  The error due to the omission of the charmed sea quark is listed separately after the total because the estimation of this error is far more rough and less quantitative than all others considered.  Further, if more reliable estimates of charmed-sea-quark effects become available in the future, this separation will enable the errors on our results to be easily adjusted {\it a posteriori}.

\begin{table}[tp]
\caption{Total error budgets for $D$-mixing matrix elements.  Entries are in percent. The uncertainty due to the omission of the charmed sea quark is listed separately because the estimated error is significantly less quantitative than that from the other contributions. 
\label{tbl:ME_total_error}
}
\begin{tabular}{c@{\quad}c@{\quad}c@{\quad}c@{\quad}c@{\quad}c@{\quad}c@{\quad}}\hline\hline
                          & Fit total & $r_1$ & FV & EM & Total & Charm sea \\
\hline
$\langle\op_1\rangle$ & 6.4 & 2.1 &0.1 & 0.2 & 6.8 & 2.0\\
$\langle\op_2\rangle$ & 4.0 & 2.1 &0.3 & 0.2 & 4.5 & 2.0\\
$\langle\op_3\rangle$ & 6.3 & 2.1 &0.3 & 0.2 & 6.6 & 2.0\\
$\langle\op_4\rangle$ & 4.2 & 2.1 &0.2 & 0.2 & 4.7 & 2.0\\
$\langle\op_5\rangle$ & 6.5 & 2.1 &0.2 & 0.2 & 6.8 & 2.0\\
\hline\hline
\end{tabular}
\end{table}

\section{Results}
\label{results}

Here we present our final results with complete error budgets including all sources of uncertainty considered in the previous
section. We first give results for the local neutral $D$-meson mixing matrix elements in Sec.~\ref{subsec:MEResults}. In Sec.~\ref{subsec:Pheno}, we discuss how to obtain bounds on generic sources of new physics and also illustrate how to apply our
results to a specific model. In addition, Appendix~\ref{app:MEcorrelations} provides tables of correlations between our matrix-element results, and between our bag-parameter results, so that they can be employed in future phenomenological studies.

\subsection{Matrix elements}
\label{subsec:MEResults}

\begin{table}[tbp]
\caption{$D$-meson mixing matrix elements $\left<\op_i\right>$ in the $\overline{\text{MS}}-\text{NDR}$ scheme at $\mu=3~\text{GeV}$ with total uncertainties. The first uncertainty is the error labeled ``Total'' in Table~\ref{tbl:ME_total_error}, while the second is the estimated error from quenching the charm sea quark. Results are shown for both the BBGLN~\cite{Beneke:1998sy} and BMU~\cite{Buras:2000if} evanescent-operator choices. Entries are in~GeV$^4$.
\label{tbl:result_ME}}
\begin{tabular}{c@{\quad}c@{\quad}c@{\quad}c@{\quad}c@{\quad}c@{\quad}}\hline\hline
                          & $\left<\op_1\right>$ & $\left<\op_2\right>$  & $\left<\op_3\right>$
                          & $\left<\op_4\right>$  & $\left<\op_5\right>$ \\
\hline
BBGLN & 0.0805(55)(16) & -0.1561(70)(31) & 0.0464(31)(9) & 0.2747(129)(55) & 0.1035(71)(21) \\
BMU   & 0.0806(54)(16) & -0.1442(66)(29) & 0.0452(30)(9) & 0.2745(129)(55) & 0.1035(71)(21) \\
\hline\hline
\end{tabular}
\end{table}

Table~\ref{tbl:result_ME} presents our final results for the relativistically normalized $D$-meson mixing hadronic matrix elements of operators $\op_i$ ($i = 1$--5) in Eqs.~(\ref{eq:O1def})--(\ref{intro_bsm}) including statistical and all sources of systematic uncertainty. We give the matrix elements in the $\overline{\text{MS}}$-NDR scheme at the scale $\mu=3~\text{GeV}$ obtained with the BBGLN~\cite{Beneke:1998sy} and BMU~\cite{Buras:2000if} choices of evanescent operators.
Although the choice of evanescent operators affects only the renormalization of operators $\op_2$ and $\op_3$, the numerical results for the other three matrix elements $\left<\op_{1,4,5}\right>$ differ slightly for the BBGLN and BMU schemes, because they are all obtained simultaneously with correlations in the joint chiral-continuum fit. The correlation matrix between the five matrix elements is given in Appendix~\ref{app:MEcorrelations}, Table~\ref{tbl:result_ME_corr}. We quote the uncertainty due to the omission of charm sea quarks separately, because the estimate is semi-quantitative. Final results for the five matrix elements are also provided as supplemental material~\cite{DmixSupplement} with double-precision.

\begin{figure*}[tbp]
	\centering
	\includegraphics[width=\linewidth]{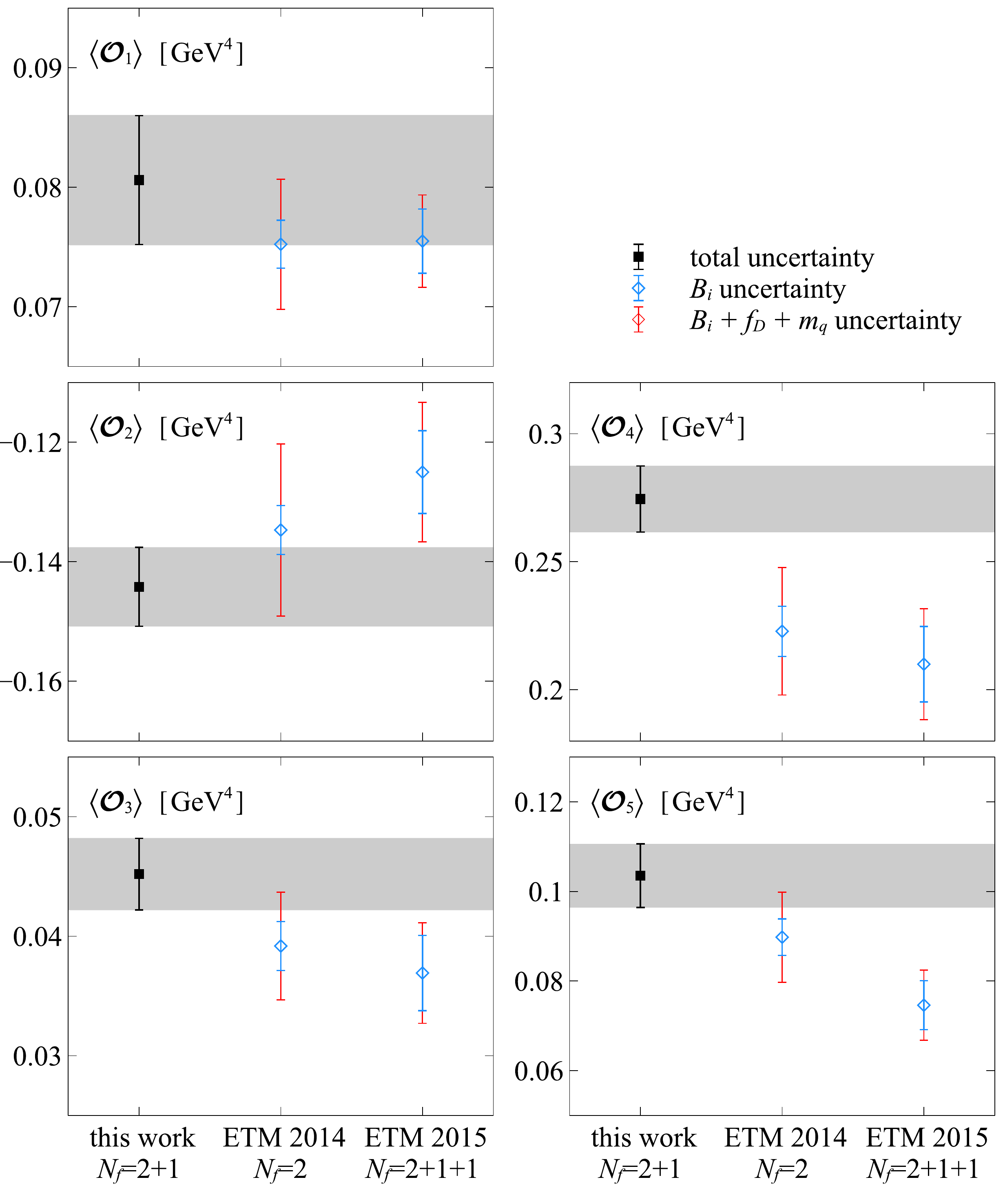}
    \caption{Comparison of the three-flavor $D$-mixing matrix elements obtained in this work (filled symbols) with the two- and four-flavor results from the ETM Collaboration~\cite{Carrasco:2014uya,Carrasco:2015pra} (unfilled symbols). For the ETM results, we have converted their quoted bag parameters to matrix elements using their two- and four-flavor quark masses and decay constants from Refs.~\cite{Blossier:2010cr,Carrasco:2013zta,Carrasco:2014cwa,Carrasco:2014poa}. The total uncertainty quoted from this work does not include the error from quenching the charm sea quark. On the ETM points, the larger red error bars include the uncertainties from $f_D$ and $m_q$ in quadrature, while the smaller blue error bars omit those uncertainties. ETM's two-flavor results do not include an estimate of the error due to quenching the strange sea quark.}
	\label{fig:DmixMECompare}
\end{figure*}

Figure~\ref{fig:DmixMECompare} compares our $D$-mixing matrix-elements with the lattice-QCD results obtained by the ETM
Collaboration using $N_f=2$~\cite{Carrasco:2014uya} and $N_f=2+1+1$~\cite{Carrasco:2015pra} twisted-mass fermions.
ETM presents values for bag parameters, which we then convert to matrix elements using Eq.~(3) of Ref.~\cite{Carrasco:2014uya}, with their own $N_f=2$ and $N_f=2+1+1$ calculations of the quark masses~\cite{Blossier:2010cr,Carrasco:2014cwa} and decay constants~\cite{Carrasco:2013zta,Carrasco:2014poa}. In each panel, the red error bars are obtained by propagating the decay-constant and quark-mass uncertainties in quadrature, while the blue error bars omit those uncertainties.
This second approach may provide a rough picture of the errors on the matrix elements, given typical correlations between the 
three- and two-point correlation functions in the numerator and denominator. Our results for $\langle\op_1\rangle$, $\langle\op_2\rangle$, $\langle\op_3\rangle$ agree with the matrix elements converted from ETM's bag parameters to within about
1--2$\sigma$, but those for $\langle\op_4\rangle$ and $\langle\op_5\rangle$ differ by 1.7--3.3$\sigma$ (assuming ETM's quark-mass and decay-constant errors to be negligible). We find, however, that different choices for the quark masses and decay constants yield considerable variations in the converted matrix elements.  If we convert the ETM bag parameters using the average quark masses and decay constants from the PDG~\cite{Rosner:2015wva,Olive:2016xmw}, we obtain matrix element results that agree with ours within 0.25--1.52$\sigma$ for all five operators.  Use of the averages from the Flavor Lattice Averaging Group~\cite{Aoki:2016frl} instead yields matrix-element results that lie in between these two determinations.  We are planning to perform a correlated, combined analysis of the $D$-mixing matrix elements from this work with our collaboration's $D$-meson decay constants calculated using the same lattice ensembles and parameters~\cite{Bazavov:2013wia,Kronfeld:2015xka}.  We will present the resulting the $D$-meson bag parameters in our forthcoming decay-constant paper.  This will enable a more direct comparison with ETM's results. 

Similar tensions to those shown in Fig.~\ref{fig:DmixMECompare} have been observed for the analogous neutral-kaon-mixing bag parameters $B_4$ and $B_5$~\cite{Bertone:2012cu,Carrasco:2015pra,Jang:2015sla,Garron:2016mva}, which the authors of Ref.~\cite{Garron:2016mva} attribute to the choice of intermediate renormalization scheme. Results obtained in Ref.~\cite{Garron:2016mva} using the symmetric regularization-independent momentum-subtraction (RI-SMOM) scheme~\cite{Sturm:2009kb} are in good agreement with an independent calculation that employs one-loop mean-field improved lattice
perturbation theory \cite{Jang:2015sla}, but differ substantially with results obtained with the RI-MOM~\cite{Martinelli:1994ty}
renormalization scheme. This discrepancy is attributed to underestimated systematic errors present in the RI-MOM scheme~\cite{Garron:2016mva,Sturm:2009kb}. We note that operator $\op_4$ has the largest negative anomalous dimension of the five $\Delta C=2$ operators, by about 50\%, and is therefore most sensitive to running of the renormalization scale. Because $\op_4$ and $\op_5$ mix under renormalization, this also affects the results for $\langle \op_5\rangle$. Thus, $\op_4$ and $\op_5$ are likely to be the most sensitive to difficulties in renormalization.

\subsection{Implications for new physics}
\label{subsec:Pheno}

As discussed in Sec.~\ref{theory_background}, neutral $D$-meson mixing is a sensitive probe of local $\Delta C=2$ interactions from physics beyond the Standard Model that contribute to the quantity $M_{12}$. The phase $\phi_{12}$ is particularly sensitive to new physics due to the CKM suppression of its contribution from the Standard Model. Assuming that any new physics does not change the phase of $\Gamma_{12}$, which is true in many BSM models~\cite{Golowich:2009ii}, then the imaginary part of $M_{12}$ gives the most sensitive constraint on new physics. Here we use our matrix-element results to bound the scale of new physics in a generic model that alters the high-scale Wilson coefficients and in a specific flavor-violating Higgs model with tree-level flavor-changing neutral currents.

A general new-physics model will give nonzero values for some of the Wilson coefficients $C_i$ at a high scale $\LambdaNP$.
To evaluate the contribution to $M_{12}$ using Eq.~(\ref{eq:M12NP}), we must run the Wilson coefficients down to the scale
$\mu=3$~GeV at which our matrix elements are renormalized. We use the one-loop running derived in Ref.~\cite{Golowich:2009ii}, which works in a different four-fermion operator basis than the one used here.  Fierz identities relate the two bases~\cite{Bouchard:2011yia}:
\begin{align}
    Q_1 &= \op_1, \\
    Q_2 &= -2\op_5, \\
    Q_3 &= \op_4, \\
    Q_4 &= \op_2, \\
    Q_5 &= -4\op_3 - 2\op_2.    
\end{align}
The remaining operators $Q_6$, $Q_7$ and $Q_8$ are parity conjugates of $Q_1$, $Q_4$ and $Q_5$. Applying this change of basis to the formulas given in Appendix~A of Ref.~\cite{Golowich:2009ii} yields the renormalization equations in our basis, which we then use directly. The operator running depends on the value of the strong coupling $\alpha_s$ at intermediate scales. We use the \texttt{RunDec} Mathematica library \cite{Chetyrkin:2000yt} to compute $\alpha_s$ with four-loop running, including quark decoupling effects.

\begin{figure}
    \centering
    \includegraphics[width=0.7\textwidth]{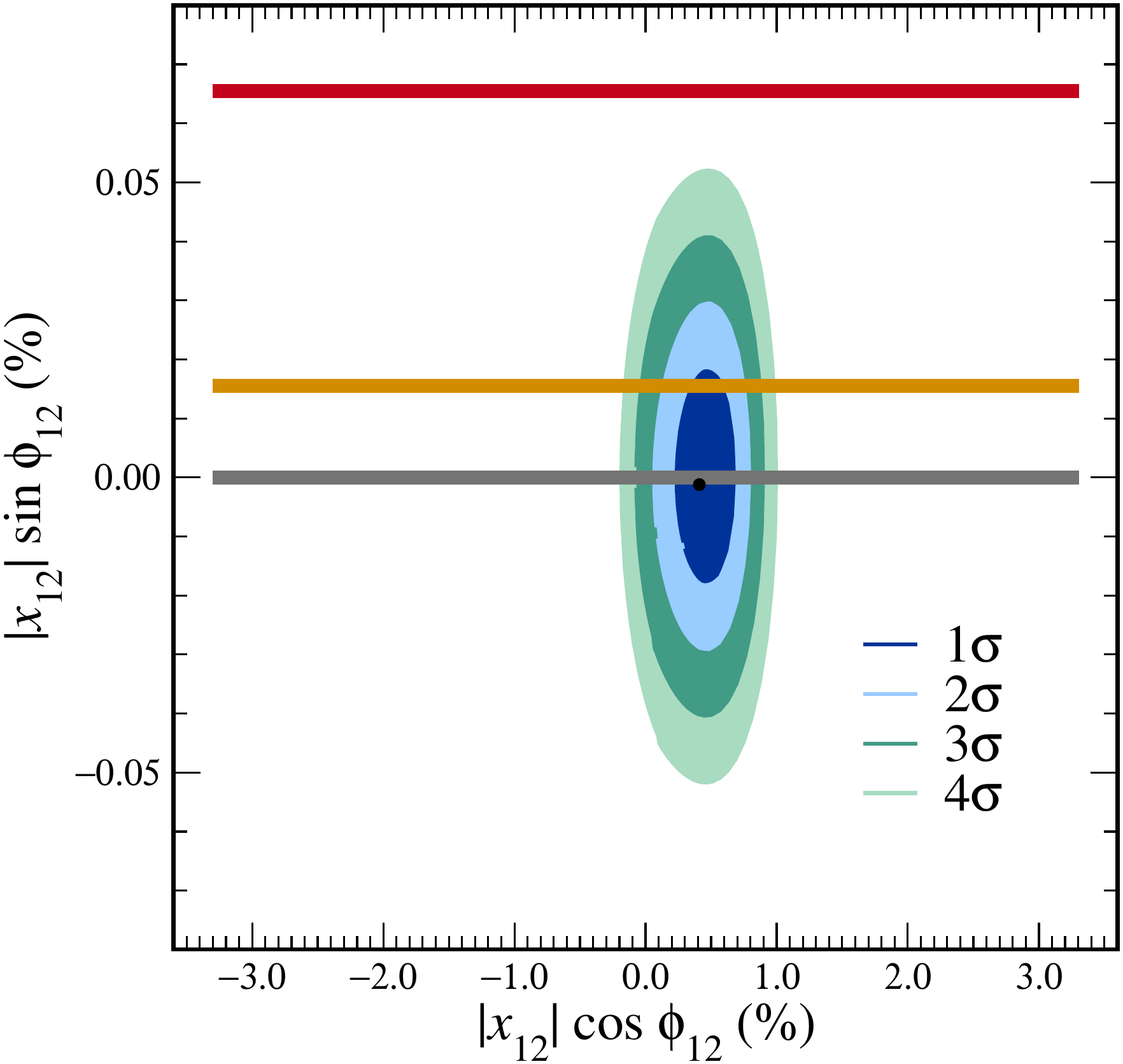}
    \caption{$|x_{12}|e^{i\phi_{12}}$ plotted as a complex number. With no new physics, the Standard-Model estimate (gray bar) is compatible with the experimental best-fit contours (blue regions). The other two bars show predictions for the complex $x_{12}$ in two specific new-physics scenarios, corresponding to the simple ``$\op_5$-only" model described in the text. The gold bar shows the SM+NP region with model parameter $\LambdaNP=40\,000$~TeV, while the red bar shows the choice $\LambdaNP=18\,000$~TeV. The former value is compatible with current experimental bounds, while the latter is ruled out.}.
    \label{fig:x12}
\end{figure}

To illustrate the use of our results in constraining physics beyond the Standard Model, we first consider a simple new-physics model which gives rise only to operator $\op_5$ at an ultraviolet scale $\LambdaNP$, with purely imaginary (and hence CP-violating) Wilson coefficient
\begin{equation}
    \Im C_5^{\text{NP}}(\LambdaNP) = \frac{1}{\LambdaNP^2}.
\end{equation}
Running down to 3~GeV, we find 
\begin{align}
    \Im C_4^{\text{NP}}(3\ \text{GeV}) &= \frac{r^{-4} - r^{1/2}}{3 \LambdaNP^2}, \\ 
    \Im C_5^{\text{NP}}(3\ \text{GeV}) &= \frac{r^{1/2}}{\LambdaNP^2},
\end{align}
where 
\begin{equation}
    r = \left(\frac{\alpha_s(\LambdaNP)}{\alpha_s(m_t)} \right)^{2/7}
        \left(\frac{\alpha_s(m_t)}{\alpha_s(m_b)} \right)^{6/23}
        \left(\frac{\alpha_s(m_b)}{\alpha_s(3\ \textrm{GeV})} \right)^{6/25},
\end{equation}
and all other Wilson coefficients at 3~GeV remain zero.  Thus, operators $\op_4$ and $\op_5$ contribute a purely imaginary term to
\begin{equation}
    x^\text{NP}_{12} = \frac{1}{M_D \Gamma_D} \sum_i C_i^{\text{NP}}(\text{3~GeV})\langle\op_i\rangle
\end{equation}
from new physics.
(Here, $x^\text{NP}_{12}$ is defined as in Sec.~\ref{theory_background}, $x^\text{NP}_{12}=M^\text{NP}_{12}/\Gamma_D$.) We can now determine the constraint on $\LambdaNP$ in this model by summing the Standard Model and new physics contributions, and then comparing to experimental bounds, as depicted in Fig.~\ref{fig:x12}. The gray box shows the region in which the Standard Model value for $x_{12}$ is expected to lie. The gold box includes a contribution from our simple scenario with $\LambdaNP=40\,000$~TeV; this level of new contribution is consistent with the experimental bounds at 1$\sigma$.
Finally, the red box shows a new-physics contribution with $\LambdaNP = 18\,000$~TeV, which is ruled out by experiment with very high confidence. \footnote{Note that if the Wilson coefficient $C_5^{\text{NP}}(\LambdaNP)$ were purely real instead of purely imaginary, it would have to be much larger in magnitude to give a SM + NP region that lies outside the experimental contour, leading to a much weaker bound (on the order of 2000~TeV in this simple scenario).}

We now apply this technique to place bounds on the scale of generic new-physics contributions to each operator. In this case, the Wilson coefficients are of the form
\begin{equation}
    C_i^\text{NP}(\LambdaNP) = \frac{F_i L_i}{\Lambda_{i,\text{NP}}^2},
\end{equation}
where $F_i$ and $L_i$ are flavor and loop-counting factors \cite{Bona:2007vi}. In order to obtain bounds on $\LambdaNP$ individually by operator, we set one $C_i^\text{NP}(\Lambda_{i,\text{NP}})=F_i L_i/\Lambda_{i,\text{NP}}^2$ at a time, with all other Wilson coefficients set to zero at the high scale. We then determine the value of $\Lambda_{i,\text{NP}}$ for which the new-physics prediction for $|x_{12}| e^{i\phi_{12}}$ is inconsistent with the experimental bound at 95\% confidence, including the uncertainty in the matrix elements for $\left< \op_i \right>$. Here we assume that $\Im F_i L_i > 0$, but taking the other sign gives a nearly identical constraint. (The experimental bound is $\Im x_{12}<0.0289\%$ ($\Im x_{12}>-0.0285$) in the positive (negative) imaginary direction.)

Using our matrix elements in the BBGLN scheme, we obtain

\begin{align}
    \Lambda_{1,\text{NP}} &\gtrsim \left(\Im F_1L_1\right)^{1/2} \times \hphantom{2}7\,630~\text{TeV}, \label{eq:NP1} \\
    \Lambda_{2,\text{NP}} &\gtrsim \left(\Im F_2L_2\right)^{1/2} \times 24\,100~\text{TeV}, \\
    \Lambda_{3,\text{NP}} &\gtrsim \left(\Im F_3L_3\right)^{1/2} \times 23\,100~\text{TeV}, \\
    \Lambda_{4,\text{NP}} &\gtrsim \left(\Im F_4L_4\right)^{1/2} \times 48\,500~\text{TeV}, \\
    \Lambda_{5,\text{NP}} &\gtrsim \left(\Im F_5L_5\right)^{1/2} \times 26\,900~\text{TeV}, \label{eq:NP5}
\end{align}
Note that these bounds are from the imaginary part of $x_{12}$ only; for new physics with $\Im F_iL_i\approx0$, the constraint on $\LambdaNP$ from the $\Re x_{12}$ will dominate.

Our bounds in Eqs.~(\ref{eq:NP1})--(\ref{eq:NP5}) are stronger than those quoted by the ETM Collaboration~\cite{Carrasco:2014uya}, in part because we use more recent, tighter experimental bounds. For operators $\op_3$ and $\op_5$ in particular, our constraints on $\LambdaNP$ are much higher (by factors of roughly 7 and 3, respectively). These two operators mix strongly with $\op_2$ and $\op_4$, respectively, such that their bounds stem principally from $C_2(3~\text{GeV})\op_2(\text{3~GeV})$ and $C_4(3~\text{GeV})\op_4(\text{3~GeV})$. If we artificially set $C_2(3~\text{GeV})=0$, then we obtain much weaker bounds of $\Lambda_{3,\text{NP}} \gtrsim 3\,330$~TeV and $\Lambda_{5,\text{NP}}\gtrsim 9\,700$~TeV, respectively, which are close to the values quoted by ETM in Ref.~\cite{Carrasco:2014uya}.

\begin{figure}
	\centering
	\includegraphics[width=0.7\textwidth]{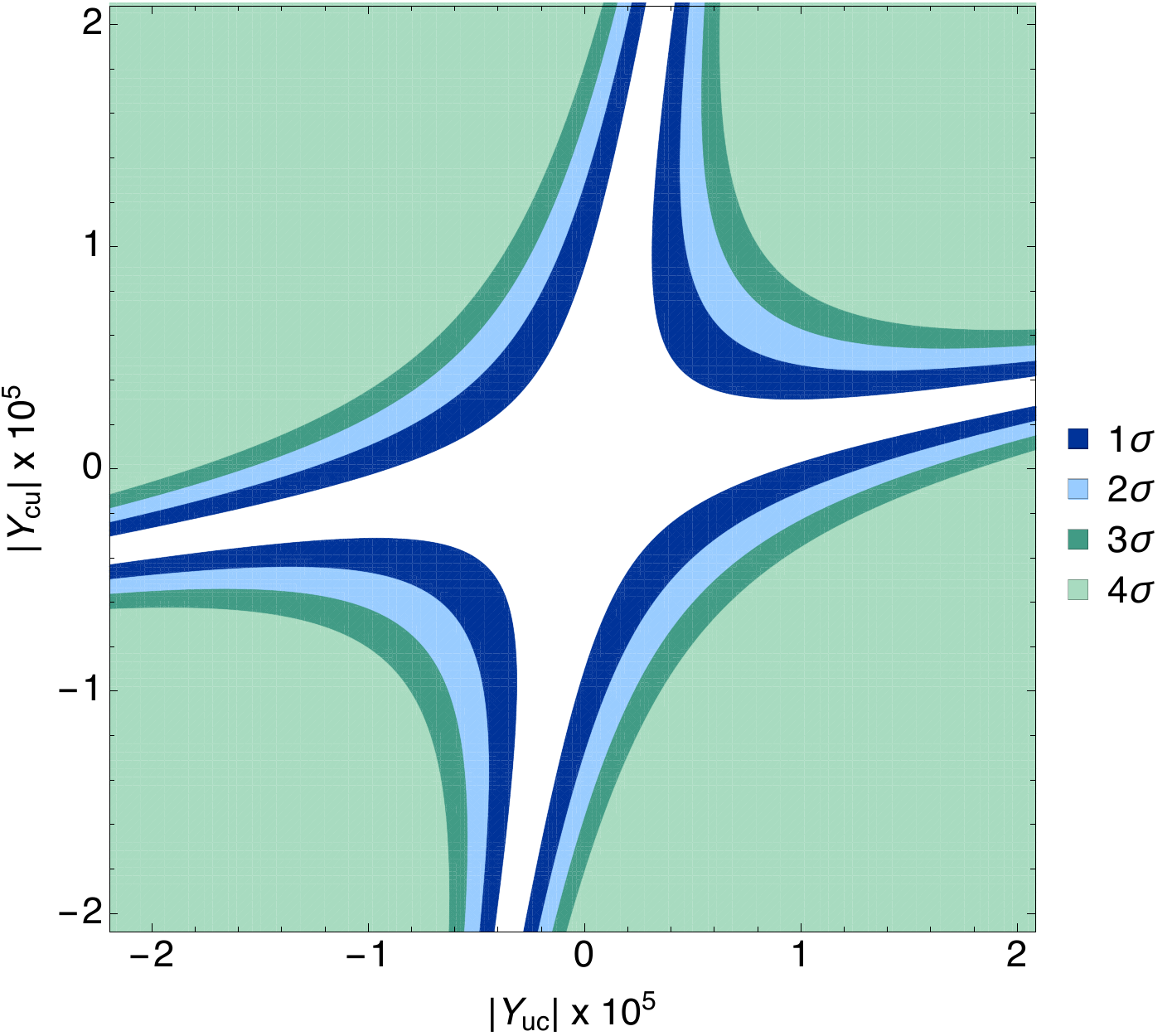}
    \caption{Exclusion contours from $D^0$ mixing on the flavor-violating Higgs couplings $|Y_{uc}|$ and $|Y_{cu}|$ in the model of Ref.~\cite{Harnik:2012pb}, with the Yukawa coupling phases marginalized over as described in the text.}
	\label{fig:x12_Higgs}
\end{figure}

To further illustrate the use of our results in constraining new physics, we examine a specific model in which the Standard Model
Higgs boson has flavor-violating couplings to quarks and leptons~\cite{Harnik:2012pb}. A Higgs coupling of the form $Y_{uc} h \bar{u}_L c_R + Y_{cu} h \bar{c}_L u_R$ will induce $\Delta C = 2$ four-fermion interactions at low energy. After integrating out the Higgs boson~$h$, the effective Hamiltonian is
\begin{equation}
    \mathcal{H}^\text{NP}_{\Delta C=2} = C_2^{uc}(m_h) \op_2 + \tilde{C}_2^{uc}(m_h) \tilde{\op}_2 + C_4^{uc}(m_h) \op_4,
\end{equation}
where the Wilson coefficients at the scale $m_h$ are given by
\begin{align}
    C_2^{uc}(m_h) &= -\frac{Y_{uc}^*{}^2}{2m_h^2}, \\
    \tilde{C}_2^{uc}(m_h) &= -\frac{Y_{cu}^2}{2m_h^2}, \\
    C_4^{uc}(m_h) &= -\frac{Y_{cu} Y_{uc}^*}{m_h^2}.
\end{align}
We write the two Yukawa couplings as $Y_{uc} = |Y_{uc}| e^{i\phi_{uc}}$, and similarly for $Y_{cu}$. Taking the Wilson coefficients to be purely imaginary and comparing the resulting $\Im x_{12}$ to the experimental bounds, we find from the $\op_2$ and $\op_4$ terms separately the bounds
\begin{align}
    |Y_{uc}|^2 + |Y_{cu}|^2  &\lesssim 1.04 \times 10^{-10}, \label{eq:Y1} \\
    |Y_{cu} Y_{uc}|          &\lesssim 2.15 \times 10^{-11}. \label{eq:Y2}
\end{align}
To be somewhat more general, we can instead marginalize over the phases in the couplings (integrating from 0 to $\pi/2$) and obtain exclusion contours in the $|Y_{uc}|$-$|Y_{cu}|$ plane, again solely from the stronger experimental bound on $\Im x_{12}$. The contours are shown in Fig.~\ref{fig:x12_Higgs}. Our bounds on the combinations of Yukawa couplings in Eqs.~(\ref{eq:Y1})--(\ref{eq:Y2}) are tighter than those in Ref.~\cite{Harnik:2012pb} by over an order of magnitude.  The improvement stems primarily from the use of newer experimental measurements than employed by the authors of Ref.~\cite{Harnik:2012pb}, who relied on the same data as the original model-independent analysis of $D$-meson mixing~\cite{Ciuchini:2007cw}. We also explicitly run the Wilson coefficients from $m_h$ down to 3~GeV in order to compare to experiment, rather than attempting to obtain ``model-independent" bounds on the Wilson coefficients at a generic high scale.

\section{Summary and outlook}
\label{sec:conclusions}

In this paper we have presented a three-flavor lattice-QCD calculation of the neutral $D$-meson mixing matrix elements of all five $\Delta C=2$ dimension six local four-fermion operators. We obtain uncertainties comparable to those from earlier $N_f=2$ and $2+1+1$ calculations by the European Twisted Mass Collaboration~\cite{Carrasco:2014uya,Carrasco:2015pra}.
Our results for $\langle D^0|\op_i|\bar{D}^0\rangle$ ($i=1$--3) agree with those of ETM to within about 1--2 standard deviations, but those for $i=4$ and 5 differ more significantly. These short-distance matrix elements are needed to evaluate the first term on the right hand side of Eq.~(\ref{eq:M12G12}), and can be combined with experimental measurements of the $D$-mixing parameters to yield useful constraints on theories beyond the Standard Model with sizable CP violation because $\phi_{12}$ is very small in the Standard Model. To illustrate the utility of our matrix-element results, we place bounds on
generic new high-scale physics that would give rise to local $\Delta C=2$ interactions, finding $\LambdaNP{(\Im F_i L_i)^{-1/2}}\gtrsim10$--$50\times10^3$~TeV for the five local $\Delta C=2$ operators. These results are more stringent than previous bounds, in part because we use the latest experimental measurements, and in part because of the way we introduce new physics at the high scale, $\LambdaNP$, and run down to 3~GeV.

The long-distance contributions described by the second term on the right hand side of Eq.~(\ref{eq:M12G12}) are expected to
dominate the Standard Model prediction of the $D^{{0}}$-meson mixing observables, $\Delta M$ and $\Delta\Gamma$,
but their size is not well known because they are difficult to calculate. Thus, despite the current precision of experimental measurements---and forthcoming improvements from the BES~III, LHCb, and Belle~2 experiments---our results and those of Refs.~\cite{Carrasco:2014uya,Carrasco:2015pra} will suffice for phenomenology until better methods for the long-distance contributions become available.

Clearly, reliable methods to compute the long-distance parts of $M_{12}$ and $\Gamma_{12}$ are needed. These must account for contributions from intermediate multi-hadron states that can propagate over  hadronic distances. Fortunately, the theoretical framework for obtaining transition amplitudes, scattering lengths, and phase shifts for two-hadron systems in a finite spatial volume is already well developed~\cite{Luscher:1986pf,Luscher:1990ux,Rummukainen:1995vs,Lellouch:2000pv,Kim:2005gf,Christ:2005gi}, and the number of lattice-QCD calculations of these systems is growing rapidly.
The first lattice-QCD calculations for systems with many open channels have been recently performed by the Hadron Spectrum
Collaboration~\cite{Dudek:2014qha,Wilson:2014cna,Dudek:2016cru,Wilson:2015dqa,Moir:2016srx}. These techniques have been extended to neutral kaon mixing, and the first lattice-QCD calculations have recently become available~\cite{Bai:2014cva,Christ:2015pwa}. There, long-distance effects are easier to quantify, because phase space suppresses all but the two-pion intermediate states. Efforts are underway by several groups to develop an analogous theoretical framework for three-hadron systems~\cite{He:2005ey,Hansen:2012tf,Briceno:2012yi,Polejaeva:2012ut,Briceno:2012rv,Briceno:2014oea,Briceno:2014uqa,Hansen:2014eka,Briceno:2015csa,Hansen:2015zga,Hansen:2016fzj,Hansen:2016ync,Briceno:2017tce}, which are relevant for $D^0$ mixing. More recently, Hansen, Meyer, and Robaina proposed a new method to extract the spectral function for multi-hadron transition rates from finite-volume four-point correlation functions~\cite{Hansen:2017mnd}. In the coming years, these new approaches may be implemented in numerical lattice-QCD calculations and lead to more quantitative estimates of long-distance contributions to neutral $D$-meson mixing.

\begin{acknowledgments}

We thank Alan Schwartz, Vittorio Lubicz, Silvano Simula, Roni Harnik, and Joachim Kopp for useful correspondence.
Computations for this work were carried out with resources provided by the USQCD Collaboration, the National
Energy Research Scientific Computing Center and the Argonne Leadership Computing Facility, which are funded
by the Office of Science of the U.S.\ Department of Energy; and with resources provided by the National
Institute for Computational Science and the Texas Advanced Computing Center, which are funded through the
National Science Foundation's Teragrid/XSEDE Program.
This work was supported in part by the U.S.\ Department of Energy under Grants
No.~DE-FG02-91ER40628 (C.B.)
No.~DE-FC02-06ER41446 (C.D.)
No.~DE{-}SC0010120 (S.G.),
No.~DE{-}SC0010005 (E.T.N.), No.~DE-FG02-91ER40661 (S.G., R.Z.), No.
DE-FG02-13ER42001 (C.C.C., D.D., A.X.K.), No.
DE{-}SC0015655 (A.X.K.), No.~DE-FG02-13ER41976 (D.T.); by the U.S.\ National Science Foundation under Grants No. PHY10-67881 and
No.~PHY14-14614 (C.D.), No.~PHY14-17805~(D.D., J.L.), and No.~PHY13-16748 and No.~PHY16-20625 (R.S.); by the Fermilab Fellowship in Theoretical
Physics (C.M.B., C.C.C.); by the URA Visiting Scholars' program (C.M.B., C.C.C., D.D., A.X.K.); by the MINECO (Spain) under Grant No.~FPA2013-47836-C-1-P (E.G.); by the Junta de Andaluc\'ia (Spain) under Grants No.~FQM-101 and No.~FQM-6552 (E.G.); by the European
Commission (EC) under Grant No.~PCIG10-GA-2011-303781 (E.G.); by the German Excellence Initiative and the European Union Seventh
Framework Program under Grant Agreement No.~291763 as well as the European Union's Marie Curie COFUND program (A.S.K.).
Brookhaven National Laboratory is supported by the Department of Energy under Contract No.\ DE{-}SC0012704.
Fermilab is operated by Fermi Research Alliance, LLC, under Contract No.\ DE-AC02-07CH11359 with the United States Department of
Energy, Office of Science, Office of High Energy Physics.
This document was prepared by the Fermilab Lattice and MILC Collaborations using the resources of the Fermi National Accelerator Laboratory (Fermilab), a U.S.\ Department of Energy, Office of Science, HEP User Facility.

\end{acknowledgments}

\appendix
\section{Correlations among \boldmath$r_1/a$ data}
\label{app:r1acorrelations}
Table~\ref{tab:r1a_correlation} provides the correlations among values of the relative scale $r_1/a$ on the ensembles, which are needed to convert quantities from lattice units to $r_1$ units prior to performing the chiral-continuum extrapolation. Correlations between $r_1/a$ arise after performing a smoothing fit simultaneously to all ensembles.

\begin{table}
\centering
    \begin{sideways}
    \begin{minipage}{0.93\textheight}
    \caption{Correlations between the relative scale $r_1/a$ on the ensembles used in this work. The central values and errors are given in Table~\ref{tab:MILC_ensemble}. Entries are symmetric across the diagonal.}
    \label{tab:r1a_correlation}
    \begin{tabular}{lcrrrrrrrrrrrrrrr}
\hline
\hline
$\approx a$ (fm) &  & \multicolumn{4}{c}{0.12}  & \multicolumn{5}{c}{0.09} & \multicolumn{4}{c}{0.06} & \multicolumn{1}{c}{0.045} & \multicolumn{1}{c}{0} \\
& $m_l^\prime / m_s^\prime $ & 0.10 & 0.14 & 0.20 & 0.40 & 0.05 & 0.10 & 0.14 & 0.20 & 0.40 & 0.10 & 0.14 & 0.20 & 0.40 & 0.20 & 0.40 \\
\hline
0.12   & 0.10 & $1.000$ &              &              &             &              &              &              &              &              &              &              &              &              &              &\\ 
          & 0.14 & $1.000$ &$1.000$ &              &             &              &              &              &              &              &              &              &              &              &              &\\ 
          & 0.20 & $1.000$ &$1.000$ &$1.000$ &             &              &              &              &              &              &              &              &              &              &              & \\
          & 0.40 & $0.978$ &$0.978$ &$0.978$ &$1.000$&              &              &              &              &              &              &              &              &              &              & \\
0.09   & 0.05 & $-0.177$&$-0.177$&$-0.177$&$0.006$&$1.000$ &              &              &              &              &              &              &              &              &              &\\ 
          & 0.10 & $-0.174$&$-0.174$&$-0.174$&$0.008$&$1.000$ &$1.000$ &              &              &              &              &              &              &              &              &\\
          & 0.14 & $-0.170$&$-0.170$&$-0.170$&$0.010$&$0.999$ &$1.000$ &$1.000$ &              &              &              &              &              &              &              & \\
          & 0.20 & $-0.165$&$-0.166$&$-0.166$&$0.013$&$0.998$ &$0.999$ &$1.000$ &$1.000$ &              &              &              &              &              &              & \\ 
          & 0.40 & $-0.139$&$-0.139$&$-0.139$&$0.032$&$0.987$ &$0.990$ &$0.993$ &$0.995$ &$1.000$ &              &              &              &              &              & \\
0.06   & 0.10 & $0.538$ &$0.538$ &$0.538$ &$0.410$&$-0.441$&$-0.425$&$-0.409$&$-0.391$&$-0.306$&$1.000$ &              &              &              &              & \\
          & 0.14 & $0.540$ &$0.540$ &$0.540$ &$0.411$&$-0.446$&$-0.430$&$-0.414$&$-0.396$&$-0.312$&$1.000$ &$1.000$ &              &              &              & \\ 
          & 0.20 & $0.541$ &$0.541$ &$0.541$ &$0.412$&$-0.451$&$-0.436$&$-0.419$&$-0.401$&$-0.317$&$1.000$ &$1.000$ &$1.000$ &              &              & \\
          & 0.40 & $0.543$ &$0.544$ &$0.544$ &$0.413$&$-0.460$&$-0.445$&$-0.429$&$-0.411$&$-0.327$&$1.000$ &$1.000$ &$1.000$ &$1.000$ &              & \\
0.045 & 0.20 & $0.572$ &$0.572$ &$0.572$ &$0.428$&$-0.583$&$-0.570$&$-0.555$&$-0.540$&$-0.463$&$0.983$ &$0.984$ &$0.985$ &$0.987$ &$1.000$ &\\
0        & 0.40 & $-0.155$&$-0.155$&$-0.155$&$0.029$&$0.994$ &$0.994$ &$0.995$ &$0.995$ &$0.988$ &$-0.426$&$-0.431$&$-0.436$&$-0.445$&$-0.569$&$1.000$\\
\hline\hline
\end{tabular}
\end{minipage}
\end{sideways}
\end{table}

\section{Priors for two-point and three-point fits}
\label{app:2pt3ptpriors}

Table~\ref{tab:corrparams} provides the priors employed in the joint two- and three-point correlator fits discussed in Sec.~\ref{sec:correlator_analysis}. The parameters $E_0$ and $Z_n$ are defined in Eq.~(\ref{eq:twoptfitfunction}), $\Delta_{k,j}$ is defined in Eq.~(\ref{eq:Esplit}), while the parameters $\mathcal{Z}^{\latop_i}_{nm}$, $i=1$--5, are defined in Eqs.~(\ref{eq:threeptfitfunction})--(\ref{eq:3ptAmpDef}). For the energy splitting the prior is defined such that $\exp(\Delta_{k,j})$ is Gaussian distributed.

\begin{table*}[tbp]
\caption{Priors employed in correlator fits in lattice-spacing units. 
Each prior is a Gaussian distribution with a central value and a one-sigma width, given in parentheses next to each central value.
The uppermost panel shows constraints on energies of the two- and three-point correlators.
The next panel gives constraints on the  amplitudes of the two-point correlators.
The lower two panels show additional constraints on the amplitude of the three-point correlators. All priors are in lattice units.}
\label{tab:corrparams}
\begin{tabular}{c@{\quad}c@{\quad}c@{\quad}c@{\quad}}
\hline
\hline
 $\approx a$ (fm) & $E_0$ & $\exp(\Delta_{1,0})$ & $\exp(\Delta_{k,k-2})$  \\
 \hline\\[-4.5mm]
$0.12$ & $1.0(0.1)$ & $-1.2(0.5)$ & $-1.0(1.5)$ \\
$0.09$ & $0.75(0.05)$ & $-1.6(0.5)$ & $-1.2(2.0)$ \\
$0.06$ & $0.56(0.04)$ & $-2.0(1.0)$ & $-1.6(2.0)$ \\
$0.045$ & $0.42(0.04)$ & $-2.3(1.0)$ & $-1.9(2.0)$ \\
\hline\\[-4.0mm]
 $\approx a$ (fm) & $Z_0$ & $Z_1$ & $Z_n$\\
\hline\\[-4.5mm]
$0.12$ &  $1.45(0.08)$ & $0.7(0.5)$ & $0.7(1.0)$\\
$0.09$ & $1.42(0.08)$ & $0.7(0.5)$ & $0.7(1.0)$\\
$0.06$ & $1.37(0.1)$ & $0.7(0.5)$ & $0.7(1.0)$\\
$0.045$ & $1.36(0.1)$ & $0.7(0.5)$ & $0.7(1.0)$\\
\hline\\[-4.0mm]
  $\approx a$ (fm) & $\mathcal{Z}^{\latop_1}_{00}$ & $\mathcal{Z}^{\latop_2}_{00}$ & $\mathcal{Z}^{\latop_3}_{00}$\\
\hline\\[-4.5mm]
$0.12$ & $0.025(1.0)$ & $-0.055(0.02)$ & $0.012(0.006)$\\
$0.09$ & $0.009(0.003)$ & $-0.02(0.006)$ & $0.005(0.002)$ \\
$0.06$ & $0.003(0.0015)$ & $-0.007(0.003)$ & $0.0017(0.0005)$\\
$0.045$ & $0.0013(0.0008)$ & $-0.0035(0.0015)$ & $0.0008(0.0003)$\\
\hline\\[-4.0mm]
 $\approx a$ (fm) & $\mathcal{Z}^{\latop_4}_{00}$ & $\mathcal{Z}^{\latop_5}_{00}$ & $\mathcal{Z}^{\latop_i}_{nm}$ \\
\hline\\[-4.5mm]
$0.12$ & $0.1(0.03)$ & $0.04(0.015)$ & $0.0(0.1)$\\
$0.09$ & $0.04(0.01)$ & $0.016(0.005)$ & $0.0(0.05)$\\
$0.06$ & $0.015(0.005)$ & $0.006(0.002)$ & $0.0(0.02)$\\
$0.045$ & $0.007(0.002)$ & $0.0027(0.0008)$ & $0.0(0.01)$\\
\hline
\hline
\end{tabular}
\end{table*}

\section{Correlations among matrix-element results}
\label{app:MEcorrelations}
Table~\ref{tbl:result_ME_corr} provides the correlations among the $D$-meson mixing matrix elements for all five operators,
to enable their use in future phenomenological studies.

\begin{table}[tbp]
\caption{Correlations between the $D$-meson mixing matrix elements in the $\overline{\rm{MS}}$-NDR-BBGLN scheme given in Table~\ref{tbl:result_ME}; entries are symmetric across the diagonal.  Correlations for the BMU scheme differ by less than 4\%.   The correlations include contributions from statistics and all systematics except the ``charm sea" error, which is less well quantified. We suggest that an error of 2\% be taken on all sums or differences of matrix elements and 0.5\% on all ratios.}
\label{tbl:result_ME_corr}
\begin{tabular}{c|@{\quad}c@{\quad}c@{\quad}c@{\quad}c@{\quad}c@{\quad}}\hline\hline
& $\left<\op_1\right>$ & $\left<\op_2\right>$ & $\left<\op_3\right>$ & $\left<\op_4\right>$ & $\left<\op_5\right>$\\\hline
$\left<\op_1\right>$ & 1.0 &  &  &  &  \\
$\left<\op_2\right>$ & $-$0.2323 & 1.0 & & & \\
$\left<\op_3\right>$ & 0.0864 &  $-$0.2513 & 1.0 &  & \\
$\left<\op_4\right>$ & 0.2153 & $-$0.3334 & 0.2163 & 1.0 &  \\
$\left<\op_5\right>$ & 0.1865 & $-$0.2246 & 0.1384 & 0.2574 & 1.0 \\
\hline\hline
\end{tabular}
\end{table}

\newpage
\bibliographystyle{apsrev4-1}
\bibliography{bibliography}

\end{document}